# Genomic regulatory architecture of human embryo retroviral LTR elements affecting evolution, development, and pathophysiology of Modern Humans.

Gennadi V. Glinsky[1]

[1] Institute of Engineering in Medicine

University of California, San Diego

9500 Gilman Dr. MC 0435

La Jolla, CA 92093-0435, USA

Correspondence: gglinskii@ucsd.edu

genlighttech@gmail.com

Web: Guennadi V Glinskii (ucsd.edu)

## Abstract

Evolution created two distinct families of pan-primate endogenous retroviruses, namely HERVL and HERVH, which infected primates' germline, colonized host genomes, and evolved into the global retroviral genomic regulatory dominion (GRD) operating during human embryogenesis. Retroviral GRD constitutes 8839 highly conserved fixed LTR elements linked to 5444 down-stream target genes forged by evolution into a functionally-consonant constellation of 26 genome-wide multimodular genomic regulatory networks (GRNs), each of which is defined by significant enrichment of numerous single gene ontology (GO)-specific traits. Locations of GRNs appear scattered across chromosomes to occupy from 5.5%-15.09% of the human genome. Each GRN harbors from 529-1486 human embryo retroviral LTRs derived from LTR7, MLT2A1, and MLT2A2 sequences that are quantitatively balanced according to their genome-wide abundance. GRNs integrate activities from 199-805 down-stream target genes, including transcription factors, chromatin-state remodelers, signal-sensing and signal-transduction mediators, enzymatic and receptor binding effectors, intracellular complexes and extracellular matrix elements, and cell-cell adhesion molecules. GRN's compositions consist of several hundred to thousands smaller GO enrichment-defined genomic regulatory modules (GRMs), each of which combines from a dozen to hundreds LTRs and down-stream target genes. Overall, this study identifies 69,573 statistically significant retroviral LTR-linked GRMs (Binominal FDR q-value < 0.001), including 27,601 GRMs validated by the single GO-specific directed acyclic graph (DAG) analyses across 6 GO annotations. Gene Set Enrichment Analyses (GSEA) of retroviral LTRs down-stream targets employing >70 genomics and proteomics databases (DB), including a large panel of DB developed from single-cell resolution studies of healthy and diseased human's organs and tissues, demonstrate that genes assigned to distinct GRNs and GRMs appear to operate on individuals' life-span timescale along specific phenotypic avenues selected from a multitude of GO-defined and signaling pathways-guided frameworks to exert profound effects on patterns of transcription, protein-protein interactions (PPI), developmental phenotypes, physiological traits, and pathological conditions of Modern Humans. GO analyses of Mouse phenotype DB and GSEA of the MGI Mammalian Phenotype Level 4 2021 DB revealed that down-stream regulatory targets of human embryo retroviral LTRs are enriched for genes making essential contributions to development and functions of all major tissues, organs, and organ systems, that were corroborated by documented numerous developmental defects in a single gene KO models. Genes comprising candidate down-stream regulatory targets of human embryo retroviral LTRs are engaged in PPI networks that have been implicated in pathogenesis of human common and rare disorders (3298 and 2071 significantly enriched records, respectively), in part, by impacting PPIs that are significantly enriched in 1783 multiprotein complexes recorded in the NURSA Human Endogenous Complexome DB and 6584 records of virus-host PPIs documented in Virus-Host PPI P-HIPSTer 2020 DB. GSEA-guided analytical inference of the preferred cellular targets of human embryo retroviral LTR elements supported by analyses of genes with species-specific expression mapping bias in Human-Chimpanzee hybrids identified Neuronal epithelium, Radial Glia, and Dentate Granule Cells as cell-type-specific marks within a Holy Grail sequence of embryonic and adult neurogenesis. Decoded herein the hierarchical super-structure of retroviral LTR-associated GRD and GRNs represents an intrinsically integrated developmental compendium of thousands GRMs congregated on specific genotype-phenotype associations. Highlighted specific GRMs may represent the evolutionary selection units which are driven by inherent genotype-phenotype associations affecting primate species' fitness and survival by exerting control over mammalian offspring survival genes implicated in reduced fertility and infertility phenotypes. Mechanistically, programmed activation during embryogenesis and ontogenesis of genomic constituents of human embryo retroviral GRD coupled with targeted epigenetic silencing may guide genome-wide heterochromatin patterning within nanodomains and topologically-associated domains during differentiation, thus affecting 3D folding dynamics of linear chromatin fibers and active transcription compartmentalization within interphase chromatin of human cells.

**List of abbreviations**

HERVH, human endogenous retrovirus type H;

HERVL, human endogenous retrovirus type L;

LTR7, long terminal repeat 7;

MLT2A1, long terminal repeat MLT2A1;

MLT2A2, long terminal repeat MLT2A2;

GREAT, Genomic Regions Enrichment of Annotations Tool;

GSEA, gene set enrichment analyses;

NHP. Non-human primates;

MYA, million years;

GRD, genomic regulatory dominion;

GRN, genomic regulatory networks;

GRM, genomic regulatory module;

GRP, genomic regulatory pathways;

HSRS, human-specific regulatory sequences;

DEGs, differentially expressed genes;

ECA, extinct common ancestor;

lncRNA, long non-coding RNA;

## Introduction

In the genome of Modern Humans, there are thousands of genomic loci origin of which could be mapped with high confidence to hundreds of distinct families and subfamilies of retroviruses collectively defined as Human Endogenous Retroviruses, HERV (Bao et al., 2015; Kojima, 2018; Vargiu et al., 2016). It has been suggested that colonization of human genome by HERVs is the result of long-term reinfections during primate evolution, specifically, germ-line reinfections, rather than retrotransposition in cis or complementation in trans (Belshaw et al. 2004), indicating that an infectious pool of endogenous retroviruses has persisted within the primate lineage throughout the past 30-40 million years. The full-length retroviruses typically consist of two or three structural genes encoding viral proteins flanked by two identical regulatory sequences termed Long Terminal Repeats (LTRs). Unique structures of retroviral LTRs might provide the evolutionary advantage for hosts for the exaptation of these sequences to function as powerful genomic regulatory elements (Thompson et al. 2016). Homologous recombination mediated by LTRs frequently results in the elimination of internal open reading frames (Mager and Goodchild 1989; Belshaw et al. 2007) and yields solitary retroviral LTRs without adjacent viral structural genes. These retroviral LTRs represent endogenous genomic regulatory elements, which function as alternative promoters, transcription factor binding sites, proximal and distal long-range enhancers, and splice sites.

Selection of specific integration sites of retroviruses in the human genome may reflect an evolutionary compromise of stochastic and deterministic forces because it favors active genes and integrated retroviruses appear preferentially detected in the open chromatin regions representing hallmarks of transcriptionally active genomic loci (Schröder et al, 2002; Cereseto and Giacca, 2004; Bushman et al., 2005; Wang et al., 2007; Albanese et al., 2008), however, within this genomic context precise insertion sites of ERVs may have elements of a random and probabilistic nature. Conservative estimates indicate that there are more than 3,000 DNA sequences derived from insertions of the human endogenous retrovirus type H (HERVH), including highly conserved sequences of its promoter LTR7, that are scattered across human genome at fixed non-polymorphic locations (Thomas et al., 2018). Sequences derived from LTR7/HERVH retroviruses are among the most abundant and extensively investigated genomic regulatory elements that were originated from various HERV families. Retroviral LTR7/HERVH insertions originated from a gamma retrovirus that presumably infected primates approximately 40 million years ago (mya) and colonized the genome of the common ancestor of Apes, Old World Monkeys, and New World Monkeys (Goodchild et al., 1993; Mager and Freeman, 1995).

The long terminal repeats designated LTR7 harbor the promoter sequence of the HERVH retroviruses and the LTR7/HERVH family has been extensively investigated in the context of its regulatory functions and locus-specific differential expression in human preimplantation embryogenesis, human embryonic and pluripotent stem cells (Fort et al., 2014; Gemmell et al., 2015; 2019; Glinsky et al., 2018; Göke et al.,2015; Izsvák et al., 2016; Kelley and Rinn, 2012; Loewer et al., 2010; Lu et al., 2014; Ohnuki et al., 2014; Pontis et al., 2019; Römer et al., 2017; Santoni et al., 2012; Takahashi et al., 2021 Theunissen et al., 2016; Wang et al., 2014; Zhang et al., 2019). It has been reported that LTR7/HERVH sequences harbor binding sites for master pluripotency transcription factors OCT4, NANOG, SP1, and SOX2 which bind HERVH/LTRs and activate their expression (Glinsky, 2015; Göke et al., 2015; Ito et al., 2017; Kelley and Rinn, 2012; Kunarso et al., 2010; Ohnuki et al., 2014; Pontis et al., 2019; Santoni et al., 2012).

Human endogenous retrovirus type K [HERVK (HTML-2)] is the primate-specific retrovirus that most recently colonized primate genomes and all human-specific and human-polymorphic HERVK insertions are associated with a specific LTR subtype designated LTR5_Hs (Hanke et al., 2016; Subramanian et al., 2011). There are at least 36 non-reference polymorphic HERVK proviruses with insertion frequencies ranging from <0.0005 to >0.75 that varied by distinct human populations (Wildschutte et al., 2016). In contrast to HERVH insertions manifesting no insertional polymorphism in human population (Thomas et al., 2018), insertional polymorphism of HERVK retroviruses has been documented by the analyses of individual human genomes, including genomes from the 1000 Genomes Project and the Human Genome Diversity Project (Autio et al., 2021; Hughes, Coffin, 2004; Barbulescu et al., 1999; Belshaw et al., 2005; Shin et al., 2013; Turner et al., 2001; Wildschutte et al., 2016).

It has been reported that LTR5_Hs/HERVK manifests transcriptional and biological activities in human preimplantation embryos and in naïve hESCs (Grow et al., 2015). Notably, LTR5_Hs elements acquire enhancer-like chromatin state signatures concomitantly with transcriptional reactivation of HERVK (Grow et al., 2015). Genome-wide CRISPR-guided activation and interference experiments targeting LTR5_Hs elements demonstrated global long-range effects on expression of human genes consistent with postulated functions of LTR5_Hs as distal enhancers (Fuentes et al., 2018). Recently reported several lines of observations highlight putative mechanisms by which regulatory LTRs of distinct retroviral families may have affected phenotypic traits contributing to species segregation during primate evolution and development of human-specific phenotypic features (Glinsky, 2022). Notable findings of this category are discoveries of potential regulatory impacts of retroviral LTRs residing in the human genome on expression of mammalian offspring survival (MOS) genes as well as regulatory effects of human-specific LTR7 and LTR5_Hs loci on expression of genes encoding markers of 12 distinct cells' populations of fetal gonads and genes implicated in pathophysiology of human spermatogenesis, including Y-linked spermatogenic failure, oligo- and azoospermia. Most recently, direct experimental evidence linking specific retroviral LTR loci (LTR7Y and LTR5_Hs) to gene regulatory networks shared between human primordial germ cells and naïve pluripotent cells has been reported (Ito et al., 2022).

On a population scale, to achieve the successful colonization of host genomes and heritable transmission of retroviral sequences, retroviruses must infect the germline during the early embryogenesis prior to or at the stage of the germ cells' biogenesis. The germline infection, propagation, and stable integration of multiple copies of viral genomes into host chromosomes are required to ensure the passage of integrated viral sequences to offspring. These could be accomplished only within a relatively small window of the developmental timeline which represents an essential temporal target for viral infections to enable successful transitions of exogenous retroviruses to the state of endogenous retroviruses (ERVs) integrated into host genomes (Glinsky, 2022). Based on these considerations, it has been concluded that HERVs colonizing the human genome during the relatively narrow embryonic development window may target functionally closely related panels of genes and developmental pathways, thus potentially affecting common sets of phenotypic traits. Recently reported analyses of human embryo regulatory LTR elements (LTR7 and LTR5_Hs) support the model of cooperative phenotypic impacts on human pathophysiology exerted by genes representing down-stream regulatory targets of different families of retroviral LTR elements, such as LTR7 and LTR5_Hs, despite their markedly distinct evolutionary histories of the human genome colonization spanning millions of years (Glinsky, 2022).

During embryonic genome activation (EGA), DNA sequences derived from HERVL retroviruses are transcriptionally activated by the homeobox transcription factor DUX4 (Hendrickson et al. 2017). HERVL retroviruses harbor regulatory LTRs designated MLT2A elements, which are subdivided into two distinct subfamilies designated MLT2A1 and MLT2A2. Chromatin at MLT2A loci becomes accessible in two-cell human embryos, MLT2A elements are transcribed in four-cell human embryos, and specific MLT2A elements function as promoters and splicing sites in primate embryos (Hashimoto et al., 2021). These observations indicate that MLT2A1 and MLT2A2 elements that are derived from primate-specific HERVL retroviruses could be classified as human embryo regulatory LTRs.

To understand when HERVH and HERVL retroviruses have infiltrated the primates' germline, detailed analyses of evolutionary conservation patterns of retroviral LTR7, MLT2A1, and MLT2A2 elements were carried out in 17 species of non-human primates (NHP) vis-à-vis reference genome of present day humans. According to these analyses, HERVH and HERVL represent pan-primate endogenous retroviruses that appear to infect and colonize primate lineage at different evolutionary time points and achieved in distinct primate species quantitatively different levels of genomic amplification during evolution. To delineate connectivity patterns between 8839 loci encoded by the LTR7 and MLT2A elements and the phenotypic pleiotropy of putative regulatory functions of distinct families of human embryo retroviral LTRs, the GREAT algorithm was employed to identify and characterize LTR-linked genes. Comprehensive Gene Set Enrichment Analyses (GSEA) of 5444 LTR-linked genes were performed to infer potential phenotypic impacts of genomic regulatory networks governed by human embryo retroviral LTRs. Observations reported in this contribution support the hypothesis that evolution of human embryo retroviral LTR elements created the global genomic regulatory dominion (GRD) consisting of 26 genome-wide genomic regulatory networks (GRNs) defined by the significant enrichment of specific gene ontology terms. Retroviral LTR-associated GRD and GRNs could be viewed as an intrinsically integrated developmental compendium of 69,573 genomic regulatory modules (GRMs) defined by significant enrichment of specific genotype-phenotypic trait associations. Many GRMs may represent the evolutionary selection units driven by inherent genotype-phenotype associations affecting mammalian species' fitness and survival with the focal selection point on nervous system development and functions. Collectively, these findings demonstrate that despite markedly distinct evolutionary histories of different families of human embryo retroviral LTRs, genes representing down-stream regulatory targets of LTR7 and MLT2A elements exert the broad cooperative phenotypic impacts on development, physiology, and pathology of Modern Humans, apparently encompassing diverse cellular compositions and structural-functional architectures of all tissues, major organs, and organ systems of human body.

## Results

### Evolutionary patterns of origin, expansion, conservation, and divergence of pan-primate endogenous retroviruses encoding human embryo retroviral LTR elements.

Evolutionary timelines of transitions of retroviruses from the state of exogenous infection agents to endogenous retroviral sequences integrated into host genomes could be inferred from the comparative quantitative analyses of highly conserved orthologous retrovirus-derived Long Terminal Repeat (LTR) loci in genomes of multiple distinct primate species which diverged from the last extinct common ancestor (ECA) at distinct time points during primates' evolution (Glinsky, 2022). To accomplish this task, a total of 8839 fixed non-polymorphic sequences of human embryo retroviral LTRs residing in genomes of Modern Humans (hg38 human reference genome database) were retrieved and the number of highly conserved orthologous loci in genomes of sixteen non-human primates (NHP) were determined (**Figure 1**). Present analyses were focused on human embryo retroviral LTR loci (see Introduction) the evolutionary origins of which could be mapped to genomes of Old World Monkeys (OWM), indicating that these genomic loci could be defined as highly-conserved pan-primate regulatory sequences because they have been preserved in primates' genomes for ~30 MYA. Three distinct LTR families meet these criteria, namely MLT2A1 (2416 loci), MLT2A2 (3069 loci), and LTR7 (3354 loci). It has been observed that MLT2A1, MLTA2, and LTR7 appear to have similar evolutionary histories defined by evolutionary patterns of expansion, conservation, and divergence during primate evolution (**Figure 1A**). However, they have clearly distinguishable histories of timelines of evolutionary origins, which are defined herein as the estimated timelines when retroviruses infect primates' germlines and transition from the state of exogenous infection agents to endogenous retroviral sequences integrated into host genomes (**Figure 1A**). The consistent earliest presence of highly-conserved pan-primate MLT2A1 and MLT2A2 loci could be mapped to genomes of both New World Monkeys (NWM) and OWM, indicating that both MLT2A1/HERVL and MLT2A2/HERVL retroviruses have infected primates' germlines before the segregation of OWM and NWM lineages (**Figure 1A**). In contrast, the consistent earliest presence in primates' genomes of LTR7 loci could be mapped to genomes of OWM (**Figure 1A**), suggesting that LTR7/HERVH retroviruses have entered germlines of the primate lineage after the separation of the NWM lineage (Glinsky, 2022). Pan-primate endogenous MLT2A1/HERVL, MLT2A2/HERVL, and LTR7/HERVH retroviruses appear to achieve a significant expansion in genomes of Great Apes, ranging from ~2-fold expansion for MLT2A2 loci to ~5-fold expansion for LTR7 loci (**Figure 1A**). In contrast to highly-conserved pan-primate human embryo retroviral LTRs, the consistent earliest presence of highly-conserved LTR5_Hs loci has been observed in the Gibbon's genome (Glinsky, 2022), in accord with the hypothesis that LTR5_Hs/HERVK retroviruses successfully colonized germlines of the primate lineage after the segregation of Gibbons' species and subsequently underwent a marked expansion in genomes of Great Apes.

In agreement with the previously reported findings for highly-conserved LTR7 and LTR5_Hs loci (Glinsky, 2022), timeline sequences of quantitative expansion of human embryo pan-primate retroviral LTRs in different primate species appear to replicate the consensus evolutionary sequence of increasing cognitive and behavioral complexities during primate evolution, which seems particularly striking for LTR7 loci (**Figure 1A**).

One of notable findings of LTR loci sequence conservation analyses is the observation that hundreds of highly-conserved pan-primate human embryo retroviral LTRs have DNA sequences divergent by more than 5% from orthologous sequences in genomes of Modern Humans in genomes of our closest evolutionary relatives, Chimpanzee and Bonobo. It was of interest to determine how many of fixed non-polymorphic human embryo retroviral LTR loci could be defined as candidate human-specific (HS) genomic regulatory elements compared to both Chimpanzee and Bonobo genomes. Setting a selection threshold at <10% of sequence identity, there are 45; 35; and 175 sequences that could be classified as human-specific MLT2A2, MLT2A1, and LTR7 loci, respectively (**Figure 1B**). A majority of candidate HS retroviral LTR loci of human embryo could be classified as bona fide human-specific insertions because they did not intersect any chains in genomes of both Chimpanzee and Bonobo.

Results of previously reported evolutionary conservation analyses indicated that nearly half (84/175; 48%) of candidate HS-LTR7 loci have been identified as highly-conserved orthologous sequences in genomes of OWM, Gibbon, Orangutan, and Gorilla (Glinsky, 2022), consistent with the hypothesis that these HS-LTR7 loci were not retained in genomes of Chimpanzee and Bonobo but preserved in genomes of Modern Humans. Similar to HS-LTR7 loci, the patterns of failure to retain human-specific DNA sequences have been observed in genomes of both Chimpanzee and Bonobo for candidate HS regulatory LTR elements encoded by MLT2A1 and MLT2A2 loci (**Figure 1B**). Collectively, these observations strongly argue that species-specific loss of highly-conserved pan-primate regulatory DNA sequences encoded by retroviral LTRs may have contributed to the emergence during primate evolution of human-specific features of genomic regulatory networks (GRN).

**Identification of retroviral LTR's down-stream target genes and analysis of potential phenotypic impacts of human embryo retroviral LTR elements on pathophysiology of Modern Humans.**

One of the main mechanisms defining biological functions of retroviral LTRs is their activity as distal genomic regulatory elements (GRE) affecting transcription of down-stream target genes. Therefore, the inference of potential biological functions of human embryo retroviral LTR elements could be made based on evaluations of experimentally documented biological activities of their down-stream target genes. To this end, the Genomic Regions Enrichment of Annotations Tool (GREAT) algorithm was employed (Methods) to identify putative down-stream target genes of human embryo retroviral LTRs. Concurrently with the identification of putative regulatory target genes of GREs, the GREAT algorithm performs stringent statistical enrichment analyses of functional annotations of identified down-stream target genes, thus enabling the inference of potential biological significance of interrogated GRNs. Importantly, the assignment of phenotypic traits as putative statistically valid components of GRN actions entails the assessments of statistical significance of the enrichment of both GREs and down-stream target genes.

The GREAT algorithm identified 5444 genes as potential down-stream regulatory targets of 8839 human embryo retroviral LTR elements (**Tables 1-3; Supplementary Table S1**). It has been observed that distinct families of human embryo retroviral LTR elements share overlapping sets of down-stream target genes, including 709 genes that were identified as the consensus set of candidate down-stream regulatory targets for MLT2A1; MLT2A2; and LTR7 elements (**Table 1; Supplemental Table S1**). At the statistical significance threshold of the binominal FDR Q value < 0.002 (**Table 2)**, a total of 54 significantly enriched records of biological processes appear preferentially affected by the regulatory actions of 8839 human embryo retroviral LTRs (**Table 2; Supplementary Table S1**). Cell-cell adhesion via plasma-membrane adhesion molecules (GO:0098742; 677 LTRs linked with 95 genes; binominal FDR Q value = 2.65E-15), cell-cell adhesion (GO:0098609; 1089 LTRs linked with 238 genes; binominal FDR Q value = 3.18E-05), and homophilic cell adhesion (GO:0007156; 481 LTRs linked with 62 genes; binominal FDR Q value = 1,15E-12) via plasma membrane adhesion molecules have been identified among top-ranked biological processes associated with regulatory effects of human embryo retroviral LTR elements (**Table 2; Supplementary Table S1**).

Interestingly, neuron cell-cell adhesion (GO:0007158) has been scored as one of significantly enriched cell type-specific GO categories (106 LTRs linked to 10 genes; binominal FDR Q-value = 9.09E-04). Multiple distinct GO categories related to biogenesis, structure, and functions of synapses appear preferential regulatory targets of human embryo retroviral LTRs (**Table 2; Supplementary Table S1**). These include regulation of synapse assembly (GO:0051963); regulation of synapse structure or activity (GO:0050803); synapse organization (GO:0050807); synapse assembly (GO:0007416); and synapse membrane adhesion (GO:0099560). Several findings suggest that putative regulatory actions of human embryo retroviral LTRs may affect fine-tuned biological structures and activities of synapses, including the presynaptic membrane assembly (GO:0097105; binominal FDR Q value = 7.42E-04) and organization (GO:0097090; 77 LTRs linked to 5 genes; binominal FDR Q value = 1.16E-03); postsynaptic membrane assembly (GO:0097104; 86 LTRs linked to 6 genes; binominal FDR Q value = 3.19E-04) as well as positive regulation of synapse assembly (GO:0051965; 355 LTRs linked to 40 genes; binominal FDR Q value = 4.83E-09). Notably, retroviral LTR elements and genes associated with the GO category of positive regulation of synapse assembly (GO:0051965) constitute a marked majority of retroviral LTRs and genes assigned to more general GO category of regulation of synapse assembly (GO:0051963): 91% (355/392) of LTRs and 78% (40/51) of genes, respectively (**Table 2; Supplementary Table S1**). Overall, the GO category of regulation of nervous system development (GO:0051960; binominal FDR Q value = 4,38E-02) has been linked with 1424 regulatory LTRs and 716 target genes, 333 of which have been implicated in development of nervous system (**Supplementary Table S1)**.

Collectively, these findings suggest that biological processes of nervous system development and functions, in particular, biogenesis and activity of synapses, may represent one of important evolutionary targets of regulatory actions of human embryo retroviral LTR elements. Consistent with this hypothesis, significantly enriched GO category of positive regulation of nervous system development (GO:0051962; binominal FDR Q value = 5.95E-04) has been linked with 983 regulatory LTRs and 204 target genes (**Table 2; Supplementary Table S1**), while significantly enriched GO category of glutamate receptor signaling pathway (GO:0007215; binominal FDR Q value = 3.85E-04) has been associated with 173 regulatory LTRs and 27 target genes (**Table 2; Supplementary Table S1**). In contrast, GO category of negative regulation of nervous system development (GO:0051962) has been linked with 483 regulatory LTRs and 119 target genes known to exert negative effects on development of nervous system (**Supplementary Table S1)**.

Furthermore, the validity of the concept of human embryo retroviral LTRs' regulatory impacts on nervous system development and functions is supported by the identification of numerous significantly enriched phenotypic traits of biological relevance assigned by GO Cellular Component and by GO Molecular Function annotations as well as Mouse Phenotype Single KO and Mouse Phenotype databases (**Supplementary Table S1**). Notable observations of potential biological relevance include identification of serotonin receptor signaling pathway (GO:0007210; 54 LTRs linked to 7 genes; binominal FDR Q-value = 0.00016) and Y-linked inheritance (HP:0001450; 80 LTRs linked to 14 genes; binominal FDR Q-value = 5.60E-07) that were scored by the GREAT algorithm as the significantly enriched records of GO Biological Process and Human Phenotype databases-defined phenotypic traits, respectively (**Supplementary Table S1**). Among multiple significantly enriched phenotypes' classification categories, GO Cellular Component annotations identified neuron projection (GO:0043005; 1486 LTRs linked to 386 genes) and synaptic membrane (GO:0097060; 679 LTRs linked to 132 genes), while GO Molecular Function analysis scored glutamate receptor activity (GO:0008066; 152 LTRs linked to 19 genes); neurotransmitter receptor activity (GO:0030594; 220 LTRs linked to 43 genes); neurexin family protein binding (GO:0042043; 58 LTRs linked to 7 genes); and calcium ion binding (GO:0005509; 979 LTRs linked to 243 genes) (**Supplementary Table S1**). Interestingly, structural elements of the postsynaptic membrane defined by the GO Cellular Component category (GO:0045211; 577 LTRs linked to 102 genes; binominal FDR Q-value = 1.98E-04) appears preferentially targeted by regulatory actions of human embryo retroviral LTRs compared to the presynaptic membrane features (GO:0042734; 186 LTRs linked to 36 genes; binominal FDR Q-value = 0.042) (**Supplementary Table S1**). The extended summary illustrating quantitative and qualitative distinctions of the retroviral LTR elements' associations with GO analyses-defined features of the postsynaptic membrane versus the presynaptic membrane is reported in the **Supplementary Summary S1**.

**Identification and characterization of 26 gene ontology-defined genomic regulatory networks (GRNs) facilitating functions of a genomic regulatory dominion (GRD) of human embryo retroviral LTR elements.**

Results of Gene Ontology terms' enrichment analyses indicate that a set of 5444 genes constituting candidate down-stream targets of human embryo retroviral LTRs appears segregated into multiple genomic regulatory networks (GRNs) defined by a single ontology-specific enrichment analysis (**Figure 1; Tables 2-5; Supplementary Table S1**). Identification of GRNs is based on analyses of retroviral LTRs and down-stream target genes distribution patterns associated with each of significantly-enriched terms (Binominal FDR Q value < 0.05) identified by the GREAT algorithm during analyses of GO Biological Process (227 significant terms), GO Cellular Component (47 significant terms), and GO Molecular Function (80 significant terms) databases (**Figure 1**). It has been observed that for a set of single ontology-defined significantly-enriched terms there is a significant positive correlation (**$R^2$ > 0.99**) between genome fractions occupied by genomic features assigned to a set of significantly-enriched terms (referred to as the genomic modicums) and proportions of LTR elements residing within the genomic modicums (referred to as the LTR set coverage). A compendium of retroviral LTRs and down-stream target genes assigned to the individual significantly-enriched GO terms, genomic modicums of which were occupying more than 5% of human genome and harbor at least 6% of human embryo retroviral LTRs, were designated as GRNs (**Figure 1; Tables 2-5; Supplementary Table S1**).

In total, there are 26 gene ontology-defined GRNs associated with human embryo retroviral LTRs each of which contain more than 500 retroviral LTR elements and more than 199 down-stream target genes (**Figure**

**1; Tables 3-5; Supplementary Table S1**). Genomic locations of GRNs appear scattered across chromosomes to occupy from 5.5% to 15.09% of the human genome. GRNs integrate activities from 529 to 1486 human embryo retroviral LTR elements and from 199 to 805 down-stream target genes, including transcription factors, chromatin-state remodelers, sensing and signal transduction mediators, enzymatic and receptor binding effectors, intracellular complexes and extracellular matrix elements, and cell-cell adhesion molecules. One of notable common structural features of GRNs is that numbers of retroviral LTR elements derived from LTR7, MLT2A1, and MLT2A2 sequences residing within the boundaries of GRN's genomic modicums are quantitatively balanced in strict accordance to their genome-wide abundance levels (**Table 4**).

Follow-up gene ontology enrichment analyses of 26 GRNs revealed further details of their genotype-phenotype association-defined architectural complexity: GRNs consist from several hundred to thousands quantitatively smaller gene ontology enrichment analysis-defined genomic regulatory modules (GRMs) (**Table 3; Supplementary Table S1; Supplementary Data Set S1: GRN1-GRN26**). Importantly, to increase the stringency of statistical definitions of candidate retroviral LTR-associated GRMs, the cut-off threshold of the Binominal FDR Q-value < 0.001 was applied in these series of GO terms enrichment analyses. One of the implications of this approach is that the false discovery rates (FDR) in these analyses could be estimated at less than 0.1%, thus indicating that the reported observations have more than 99.9% likelihood of being true findings. Based on the results of these analyses, it has been concluded that genomic constituents of each GRN comprise of several hundred to thousands smaller gene ontology enrichment analysis-defined GRMs, each of which contains from a dozen to hundreds LTRs and down-stream target genes. Overall, these analyses identified 69,573 statistically significant retroviral LTR-linked GRMs (**Tables 3-4; Supplementary Table S1; Supplementary Data Set S1: GRN1-GRN26**), including 27,601 GRMs validated by the single ontology-specific directed acyclic graph (DAG) analyses across 6 gene ontology annotations databases (**Table 5; Supplementary Table S1; Supplementary Data Set S1: GRN1-GRN26**). It has been determined that within each GRN there is a significant positive correlation between the human genome fractions occupied by the GRMs and LTR set coverage for the GRMs (**Supplementary Summary S2**).

These observations suggest that human embryo retroviral LTRs may exert regulatory effects on thousands distinct GRMs containing functionally and/or structurally-related sub-sets of genes affecting multiple facets of Modern Humans' developmental processes as well as physiological and pathological traits, in particular, a constellation of biological processes, cellular components, and molecular activities affecting development and functions of the nervous system.

**Selected examples of GRMs and associated phenotypic traits identified by GO analyses of specific retroviral LTR-governed GRNs.**

Potential phenotypic impacts on human pathophysiology of human embryo retroviral LTR-governed GRNs could be inferred by analyses of significantly enriched genotype-phenotype associations annotated to their composite GRMs. The GRN "Regulation of Membrane Potential" have the largest number of significantly-enriched GRMs and associated phenotypic traits identified by GO analyses of the Human Phenotype database (**Tables 3-5; Supplementary Data Set S1**). A set of 79 retroviral LTRs linked to 10 genes have been identified that were annotated to GRM X-linked inheritance (HP:0001417) and GRM Gonosomal inheritance (HP:0010985) at Binominal FDR Q-values of 2.82E-29 and 1.86E-25, respectively. Multiple significantly-enriched GRMs are associated with human phenotypes of major clinical importance. These include: Autism (HP:0000717; 57 LTRs linked to 7 genes; Binominal FDR Q-value = 3.39E-33); and Autistic behavior (HP:0000729; 65 LTRs linked to 11 genes; Binominal FDR Q-value = 2.10E-26); Parkinsonism (HP:0001300; 26 LTRs linked to 8 genes; Binominal FDR Q value = 1.71E-11); Dementia (HP:0000726; 26 LTRs linked to 7 genes; Binominal FDR Q-value = 1.20E-09); Intellectual disability (HP:0001249; 179 LTRs linked to 43 genes; Binominal FDR Q value = 4.28E-31) and a severe form of Intellectual disability (HP:0010864; 52 LTRs linked to 8 genes; Binominal FDR Q value = 2.13E-24); Syncope (HP:0001279; 35 LTRs linked to 11 genes; Binominal FDR Q value = 5.46E-24); Abnormality of cardiovascular system electrophysiology (HP:0030956; 64 LTRs linked to 22 genes; Binominal FDR Q value = 1.96E-26); and Seizures (HP:0001250; 160 LTRs linked to 36 genes; Binominal FDR Q value = 5.19E-33). The extended summary illustrating selected findings derived from GO analyses of the Human Phenotype database related to human disorders is reported in the **Supplementary Summary S3**.

GO Biological Process analyses identified multiple GRMs representing common significantly-enriched phenotypic traits associated with numerous GRNs. Several GRMs common for the GRN "Synapse Part" and the GRN “Postsynapse” include regulation of amyloid precursor protein catabolic process; regulation of beta-amyloid formation; positive regulation of beta-amyloid formation; response to beta-amyloid; and cellular response to beta-amyloid.

GRNs “Synaptic Membrane” and “Postsynaptic Membrane” have several common GRMs, namely regulation of amyloid precursor protein catabolic process; regulation of beta-amyloid formation; and positive regulation of beta-amyloid formation. Similarly, common GRMs for the GRN “Trans-synaptic Signaling” and the GRN “Chemical Synaptic Transmission” are beta-amyloid metabolic process; response to beta-amyloid; and cellular response to beta-amyloid; while the GRM “Regulation of Membrane Potential” harbors a set of functionally-related GRMs, namely regulation of beta-amyloid formation; positive regulation of beta-amyloid formation; response to beta-amyloid; and cellular response to beta-amyloid.

GO Molecular Function analyses identified several GRMs representing common significantly-enriched phenotypic traits associated with seven GRNs, namely **“**Synapse Part”; “Postsynapse”; “Synaptic Membrane”; “Postsynaptic Membrane”; “Trans-synaptic Signaling”; “Chemical Synaptic Transmission”; “Regulation of Membrane Potential”. A set of common GRMs shared by these GRNs include beta-amyloid binding; ionotropic glutamate receptor activity; kainate selective glutamate receptor; NMDA glutamate receptor; AMPA glutamate receptor. Interestingly, the GO Molecular Function terms ionotropic glutamate receptor activity; kainate selective glutamate receptor; NMDA glutamate receptor; AMPA glutamate receptor represent the exactly same selection of phenotypic traits that were identified as down-stream regulatory targets of human-specific LTR7 and LTR5_Hs loci (Glinsky, 2022).

Overall, genes linked to phenotypic traits reflecting various aspects of physiological and pathological processes associated with the biogenesis, metabolism, and functions of the amyloid precursor protein and beta-amyloid peptides appear to represent one of the down-stream targets of preference during the evolution of human embryo retroviral LTR elements.

**Validation of GRNs and GRMs governed by human embryo retroviral LTR elements employing the directed acyclic graph (DAG) analysis.**

It was important to confirm the validity of statistical definitions of GRNs and GRMs based on the binominal and hypergeometrc FDR Q values (Methods) employing independent analytical approaches. To this end, a directed acyclic graph (DAG) test based on the enriched terms from a single ontology-specific table generated by a GREAT algorithm has been employed (**Figures 2-3; Table 3; Supplementary Data Sets S1: GRN1-GRN26**). Patterns and directions of connections between significantly enriched GO modules are guided by the experimentally-documented temporal logic of developmental processes and structural/functional relationships between gene ontology enrichment analysis-defined statistically significant terms. Therefore, only statistically significant GRMs from a single gene ontology-specific table generated by the GREAT algorithm that manifest connectivity patterns defined by experimentally documented developmental and/or structure/function/activity relationships are deemed valid observations and visualized as a consensus hierarchy network of the ontology-specific DAGs (**Figures 2-3; Tables 3-5; Supplementary Data Sets S1: GRN1-GRN26**). Based on these considerations, the DAG algorithm draw the developmental and structure/function/activity relationships-guided hierarchy of connectivity between statistically significant gene ontology enriched GRMs. Visualization snapshots of representative examples illustrating the individual ontology-specific DAGs are reported in the **Figures 2-3** and comprehensive reports of quantitative and qualitative components of these analyses are presented in **Tables 3-5; Supplementary Data Sets S1: GRN1-GRN26**.

GO analyses of Mouse phenotype databases and GSEA of the MGI Mammalian Phenotype Level 4 2021 database revealed that genes comprising down-stream regulatory targets of human embryo retroviral LTRs affect development and functions of essentially all major tissues, organs, and organ systems. These include central and peripheral nervous systems, cardio-vascular (circulatory) and lymphatic systems, gastrointestinal (digestive) and urinary (excretory) systems, musculoskeletal and respiratory systems, endocrine and immune systems, as well as integumentary and reproductive systems. Analyses of single gene KO models revealed profound developmental defects and numerous examples of organogenesis failures, indicating that activities of retroviral LTR down-stream target genes are essential for development and functions of major organs and organ systems.

Of potentially major interest are observations that all 26 GRNs governed by human embryo retroviral LTR elements harbor significantly enriched records of GRMs annotated to multiple distinct GO terms of Mammalian Offspring’s Survival (MOS) phenotypes (**Table 6; Supplementary Data Set S1: GRN1-GRN26**). For example, GRN2 “Regulation of nervous system development” (GO:0051960) harbors numerous significantly enriched GRMs annotated to distinct MOS phenotypes, including: neonatal lethality, complete penetrance and neonatal lethality, incomplete penetrance; lethality at weaning, complete penetrance; perinatal lethality, complete penetrance; postnatal lethality, complete penetrance and postnatal lethality, incomplete penetrance; embryonic lethality during organogenesis and embryonic lethality during organogenesis, complete penetrance; lethality throughout fetal growth and development and lethality throughout fetal growth and development, incomplete penetrance; lethality at weaning; perinatal, neonatal, and postnatal lethality phenotypes (**Table 6; Supplementary Data Set S1: GRN2**). Other notable GRMs of retroviral LTR-linked GRNs that were annotated to significantly enriched MOS phenotypes include: embryonic lethality between implantation and placentation, complete penetrance; embryonic lethality between implantation and somite formation, incomplete penetrance; embryonic lethality between somite formation and embryo turning, incomplete penetrance; prenatal lethality prior to heart atrial septation; embryonic lethality, complete penetrance (**Table 6; Supplementary Data Set S1: GRN1-GRN26**).

Collectively, these observations support the conclusion that GRNs comprising the GRD of human embryo retroviral LTR elements could be viewed as an intrinsically integrated compendium of thousands genotype-phenotypic trait associations-defined GRMs, many of which may represent the evolutionary selection units defined by inherent genotype-phenotype associations affecting species’ fitness and survival.

**Gene Set Enrichment Analyses (GSEA) of down-stream target genes of human embryo retroviral LTR elements revealed global multifaceted impacts on physiological traits, developmental phenotypes, and pathological conditions of Modern Humans.**

To gain further insights into architectural frameworks and functional features of the GRD associated with human embryo retroviral LTR elements, the Enrichr bioinformatics platform was utilized to carry out the comprehensive series of GSEA employing more than 70 genomics and proteomics databases (Methods). GSEA were executed by imputing 5444 candidate down-stream regulatory target genes of human embryo retroviral LTR elements and retrieving significantly enriched records capturing phenotypic traits affected by subsets of candidate LTR-regulated genes (**Tables 7-8; Supplementary Figure S1; Supplementary Tables S1-S2**). These analyses documented global multifaceted impacts of genes representing putative down-stream regulatory targets of human embryo retroviral LTR elements on physiological traits, developmental phenotypes, and pathological conditions of Modern Humans (**Table 7; Supplementary Figure S1; Supplementary Table S1**).

Broadly, identified in this study retroviral LTR target genes have been implicated in pathogenesis and manifestation of a large spectrum of human common, orphan, and rare disorders; they are engaged in numerous developmental and regulatory pathways contributing to differentiation and specialization processes across the hierarchy of human organs, tissues, cells, and sub-cellular structures; have been shown to affect phenotypic traits at the multiple levels of a cellular molecular composition and regulatory infrastructure, including signal transduction triggered by numerous endogenous ligands and multiple families of transmembrane receptors; structural/functional organization of mitochondria, endoplasmic reticulum, and nucleus; activities of transcription factors and chromatin remodeling effectors. Analyzed herein retroviral LTR target genes have been linked with the human aging phenotype in gene expression profiling experiments and with a multitude of both physiological and pathological human phenotypes in genotype-phenotype association studies across various analytical platforms and databases, in particular, human disease associations' databases (**Table 7; Supplementary Figure S1; Supplementary Table S1**).

Protein products of retroviral LTR down-stream target genes have been documented to participate in protein-protein interactions of endogenous multiprotein complexes as well as host-pathogen interactions during viral and microbial infections. GSEA of retroviral LTR down-stream target genes employing the OMIM Expanded database of protein-protein interactions (PPI) of human disease-causing genes identified 34 retroviral LTR target genes, protein products of which appear engaged in PPIs with protein products of disease-causing genes implicated in development of a wide spectrum of human disorders (**Supplementary Figure S1; Supplementary Table S1**). Genes comprising the 34-gene signature of retroviral LTR targets are significantly enriched in 1387 records of human diseases (GSEA of the DisGeNET database; adjusted p-value $< 0.05$). These observations prompted in depth follow-up investigations of a regulatory network comprising 88 retroviral LTR elements (27LTR7; 35 MLT2A1; and 26 MLT2A2 sequences), which were mapped as putative up-stream regulator elements of the 34 genes.

**Characterization of 88 human embryo retroviral LTRs comprising putative up-stream regulatory elements of PPI networks of human disease-causing genes.**

Results of the DAG analyses of 88 human embryo retroviral LTR elements revealed highly complex hierarchical networks of significantly enriched phenotypic traits identified by analyses of GO Biological Process and Mouse Phenotype databases, while analyses of GO Cellular Component and GO Molecular Function documented relatively simple architectures of significantly enriched phenotypic traits (**Figure 4; Supplementary Table S1; Supplementary Data Set S2**). Overall, 1533 significantly enriched records linked to down-stream target genes of 88 retroviral LTR elements have been identified at the statistical threshold of the Binominal FDR Q-value < 0.001. These include 432; 36; 42; 58; 243; and 722 significantly enriched phenotypic traits identified by Gene Ontology analyses of GO Biological Process; GO Cellular Component; GO Molecular Function; Human Phenotype; Mouse Phenotype Single KO; and Mouse Phenotype databases, respectively (**Supplementary Data Set S2**).

Of note, GO enrichment analyses of 88 retroviral LTR elements employing the Mouse Phenotype Single KO database identified numerous significantly records of genes annotated to distinct mammalian offspring's survival (MOS) phenotypes. The list of enriched GO terms of this category includes several prenatal and perinatal lethality phenotypes [perinatal lethality (MP:0002081); perinatal lethality, complete penetrance (MP:0011089); perinatal lethality, incomplete penetrance (MP:0011090); prenatal lethality, complete penetrance (MP:0011091); prenatal lethality, incomplete penetrance (MP:0011101)]; distinct embryonic lethality phenotypes [embryonic lethality, complete penetrance (MP:0011092); embryonic lethality between implantation and somite formation (MP:0006205); embryonic lethality during organogenesis MP:0006207); embryonic lethality between implantation and placentation (MP:0009850); and the premature death phenotype (MP:0002083).

GSEA of 34 genes identified hundreds of significantly enriched phenotypic traits across numerous genomics and proteomics databases, representative examples of which are reported in the **Figure 5;**

**Supplementary Figure S1; Supplementary Table S1**. In multiple instances numbers of significantly enriched phenotypic records linked with 34 genes are markedly higher compared to the parent set of 5444 genes (**Figure 5B; Supplementary Figure S1; Supplementary Table S1)**. For example, there are 121 and 124 significantly enriched records of 34 genes identified by GSEA of the Transcription Factors PPI database and the Hub Proteins PPI database, respectively, while there is only one significantly enriched record in each database identified by GSEA of 5444 genes (**Figure 5B**). Similarly, there are 1387 and 477 significantly enriched records of human diseases identified by GSEA of 34 genes and 5444 genes, respectively, employing the DisGeNET database (**Supplementary Figure S1; Supplementary Table S1)**. Apparent qualitative distinctions were noted between top 10 significantly enriched records of human diseases: all top 10 GSEA records of 34 genes' enrichment were various records of malignant neoplasms, while top 10 records linked with GSEA of 5444 genes were neurodevelopmental and substance abuse disorders (**Supplementary Figure S1; Supplementary Table S1)**. In contrast to common human diseases, 78 and 1095 significantly enriched records of human orphan disorders have been identified by GSEA of 34 genes and 5444 genes, respectively, employing the Orphanet Augmented 2021 database (**Supplementary Figure S1; Supplementary Table S1)**.

Follow-up analyses revealed that overall there are 206 human genes encoding proteins manifesting significantly enriched records of protein-protein interactions (PPIs) with protein products encoded by 34 genes (**Figures 5-6; Supplementary Table S1; Supplementary Summaries S4-S5**). Because protein products of 34 genes manifest significant enrichment of PPIs with human disease-causing genes (**Figure 5; Supplementary Figure S1; Supplementary Table S1; Supplementary Summary S4**), these observations suggest that a network of proteins encoded by 240 genes may be engaged in PPIs with human disease-causing genes as well. This hypothesis was tested by performing GSEA of 240 genes comprising the putative PPI network of human disease-causing genes (**Figures 5-6; Supplementary Table S1; Supplementary Summaries S4-S5**).

GSEA of 240 genes consistently demonstrated that numbers of significantly enriched phenotypic traits linked with 240 genes are markedly higher compared to either 34 genes or the parent set of 5444 genes (**Figures 5-6; Supplementary Table S1; Supplementary Summaries S4-S5)**. GSEA employing the DisGeNET database identified 3298; 1387; and 477 significantly enriched records of human diseases linked with 240 genes, 34 genes, and 5444 genes, respectively (**Figure 5; Supplementary Figure S1; Supplementary Table S1; Supplementary Summary S4)**. Qualitatively distinct patterns noted between top 10 significantly enriched records of human diseases associated with 34 genes versus 5444 genes (**Supplementary Figure S1; Supplementary Table S1)** were recapitulated by GSEA of 240 genes: all top 10 GSEA records linked with 240 genes were multiple records of malignant neoplasms, while top 10 records linked with GSEA of 5444 genes were neurodevelopmental and substance abuse disorders (**Figure 5; Supplemental Table S1; Supplementary Summary S4**). In contrast to common human diseases, GSEA employing the Orphanet Augmented 2021 database identified 199 and 1095 significantly enriched records of human orphan disorders associated with 240 genes and 5444 genes, respectively (**Figure 5; Supplementary Table S1; Supplementary Summary S4)**.

Particularly intriguing observations were recorded during GSEA of proteomics databases focused on the molecular architecture of a large compendium of endogenous PPI networks. There are 268 and 228 significantly enriched records of subsets of 240 genes identified by GSEA of the Transcription Factors PPI database and the Hub Proteins PPI database, respectively, while there are 121 and 124 significantly enriched record in corresponding databases identified by GSEA of 34 genes (**Figures 5-6; Supplementary Table S1; Supplementary Summary S4-S5**). Similarly, there are 606 and 62 significantly enriched records of 240 genes identified by GSEA of the CORUM Protein Complexes database and the SILAC Phosphoproteomics database, respectively, while there are 143 and 22 significantly enriched record in corresponding databases identified by GSEA of 34 genes (**Figure 6; Supplementary Table S1; Supplementary Summary S5**). Strikingly, GSEA employing the NURSA Human Endogenous Complexome database identified 1783 significantly enriched records linked with 240 genes, while no significantly enriched traits were recorded for GSEA of either 34 genes or 5444 genes (**Figure 6; Supplementary Table S1; Supplementary Summary S5**). These observations indicate that protein products selected from a panel of 240 genes are significantly enriched in PPIs of a marked majority of endogenous human multimolecular complexes recorded in the NURSA Human Endogenous Complexome database. It follows, that a set of 88 human embryo retroviral LTR elements may extend its regulatory impacts on vast networks of human endogenous multimolecular complexes via protein products of 34 down-stream target genes engaged in PPIs with 268 transcription factors and 228 hub proteins. Strikingly, these networks appear impacting PPIs that are significantly enriched in 6584 records of virus-host PPIs documented in Virus-Host PPI P-HIPSTer 2020 database and 2071 records of human rare disorders catalogued in the Rare Diseases AutoRIF Gene Lists database (**Supplementary Table S1)**.

Notably, GSEA of 240 genes employing the OMIM Expanded database of PPIs of human disease-causing genes further validated and expanded this concept by identification of 51 genes, protein products of which appear engaged in PPIs with protein products of disease-causing genes implicated in development of numerous human disorders of an extremely broad spectrum of clinical phenotypes (**Figure 6; Supplementary Table S1; Supplementary Summary S5**).

Collectively, present analyses suggest that a common molecular mechanism driving development of human pathologies may be associated with altered expression of proteins encoded by retroviral LTR target genes and down-stream effects on PPI networks engaging protein products of disease-causing genes. Malfunctions of endogenous multimolecular complexes tagged by PPIs of proteins encoded by retroviral LTR target genes may contribute to molecular pathogenesis of an exceedingly large number of human diseases, which exhibit highly diverse spectrum of clinical manifestations including malignancies of different solid organs and hematopoietic system; blood, lymphatic and endocrine system disorders; immunoinflammatory and autoimmune maladies; neurodevelopmental, neuropsychiatric, and neurodegenerative diseases.

**Genes representing common regulatory targets of human embryo retroviral LTR elements are segregated into sub-sets of genetic loci coalesced within genomic regulatory pathways of common biological functions.**

Notable common features of all identified herein GRNs governed by human embryo retroviral LTR elements are the quantitatively balanced numbers of residing within GRN's boundaries retroviral LTR elements derived from LTR7, MLT2A1, and MLT2A2 sequences (**Table 4**). These observations indicate that regulatory impacts on genes comprising constituents of corresponding GRNs may be executed by distinct families of human embryo retroviral LTR elements, which then likely to exert shared regulatory effects on corresponding phenotypic traits.

The validity of this hypothesis was tested by performing GSEA of 709 genes comprising a set of common down-stream regulatory targets of LTR7, MLT2A1, and MLT2A2 elements (**Table 1**). In these analytical experiments, GSEA-defined significantly enriched sets of genes operating within specific pathways were mapped to sub-sets of their putative up-stream regulatory elements among human embryo retroviral LTRs (**Supplementary Table S2; Supplementary Data Set S2**). Then each of these sub-sets of retroviral LTRs associated with specific pathways were subjected to GO analyses using the GREAT algorithm. Results of these analyses demonstrate that genetic loci representing common regulatory targets of human embryo retroviral LTRs appear segregated into relatively small sets of LTR/gene panels coalesced within genomic regulatory pathways of common biological functions (**Tables 8-10; Figures 7-14; Supplementary Table S2; Supplementary Data Set S2; Supplementary Figure S2**). These GSEA/GREAT analyses-defined sets of genes and associated retroviral LTR elements were designated Genomic Regulatory Pathways (GRPs).

Similar to GRNs, all identified by GSEA/GREAT analyses GRPs contain numerous retroviral LTRs of LTR7, MLT2A1, and MLT2A2 families and harbor hundreds of GO annotated GRMs. However, GRPs appear to represent quantitatively smaller genomic regulatory units compared to 26 GRNs based on the assessment of numbers of significantly enriched constituent GRMs (**Tables 3-5; Table 8; Supplementary Table S2; Supplementary Data Sets S1; S2**) as well as relatively scarce representations of constituent GRMs within DAG analyses-defined GO-specific hierarchical networks (**Figures 2-3; Figure 7-14; Supplementary Data Sets S1; S2**). Various regulatory aspects of development and functions of peripheral and central nervous systems at cell type-specific levels and distinct morphological regions appear to represent a dominant common theme shared by GRPs (**Figure 7-14; Supplementary Data Set S2**). These observations suggest that GSEA-defined GRPs, similar to GO annotations-defined GRMs and GRNs, may constitute a constellation of genomic regulatory units each of which is aligned to specialized biological pathways united by a functional synergy required to preserve the developmental and operational integrity of nervous system.

It has been noted that a compendium of GSEA/GREAT analyses-defined GRPs comprising 966 human embryo retroviral LTR elements and 255 down-stream target genes (**Figure 14; Table 8; Supplementary Data Set S2**) manifests genomic features quantitatively and qualitatively resembling GRNs, including overlapping subsets of retroviral LTRs and down-stream target genes assigned by GO annotations to 26 GRNs and their constituent GRMs. Intriguing examples of these common GRMs of potential evolutionary significance include sets of retroviral LTRs/down-stream target genes annotated to significantly enriched records of mammalian offspring survival (MOS) phenotypes. GO annotations employing the Mouse Phenotype Single KO database identified the following significantly-enriched records of relevance to MOS phenotypes:

1. MP:0002082; postnatal lethality; 247 LTRs; 45 genes; Binominal FDR Q-value = 1.59E-17;
2. MP:0002081; perinatal lethality; 251 LTRs; 41 genes; Binominal FDR Q-value = 7.30E-15;
3. MP:0011089; perinatal lethality, complete penetrance; 91 LTRs; 14 genes; Binominal FDR Q-value = 6.84E-12;
4. MP:0011090; perinatal lethality, incomplete penetrance; 88 LTRs; 13 genes; Binominal FDR Q-value = 5.84E-11;
5. MP:0011086; postnatal lethality, incomplete penetrance; 149 LTRs; 30 genes; Binominal FDR Q-value = 7.92E-10;
6. MP:0011083; lethality at weaning, complete penetrance; 33 LTRs; 5 genes; Binominal FDR Q-value = 1.43E-08;
7. MP:0011085; postnatal lethality, complete penetrance; 106 LTRs; 20 genes; Binominal FDR Q-value = 2.25E-07;
8. MP:0011110; preweaning lethality, incomplete penetrance; 94 LTRs; 14 genes; Binominal FDR Q-value = 1.16E-06;
9. MP:0001926; female infertility; 65 LTRs; 11 genes; Binominal FDR Q-value = 1.36E-06;

Collectively, these observations indicate that two critically important classes of genetic loci appear co-segregated within identified herein human embryo retroviral LTR-governed genomic regulatory units of various architecture: a). Genes essential for the integrity of nervous system development and functions; b). Genes essential for mammalian offspring survival. It seems likely that this striking co-segregation may reflect co-regulation of these two classes of genes by common sets of retroviral LTR elements. Since knockout of many genes within these retroviral LTR-governed GRNs causes offspring lethality, it is logical to expect that their sustained activity would be essential for offspring survival. Sustained activity of MOS genes within retroviral LTR-governed GRNs could be achieved either by relieving the repressor actions of retroviral LTRs or maintaining the activator effects of retroviral LTRs. Alignments of this genomic regulatory architecture of human embryo retroviral LTR-governed GRNs with control of genes essential for development and functions of nervous system may have contributed to acceleration of central nervous system's dramatic changes during primate evolution.

**Enrichment patterns of genomic loci annotated to mammalian phenotypes of reduced fertility and infertility within GRNs governed by human embryo retroviral LTR elements.**

One of the potential biological consequences of essential MOS genes being under control of human embryo retroviral LTR elements is potential effects of this genomic regulatory architecture on species' fertility phenotypes. In agreement with this hypothesis, it has been observed that GO-defined GRMs annotated to reduced fertility and/or infertility phenotypes are consistently enriched within 18 of 26 GRNs governed by human embryo retroviral LTR elements (**Supplementary Table S3**). Notably, all GRNs annotated to phenotypic traits closely related to development and functions of nervous system harbor significantly enriched GRMs of reduced fertility and/or infertility phenotypes (**Supplementary Table S3**). In contrast, GRNs that are more distantly related to or not directly associated with nervous system development and functions do not harbor GRMs of reduced fertility and/or infertility phenotypes among significantly enriched GRMs. These findings suggest that during primate evolution functionally essential fertility genes (that is, genes knockout of which causes reduced fertility and/or infertility phenotypes) and genes essential for development and functions of nervous system were placed under regulatory control of human embryo retroviral LTR elements. It follows that biologically meaningful repressor effects on both classes of genes during ontogenesis would cause defects in development and functions of nervous system as well as would result in reduced fertility and/or infertility phenotypes, thus directly affecting the survivability of the offspring lineage and evolutionary fitness of species.

It was of interest to determine whether GO analyses of human embryo retroviral LTR elements would reveal the potential genomic regulatory connectivity between functionally essential fertility genes and genes essential for mammalian offspring survival. To this end, GO annotations records of the retroviral LTR-governed

GRM of the reduced fertility phenotype (MP:0001921) were analyzed to identify the significantly enriched records of retroviral LTR elements and genes associated with Mammalian Offspring Survival (MOS) phenotypes (**Table 9**). It has been observed that one of the important features of the retroviral LTR-governed GRM of the reduced fertility phenotype is the significant enrichment of retroviral LTR elements and genes annotated by GO analyses to multiple records of MOS phenotypes (**Table 9**). These observations indicate that abrogated functions of genes essential for mammalian offspring survival (that is, genes knockout of which causes the offspring lethality phenotypes) may represent one of key genetic elements contributing to reduced fertility and/or infertility phenotypes. Based on reported herein observations, it has been concluded that during primate evolution functionally essential fertility genes, genes essential for offspring survival, and genes essential for development and functions of nervous system may have been placed under regulatory control of human embryo retroviral LTR elements. This implied genomic regulatory connectivity may constitute a critical element of evolutionary concordance linking the requirement of balanced and coherent development and functions of nervous system to the survivability of the offspring lineage and evolutionary fitness of primate species.

**High-confidence down-stream target genes of HERVs constitute a significant majority of candidate regulatory targets of human embryo retroviral LTR elements.**

Reported herein inferences of phenotypic impacts of HERVs was made based on the analyses of potential regulatory impacts of human embryo retroviral LTR elements on down-stream target genes without taking into account direct experimental evidence of gene expression changes in human cells following targeted genetic manipulations of HERVs. To address this limitation, genes expression of which is significantly altered in human cells subjected to either shRNA-mediated interference or CRISPR-guided epigenetic silencing directed at defined genetic loci of specific HERV families were identified among candidate down-stream regulatory targets of human embryo retroviral LTR elements. Then proportions of these high-confidence down-stream regulatory targets of retroviral LTRs with experimentally documented gene expression changes were calculated (**Table 10**).

Results of these analyses demonstrate that selective targeting in human cells of defined regulatory sequences encoded by HERVs caused significant expression changes of sub-sets of genes among candidate down-stream regulatory targets of human embryo retroviral LTRs (**Table 10; Supplementary Table S2**). These significant changes of gene expression were documented in multiple independent experiments employing various human cell lines, which were genetically engineered using either shRNA interference or CRISPR-guided epigenetic silencing tools tailored for selective targeting of distinct regulatory elements of HERVs (including retroviral LTRs and long non-coding RNAs, lncRNAs). While identified proportions of genes comprising high-confidence down-stream regulatory targets varied from 14% to 36% in different experimental settings, cumulative sets of identified herein high-confidence down-stream regulatory targets constitute 67% of candidate retroviral LTR-regulated genes (**Table 10; Supplementary Table S2**). Based on these observations, it has been concluded that high-confidence retroviral LTR-regulated genes defined by expression profiling experiments comprise a significant majority of 5444 genes and 709 genes, which represent, respectively, cumulative and common sets of identified herein putative down-stream regulatory targets of LTR7; MLT2A1; and MLT2A2 human embryo retroviral LTR loci.

It was of interest to determine whether the inferences of potential global phenotypic impacts of human embryo retroviral LTR elements on human development and pathophysiology could be reproduced by GSEA of high-confidence down-stream regulatory target genes of retroviral LTRs. To address this question, significantly enriched records identified by GSEA of 709 genes and 473 genes during analyses of corresponding genomics and proteomics databases were retrieved and catalogued at the statistical significance threshold of adjusted P value (FDR q value) $< 0.05$ (**Table 11; Supplementary Table S2**). Side by side comparisons of the results demonstrate that GSEA of 473 high-confidence down-stream regulatory target genes of retroviral LTRs essentially recapitulated the quantitative and qualitative GSEA patterns of 703 genes representing a common set of candidate down-stream regulatory targets of retroviral LTRs (**Table 11; Supplementary Table S2**). Therefore, the inferences of potential global phenotypic impacts of human embryo retroviral LTR elements on human development and pathophysiology that have been made based on GO-guided annotations and GSEA of candidate down-stream target genes are recapitulated by GSEA of gene expression profiling experiments-validated high-confidence down-stream regulatory target genes of HERVs.

**Human embryo regulatory LTRs derived from distinct families of pan-primate endogenous retroviruses and human-specific genomic regulatory sequences (HSRS) share common sets of candidate down-stream target genes.**

One of important conclusions that could be derived from reported herein observations is that human embryo retroviral LTR elements could have contributed to primate evolution by driving the emergence of human-specific phenotypic traits. To test this hypothesis and gauge potential impacts of retroviral LTRs on emergence of human-specific phenotypic traits, identified and functionally characterized in this study candidate down-stream regulatory targets of human embryo retroviral LTR elements were cross-referenced with previously reported down-stream target genes of 110,339 human-specific genomic regulatory sequences (HSRS), which are fixed in human population and creation of which was attributed to several distinct mechanisms (Glinsky, 2015 – 2022; Methods). For each binary category of distinct regulatory loci and corresponding down-stream regulatory targets, common sets of down-stream target genes were identified and evaluated based on two quantitative metrics: a) enrichment of observed numbers of common target genes compared to expected numbers based on the random target distribution model; b) evaluations of relative abundance profiles of common sets of down-stream target genes calculated as the percentage of all down-stream target genes. It has been surmised that identification of significantly enriched common sets of genes independently defined as putative down-stream regulatory targets of HSRS and human embryo retroviral LTRs may reveal common phenotypic traits affected by different classes of genomic regulatory elements.

Consistent with the idea of overlapping sets of down-stream targets for different classes of genomic regulatory elements, genes identified as down-stream regulatory targets of MLT2A2 and MLT2A1 loci are significantly enriched among LTR7 target genes, while genes identified as down-stream regulatory targets of LTR7 loci are significantly enriched among MLT2A2 and MLT2A1 target genes (**Figure 15; Supplementary Table S4**). Furthermore, genome-wide connectivity profiles of human embryo retroviral LTRs and putative target genes are significantly recapitulated within genomic regulatory networks presumably governed by 72,349 human-specific mutations of regulatory DNA sequences, including 17,777 fixed human-specific insertions and deletions; 35,074 fixed human-specific neuro-regulatory single nucleotide mutations shared with archaic humans; and 7,236 fixed human-specific regulatory regions devoided of archaic humans' DNA (**Figure 15; Supplementary Table S4**).

In several instances it was possible to dissect common patterns of down-stream regulatory targets' enrichment of genes linked with different families of human embryo regulatory LTRs vis-a-vis defined sets of human-specific sequences previously identified as components of different GRNs (**Figure 15; Table 12; Supplementary Table S4**). These set of analyses included the pairwise comparisons of the relative enrichment profiles of genes comprising down-stream targets of Naïve & Primed hESC functional enhancers and genes linked with 1619 human-specific Naïve & Primed hESC functional enhancers (Barakat et al., 2018; Glinsky and Barakat, 2019); genes comprising down-stream targets of transposable elements (TE) loci expressed in dorsolateral prefrontal cortex (DLPFC) of the adult human brain and genes linked with 4690 human-specific TE loci expressed in human adult DLPFC (Guffanti et al., 2018); genes identified as down-stream targets of a highly-diverse set of 59,732 HSRS and genes associated with 12,262 created de novo HSRS (Glinsky, 2020). Notably, in all instances genes comprising down-stream targets of bona fide human-specific components of genomic regulatory networks appear significantly enriched among genes defined as down-stream targets of different families of human embryo retroviral LTR elements (**Figure 15; Table 12; Supplementary Table S4**).

Identified herein consistent patterns of significant enrichment of common sets of genes comprising down-stream regulatory targets of HSRS and human embryo retroviral LTRs seem related to the 3D topological domains' architecture of linear chromatin fibers as well as chromatin state signatures. This notion is supported by observations of significant enrichments of genes associated with rapidly evolving in humans topologically associating domains (TADs) among genes comprising down-stream regulatory targets of human embryo retroviral LTRs (**Figure 15; Supplementary Table S4**). In a similar fashion, evaluation of genes associated with dynamic TADs, which are TADs undergoing marked changes of their genomic boundaries during differentiation of human cells, documented their significant enrichment among down-stream target genes of human embryo retroviral LTR elements (**Figure 15; Supplementary Table S4**). It has been observed that enrichment score values of these categories of common down-stream target genes appeared placed consistently among top-scoring significantly-enriched gene sets within panels of down-stream regulatory targets of distinct families of human embryo retroviral LTRs.

Genes identified herein as candidate down-stream regulatory targets of distinct families of human embryo retroviral LTR elements appear related to a significant degree to down-stream target genes of DNA sequences defined as genomic regulatory elements based on their chromatin state's signatures, namely 8,297 fixed human-specific regulatory regions (Marnetto et al., 2014) and 7,598 hESC developmental enhancers (Rada-Iglesias et al., 2011). Results of these analyses documented significant enrichment of genes comprising down-stream targets of chromatin state-defined regulatory DNA sequences, revealing genome-wide connectivity components of human embryo regulatory LTR elements and their putative target genes that are recapitulated, in part, within genomic regulatory networks governed by 8,297 fixed human-specific regulatory regions and 7,598 hESC developmental enhancers defined by chromatin state signatures. These findings imply that subsets of genes comprising down-stream targets of retroviral LTRs represent down-stream target genes regulated by hESC developmental enhancers and/or fixed human-specific regulatory regions implicated in control of differentiation phenotypes of human cells.

One of biological processes affecting the integrity of genomic DNA sequences during evolution is colonization and functional/structural remodeling of primates' genomes by mobile genetic elements (MGE), including various families of TE and HERVs, most of which lost their mobility and have been co-opted to perform a variety of genomic regulatory functions. L1, Alu and SVA elements are the only transposable elements that have unequivocally been shown to be currently active in humans, as shown by de novo insertions that are responsible for genetic disorders (Cordaux and Batzer, 2009). Inevitable mutational losses of essential regulatory features of transposon insertions dictate the necessity of continuous retrotransposition during evolution. Latest examples of structural/functional exaptation of MGEs likely reflecting relatively recent events of genomic evolution of present day Modern Humans have been mapped based on identification and characterization of fixed insertions of MGEs that were most recently active in human genome (Autio et al., 2021). It has been observed that down-stream target genes of 1937 genomic regulatory elements constituting fixed DNA insertions of most recently active MGEs are significantly enriched among down-stream target genes of human embryo retroviral LTRs of distinct HERV families. Overall, a majority of genes (536 of 926 genes; 58%) comprising down-stream regulatory targets of fixed MGE loci most recently active in human genome appears selected from genes identified in this study as putative down-stream targets of distinct families of human embryo retroviral LTR elements. These findings indicate that GRNs governed by different families of human embryo retroviral LTRs and insertions of MGEs that were most recently active in human genome share a common set of down-stream regulatory targets comprising 536 genes.

Comparative analyses of genome-wide connectivity maps of genomic coordinates of human embryo retroviral LTRs, human-specific regulatory elements, and their candidate down-stream target genes imply concordant patterns of their inferred phenotypic impacts and suggest a genomic regulatory mechanism contributing to emergence of human-specific traits during primate evolution. Considering evolutionary timelines of introduction and expansion of distinct HERVs' families within the primate lineage vis-à-vis human-specific genomic regulatory sequences, these observations are consistent with the idea that human embryo retroviral LTR elements have made important contribution to creation of genomic fingerprints guiding the evolution of regulatory DNA features facilitating the emergence of human-specific phenotypic traits.

**Down-stream target genes of human embryo retroviral LTR elements are significantly enriched among genes manifesting species-specific expression mapping bias in Human-Chimpanzee hybrids of induced pluripotent stem cells and brain organoids.**

Dissection of genomic and molecular anatomy of regulatory mechanisms defining development and manifestations of human-specific phenotypic traits remains a highly significant challenge despite recent significant progress in explorations of species-specific developmental features using primate-derived induced pluripotent stem cells (iPSC) and brain organoids, which were particularly useful for identification of species-specific differences of gene expression (Camp et al., 2015; Gallego Romero. et al., 2015; Glinsky, 2020; Kanton, et al., 2019; Kronenberg et al., 2017; Lancaster et al., 2013; Mora-Bermudez et al., 2016; Muchnik et al., 2019; Otani et al., 2016; Pasca et al., 2015; Pasca, 2018; Pollen et al., 2019; Prescott et al., 2015; Qian et al., 2019; Sloan et al., 2017). Most recently, human-chimpanzee hybrid iPSC (hyiPSC) has been developed and utilized to study species-specific gene expression differences in hyiPSC and hyiPS cells differentiated into region-specific brain organoids termed hybrid cortical spheroids, hyCS (Agoglia et al., 2021; Gokhman et al., 2021). Stable tetraploid hybrids of human and chimpanzee cells enabled to control for non-genetic confounding factors such as different timing of developmental dynamics and differentiation processes, micro-environmental fluctuations, differences in cellular compositions, experimental batch effects, which are limited the utility of other experimental approaches aiming to disentangle species-specific divergence of gene expression. Significantly, cross-species tetraploid hybrid cells allowed to unambiguously identify species-specific gene expression differences caused specifically by genomic *cis*-regulatory divergence between species, which include promoters and enhancers affecting only down-stream target genes residing on the same species-specific DNA molecules (Agoglia et al., 2021; Gokhman et al., 2021). Therefore, it was logical to test the hypothesis that human embryo retroviral LTR elements may contribute to emergence of human-specific phenotypic traits by evaluating their potential regulatory effects on genes manifesting species-specific expression differences in human-chimpanzee hybrid cells.

To this end, individual binary categories of 5444 down-stream regulatory targets of human embryo retroviral LTRs with three different sets of genes manifesting species-specific gene expression biases in human-chimpanzee hybrid cells were selected, then common (overlapping) sets of genes were identified and significance of enrichment was estimated of observed numbers of common genes compared to corresponding expected numbers based on the random distribution model (**Figure 16; Table 13**). Notably, consistent patterns of significant enrichment of human embryo retroviral LTR target genes among three different categories of genes manifesting species-specific biases of gene expression in human-chimpanzee hybrid cells (**Figure 16; Table 13**). There are 95 genes putatively regulated by human embryo LTRs among 374 genes with species-specific expression mapping bias in hyiPSC of human-chimpanzee hybrids (**Table 13; Figure 16**), which constitute 2.2-fold enrichment compared to expected numbers of common genes (p = 9.58E-14). Similarly, there was observed the 1.9-fold (p = 5.06E-07) enrichment of human embryo LTRs' down-stream target genes among genes exhibiting species-specific expression mapping bias in human-chimpanzee hybrids of Cortical Spheroid (Brain Organoids), hyCS (**Table 13; Figure 16**). Finally, a total of 56 down-stream target genes of human embryo retroviral LTRs constitute 30.6% of genes manifesting species allele-specific expression profiles in human-chimpanzee hybrid cells (**Table 13**), which represents a 2.7-fold enrichment ($p = 4.36E-12$) compared to expected by chance alone numbers. Examination of expression profiles of genes with species-specific expression bias in human-chimpanzee hybrids revealed the apparently discordant patterns of their expression in both hyiPSC (**Figure 16**; panels A and B) and hyCS brain organoids' settings (**Figure 16**; panels C and D) supporting the idea that species-specific differences in gene expression may contribute to phenotypic divergence of Humans and Chimpanzee in regulation and manifestation of stemness phenotype and cortical differentiation.

Overall, these analyses revealed that a significant fraction of genes with species-specific expression bias in human-chimpanzee hybrid cells represents down-stream target genes of human embryo retroviral LTRs (177 of 676 genes; 26%), indicating that regulatory effects of retroviral LTR elements may have affected species segregation during the evolution of Great Apes by contributing to phenotypic divergence of Modern Humans. It was of interest to determine what morphological structures and phenotypic traits were preferred targets of retroviral LTR elements during Modern Human divergence during primate evolution. This question was addressed by performing GSEA of identified herein 177 genes (**Supplementary Table S5**) across multiple genomic database (Methods). GSEA of 177 genes employing the ARCHS4 Human Tissues database identified 23 records with significant enrichment of human embryo retroviral LTRs' target genes (adjusted p value $< 0.05$), among which the Dentate Granule Cells were the top-scoring term (adjusted p value = 3.62E-09). A total of 53 genes linked with 223 human embryo retroviral LTRs were identified among genes constituting genetic markers of the Dentate Granule Cells (**Figure 16E**). In addition to the Dentate Granule Cells, prefrontal cortex (52 genes), cingulate gyrus (51 genes), cerebellum (48 genes), and cerebral cortex (48 genes) were identified among the top 5 scoring records (**Figure 16E**).

To gain insights into the broader developmental context of biological functions of genes associated with 223 human embryo retroviral LTR elements, the DAG analyses were carried out employing the GREAT algorithm (Methods). Notably, GO annotations of the Mouse Single KO database generated local DAG which identified abnormal synaptic transmission (MP:0003635; linked with 66 LTRs and 17 genes; Binominal FDR Q-

value = 1.49E-15); abnormal learning/memory/conditioning (MP:0002063) and abnormal cognition (MP:0014114; both terms linked with 49 LTRs and 14 genes; Binominal FDR Q-value = 7.98E-08); abnormal excitatory postsynaptic currents (MP:0002910; 19 LTRs linked with 6 genes; Binominal FDR Q-value = 2.08E-06); abnormal synaptic vesicle morphology (MP:0004769; 11 LTRs linked with 5 genes; Binominal FDR Q-value = 5.37E-06); and abnormal CNS synaptic transmission (MP:0002206; linked with 65 LTRs and 16 genes; Binominal FDR Q-value = 3.07E-17) among 250 significantly enriched functional and developmental anomalies caused by knockout of individual genes associated with 223 human embryo retroviral LTRs (**Figure 16**; panels F and G).

Taken together, reported observations indicate that genes manifesting species-specific expression bias in human-chimpanzee hybrid cells and linked by putative regulatory associations with human embryo retroviral LTR elements are implicated in higher cognitive functions of learning and memory; biological processes of CNS synaptic transmission, excitatory postsynaptic currents, and neuroactive ligands-receptors interactions; gene expression markers of numerous morphological components of human brain, including the Dentate Granule Cells, prefrontal cortex, cingulate gyrus, cerebellum, cerebral cortex, superior frontal gyrus, dorsal striatum, motor and sensory neurons, midbrain, oligodendrocytes, and neuronal epithelium (**Table 14; Figure 16; Supplementary Table S5**). Additionally, this family of genes represents gene expression markers of adult human somatic tissues (small intestine; testis; alpha cells; retina) and morphological elements of human embryogenesis namely fetal brain cortex, oocyte, and zygote (**Table 14; Supplementary Table S5**).

The hypothesis that down-stream target genes of human embryo retroviral LTRs may affect human-specific gene expression networks operating in adult human brain was further validated by observations that human embryo retroviral LTRs' target genes are significantly enriched among genes expression of which demarcates human-specific features of spatial gene expression profiles delineated by comparative analyses of eight brain regions of humans, chimpanzees, gorillas, a gibbon, and macaques (Xu et al., 2018). Overall, a significant fraction of genes comprising down-stream target of human embryo retroviral LTRs represents genes implicated in differential transcriptomic patterns of eight regions of primates' brains (1968 of 5444 genes; 36.2%; p = 1.03604E-20; hypergeometric distribution test). It has been observed that human embryo retroviral

LTRs' target genes constitute 35.2% of genes manifesting human-specific expression differences in examined eight brain regions in contrast to 25.4% genes exhibiting chimpanzee-specific expression differences (p = 0.026; 2-tail Fisher's exact test). Furthermore, detailed analyses of expression profiles of genes assigned to 26 distinct gene expression modules of primates' brain identified 3 modules (modules 22; 16; and 15; **Figure 16H**) that manifest most significant relative enrichment of human embryo retroviral LTR's target genes in contrast to 7 modules (modules 4; 7 – 11; 18) exhibiting most significant relative depletion (**Figure 16H**). Notably, two of three modules with most significant relative enrichment (human-specific module 22 and module 16) of human embryo retroviral LTRs' down-stream target genes are known to contain the disproportionally high numbers of genes with hippocampus-specific expression (Xu et al., 2018). These findings indicate that hippocampus-specific genes may constitute one of preferred regulatory targets of human embryo retroviral LTR elements.

Collectively, reported herein observations support the hypothesis that genes comprising putative down-stream regulatory targets of human embryo retroviral LTRs contributed to phenotypic divergence of Modern Humans during the primate evolution and may have impacted the emergence of uniquely humans' phenotypic traits. Identification of hippocampus and more specifically Dentate Granule Cells among significant morphological targets of preference of regulatory effects of human embryo retroviral LTR elements may assist in precise experimental definition of important morphological structures of human brain having evolutionary shaped critical impacts on human-specific features of higher cognitive functions.

## Discussion

Transitions of exogenous retroviral infections to the state of endogenous retroviruses have persisted within the primate lineage throughout the past 30-40 million years. Concurrently with transitions of exogenous retroviral infections to the state of endogenous retroviruses (ERVs), sophisticated and highly complex molecular mechanisms are co-evolved in mammalian cells to identify and control endogenous retroviral sequences (Dopkins et al., 2022). One of the outcomes of these processes was a large-scale colonization of human genome by HERVs, which expanded during evolution to represent ~8% of DNA sequences in human cells. Colonization of human genome by HERVs is likely to represent the end result of long-term germ-line reinfections during primate evolution, rather than retrotransposition in cis or complementation in trans following a single infection cycle (Belshaw et al. 2004). Resulting large-scale expansion of highly homologous sequences derived from individual retroviruses and high diversity of numerous retroviral families created formidable scientific challenges of both analytical and conceptual dimensions. In recent years, several bioinformatics pipelines have been developed and implemented which are dedicated to highly accurate identification and quantification at the level of individual genetic loci resolution of DNA and RNA sequences originated from endogenous retroviruses in various experimental settings, including DNA and RNA extracted from human tissues and cells (Guffanti et al., 2018), ChIP-seq and HiC sequencing outputs (Lerat, 2022), as well as from single-cell RNA sequencing experiments (Berrens et al., 2022; He et al., 2021). These advances facilitated significant progress in our understanding of biological roles of DNA sequences derived from individual families of HERVs and revealed novel epigenetic, transcriptional, and posttranscriptional mechanisms of their locus-specific regulation. However, advances in development and implementation of analytical and conceptual approaches aiming to dissect the potential cross-talks of distinct families of HERVs that colonized human genomes during primate evolution and their contribution to emergence of human-specific phenotypic traits remain limited.

Present analyses were focused on human embryo retroviral LTR loci defined herein as LTR loci with experimentally documented genomic regulatory functions affecting phenotypes deemed essential for human embryonic development, in particular, preimplantation stages of human embryogenesis. To this end, patterns of evolutionary origins and global effects of 8839 human embryo regulatory LTR elements derived from distinct families of pan-primate endogenous retroviruses on genome-wide transcriptional regulatory networks were evaluated with emphasis on the assessment of their potential impacts on developmental processes and pathological conditions of Modern Humans. Phenotypic patterns of regulatory effects on pathophysiology of Modern Humans were validated by differential GSEA of high confidence down-stream target genes expression of which is altered following genetic and/or epigenetic targeting of human embryo retroviral LTR elements. Collectively, present analyses demonstrate a global scale of potential phenotypic impacts of human embryo retroviral LTR elements on pathophysiology of Modern Humans.

**Sequential patterns of evolutionary origins and quantitative expansion during primate evolution of distinct families of human embryo retroviral LTR elements.**

Overall, analyzed in this contribution three distinct LTR families, namely MLT2A1 (2416 loci), MLT2A2 (3069 loci), and LTR7 (3354 loci), appear to have similar evolutionary histories defined by evolutionary patterns of expansion, conservation, and divergence during primate evolution. However, they manifest clearly discernable distinct timelines of evolutionary origins, which were defined as the estimated timelines when retroviruses infect primates' germlines and transition from the state of exogenous infection agents to endogenous retroviral sequences integrated into host genomes of primate lineage. Present analyses indicate that both MLT2A1/HERVL and MLT2A2/HERVL retroviruses have infected primates' germlines before the segregation of Old World Monkeys (OWM) and New World Monkeys (NWM) lineages. This conclusion is based on the evidence that the consistent earliest presence of highly-conserved pan-primate MLT2A1 and MLT2A2 loci has been mapped to genomes of both NWM and OWM. In contrast, the consistent earliest presence in primates' genomes of LTR7 loci could be mapped to genomes of OWM but not to genomes of NWM. These observations suggest that LTR7/HERVH retroviruses have entered germlines of the primate lineage after the separation of the OWM and NWM lineages. Pan-primate endogenous MLT2A1/HERVL, MLT2A2/HERVL, and LTR7/HERVH retroviruses appear to achieve a significant expansion in genomes of Great Apes, which is ranging from ~2-fold expansion for MLT2A2 loci to ~5-fold expansion for LTR7 loci. Interestingly, the consistent earliest presence of highly-conserved LTR5_Hs loci has been mapped to the Gibbon's genome (Glinsky, 2022), consistent with the primate evolution model stating that LTR5_Hs/HERVK retroviruses successfully colonized germlines of the primate lineage after the segregation of Gibbons' species. Subsequently, MLT2A1/HERVL; MLT2A2/HERVL; LTR7/HERVH; and LTR5_Hs/HERVK retroviruses appear to undergo a marked expansion in genomes of Great Apes. Interestingly, the consistent patterns of increasing cognitive and behavioral complexities during primate evolution have been observed on the OWM evolutionary branch, which continually evolved in Africa, but not on geographically segregated to South America the NWM branch.

Consistent with the previously reported findings for highly-conserved LTR7 and LTR5_Hs loci (Glinsky, 2022), patterns of quantitative expansion in different primate species of human embryo retroviral LTR elements appear to replicate the consensus evolutionary sequence of increasing cognitive and behavioral complexities during primate evolution. The apparent concordance of increasing cognitive and behavioral complexities during the evolution of primate species with clearly discernable evidence of sequential introduction and expansion of distinct families of retroviral LTR elements in genomes of the primate lineage suggest that human embryo retroviral LTRs originated from distinct retroviruses may have contributed to the guidance of primate evolution toward the emergence of human-specific phenotypic traits. This hypothesis seems particularly appealing when considered together with the evidence of global impacts on exceedingly broad spectrum of developmental phenotypes of genes constituting down-stream regulatory targets of human embryo retroviral LTR elements.

**Emergence during primate evolution of the global Genomic Regulatory Dominion (GRD) governed by human embryo retroviral LTR elements.**

Observations reported in this contribution provide the foundation for the hypothesis that evolution of human embryo retroviral LTR elements created the global retroviral Genomic Regulatory Dominion (GRD) consisting of 26 gene ontology enrichment-defined genome-wide Genomic Regulatory Networks (GRNs). Decoded in this study the hierarchical super-structure of retroviral LTR-associated GRD and GRNs represents an intrinsically integrated developmental compendium of 69,573 Genomic Regulatory Modules (GRMs) each of which is congregated along specific experimentally-defined genotype-phenotypic trait associations. On the levels of individual genomic loci, described in this contribution retroviral GRD integrates 8839 highly conserved LTR elements linked to 5444 down-stream target genes. These scattered across human genome individual genomic loci were forged by evolution into a functionally-consonant constellation of 26 genome-wide multimodular GRNs, each of which harbors hundreds significantly enriched single gene ontology-specific traits. Genomic coordinates of GRNs are scattered across chromosomes to occupy from 5.5% to 15.09% of the human genome. Each GRN acquired during evolution similar regulatory architecture: it harbors from 529 to 1486 human embryo retroviral LTR elements derived from LTR7, MLT2A1, and MLT2A2 sequences that are quantitatively balanced corresponding to their genome-wide abundance. GRNs integrate activities of functionally and structurally highly diverse panels of down-stream regulatory targets comprising from 199 to 805 genes. These include genes encoding transcription factors, chromatin-state remodelers, signal sensing and signal transduction mediators, enzymatic and receptor binding effectors, intracellular complexes and extracellular matrix elements, and cell-cell adhesion molecules. Compositions of individual GRNs consist of several hundred to thousands smaller gene ontology enrichment analysis-defined genomic regulatory modules (GRMs), each of which combines from a dozen to hundreds retroviral LTRs and down-stream target genes. A total of 69,573 statistically significant retroviral LTR-linked GRMs (Binominal FDR q-value < 0.001) have been identified in this contribution, including 27,601 GRMs that have been validated by the single ontology-specific directed acyclic graph (DAG) analyses across 6 gene ontology annotations databases.

**Insights into putative concurrent mechanisms of biological functions and global organism-level impacts of human embryo retroviral LTR elements.**

To corroborate and extend findings documented based on GO-guided analytical approaches, comprehensive series of Gene Set Enrichment Analyses (GSEA) of retroviral LTRs down-stream target genes were carried out employing more than 70 genomics and proteomics databases, including a large panel of databases developed from single-cell resolution studies of healthy and diseased human's organs and tissues. It has been observed that distinct GRNs and GRMs appear to operate on individuals' life-span timescale along specific phenotypic venues selected from an exceedingly broad spectrum of down-stream gene ontology-defined and signaling pathways-guided frameworks. GRNs and GRMs governed by human embryo retroviral LTRs seem functionally designed to exert profound effects on patterns of transcription, protein-protein interactions, developmental phenotypes, physiological traits, and pathological conditions of Modern Humans.

GO analyses of Mouse phenotype databases and GSEA of the MGI Mammalian Phenotype Level 4 2021 database revealed that down-stream regulatory targets of human embryo retroviral LTRs are enriched for genes that are essential for faithful execution of developmental processes and functionally-adequate states of all major tissues, organs, and organ systems. These conclusions are supported by observations documenting profound developmental defects in corresponding single gene KO mouse models. Furthermore, genes comprising candidate down-stream regulatory targets of human embryo retroviral LTRs are engaged in PPI networks that have been implicated in pathogenesis of human common and rare disorders (3298 and 2071 significantly enriched records, respectively), in part, by impacting PPIs that are significantly enriched in 1783 multiprotein complexes recorded in the NURSA Human Endogenous Complexome database and 6584 records of virus-host PPIs documented in Virus-Host PPI P-HIPSTer 2020 database.

Collectively, results of multiple analytical experiments reported in this contribution consistently demonstrate that genes comprising candidate down-stream regulatory targets of human embryo retroviral LTR elements affect development and functions of essentially all major tissues, organs, and organ systems, including central and peripheral nervous systems, cardio-vascular (circulatory) and lymphatic systems, gastrointestinal (digestive) and urinary (excretory) systems, musculoskeletal and respiratory systems, endocrine and immune systems, as well as integumentary and reproductive systems. In-depth follow-up analyses of single gene knockout models of defined retroviral LTR down-stream targets revealed a multitude of profound developmental defects and numerous examples of organogenesis failures, indicating that activities of human embryo retroviral LTR down-stream target genes are essential for development and functions of major organs and organ systems.

**Analytical inference of a cell-type-level preferred morphological targets of regulatory effects of human embryo retroviral LTR elements.**

Applications of GO annotations and GSEA approaches revealed exceedingly broad impacts of genes comprising putative regulatory targets of human embryo retroviral LTRs on the levels of tissues, organs, and organ systems of the human body. Similarly, GSEA of single-cell genomic databases (for example, PanglaoDB Augmented 2021 database; CellMarker Augmented 2021 database; Azimuth Cell Types 2021 database) documented a very large number of distinct cell types among significantly enriched records, including a multitude of specific types of neurons, radial glia and Bergmann glia cells, alpha and beta cells, oligodendrocytes, astrocytes, multiple types of stem cells and precursor cells, and anterior pituitary gland cells among others. However, targets of preference interrogations among 177 LTR target genes manifesting species-specific biases in human-chimpanzee hybrids employing the ARCHS4 Human Tissues database (**Table 14**) enabled identification of some specific cell types that appear to represent the preferred regulatory targets of human embryo retroviral LTR elements, specifically Dentate Granule Cells and neuronal epithelium. Identification of Dentate Granule Cells is of particular interest because Dentate Granule Cells have been implicated in the parallel emergence of both dynamic and stable memory engrams in the hippocampus (Hainmueller and Bartos, 2018) and the contribution of adult-born Dentate Granule Cells of Dentate Gyrus to the downstream firing dynamics of neocortical information flow to the hippocampus (McHugh et al., 2022). These latest findings are highly consistent with the critical role of the adult hippocampal neurogenesis, which occurs precisely only in the Dentate Gyrus, in higher cognitive functions, especially learning, memory, and emotions (Kempermann et al., 2015; 2023; Kuhn et al., 2018).

These cell types remain consistently among top-scoring significantly enriched records throughout the layered multistage process of GSEA-guided interrogations of different sets of retroviral LTR down-stream target genes, including 5444 genes comprising the cumulative set of retroviral LTR regulatory targets (**Supplementary Table S1**); 709 genes representing the common set of retroviral LTR regulatory targets (**Supplementary Table S2**); 473 genes comprising high-confidence down-stream regulatory targets of retroviral LTRs (**Supplementary Table S2**); and 177 retroviral LTR down-stream target genes manifesting species-specific expression bias in human-chimpanzee hybrid cells (**Table 14; Supplementary Table S5**). For example, Radial Glia Cells (78 retroviral LTR target genes; adjusted p value = 8.80E-18; PanglaoDB Augmented 2021 database); Neuronal epithelium (937 retroviral LTR target genes; adjusted p value = 2.24E-47; ARCHS4 Human Tissues database); and Dentate Granule Cells (863 retroviral LTR target genes; adjusted p value = 2.45E-28; ARCHS4 Human Tissues database), have been identified among top-scoring significantly enriched terms during GSEA of 5444 genes comprising a cumulative set of putative down-stream regulatory targets of human embryo retroviral LTR elements (**Supplementary Table S1**). It has been observed that overall human embryo retroviral LTRs' down-stream target genes constitute large fractions of all gene expression markers assigned in the respective genomic databases to corresponding cell types: they constitute 68% of genes for Radial Glia Cells; 40% of genes for Neuronal epithelium; and 37% of genes for Dentate Granule Cells.

Collectively, these observations strongly argue that specific cell types, namely neuronal epithelium, Radial Glia Cells, and Dentate Granule Cells, may represent the cellular targets of preference of genomic regulatory networks governed by human embryo retroviral LTR elements and suggest that embryonic and adult neurogenesis could have been among the principal biological processes contributing to striking phenotypic divergence of Modern Humans during the evolution of Great Apes. In agreement with this notion, GSEA of human embryo retroviral LTRs down-stream target genes comprising of sets of 709 genes; 473 genes; or 177 genes employing the GTEx Human Tissue Expression database of up-regulated genes consistently demonstrated that all top 100 significantly enriched records constitute the brain tissues of male or female individuals (**Supplementary Tables S2; S5**). In contrast, there were no significantly enriched records of human brain tissue samples identified by GSEA of human embryo retroviral LTRs down-stream target genes employing the GTEx Human Tissue Expression database of down-regulated genes.

**Insights into molecular mechanisms of global genome-wide and proteome-wide actions of human embryo retroviral LTR elements.**

Human embryo retroviral LTR elements undergo programmed epigenetic activation and silencing during human embryonic development starting from the earliest stages of preimplantation embryogenesis (Introduction). Mechanistically, activation of human endogenous retroviral LTR elements may function as epigenetic switch facilitating targeted reversible transitions of the 3D regulatory chromatin architecture within topologically associating domains from dominant single enhancer modes of active retroviral LTRs to the state of agile host super-enhancer loci of the regulatory chromatin microenvironment. Epigenetic silencing of retroviral LTR loci could create distribution patterns of silenced chromatin domains of H3K9me3 repressive histone marks. Conversely, active retroviral LTR elements may contribute to creation of transcriptionally active chromatin state by impacting deposition of H3K4me3 histone marks and expansion of broad H3K4me3 domains.

Deposition of H3K9me3 histone marks is significantly enriched on LTR retrotransposons throughout human preimplantation embryogenesis (Yu et al., 2022; Xu et al., 2022), indicating that human embryo retroviral LTR elements facilitate genome-wide heterochromatin patterning. Genes that are marked by or positioned near H3K9me3 peaks throughout the human development are related to cell fate specification, which is consistent with the idea that one of important functions of H3K9me3 chromatin modifications is to block the premature expression of genes essential for regulation of developmental stages (Yu et al., 2022; Xu et al., 2022). Embryonic stage-specific establishment of H3K9me3 histone modifications silences targeted LTR loci of different families during human embryogenesis and 8-cell-specific H3K9me3 domains demarcate enhancer-like genomic regions marked by hallmarks of enhancers H3K4me1 and H3K27Ac at the later developmental stages and in adult human somatic tissues (Xu et al., 2022).

In human blastocysts, H3K4me3/H3K9me3 and H3K4me3/H3K27me3 bivalent chromatin domains have been identified that contributed to heterochromatin patterning primed for lineage segregation and likely involved in lineage differentiation (Xu et al., 2022). It has been reported that H3K4me3/H3K9me3 bivalent histone methylation signature denotes poised differentiation master regulatory genes in trophoblast stem cells, extraembryonic endoderm stem cells, and preadipocytes (Matsumura et al., 2015; Rugg-Gunn et al., 2010). Genes occupied by H3K4me3/H3K9me3 bivalent chromatin domains were implicated in various metabolic processes and are not significantly expressed during early embryonic development (Xu et al., 2022), while nearly 30% of LTRs were occupied by H3K4me3/H3K9me3 bivalent chromatin modifications indicating that some human embryo retroviral LTR loci may be poised for activation at the later developmental stages. These observations suggest that human embryo retroviral LTR elements may contribute to creation of bivalent H3K4me3/H3K9me3 chromatin domains, thus affecting the genome-wide patterns of both H3K9me3 repressive chromatin marks and H3K4me3 active chromatin modifications. I

Interestingly, CRISPRi-mediated silencing of the LTR7Y/B loci in hESC documented significant effects on expression of eight chromatin writers (methyltransferases) and erasers (demethylases) regulating methylation status of the H3K4 histone, suggesting a mechanism of selective activation of the *KMT2D* methyltransferase gene expression by LTR7Y/B (Glinsky, data not shown and manuscript in preparation). These findings imply that human embryo retroviral LTRs, in particular, LTR7Y/B loci, may facilitate the establishment of the broad H3K4me3 chromatin domains. Consistent with this hypothesis, the KMT2D methyltransferase plays the principal role in establishing the broad H3K4me3 chromatin domains and constitutes an intrinsic components of enhancers-forming multimolecular liquid-liquid phase-separated condensates, which are particularly prominent at super-enhancers (reviewed in Kent et al., 2023). Consequently, activation/repression cycles of retroviral LTR actions may function as an epigenetic state transitions' shuttle between on/off epigenetic control of super-enhancer activities at the selected topological domains and nanodomains of the interphase chromatin. Aberrations of these epigenetic reprogramming patterns may contribute to developmental defects and pathogenesis of human disorders.

Observations reported in this contribution may have important implications for our understanding of the molecular pathophysiology of human diseases linked with malfunctions of endogenous multimolecular complexes tagged by PPI networks of protein products encoded by retroviral LTR-target genes. Present analyses suggest that one of widely shared mechanisms underlying development of a broad spectrum of human pathologies of common, rare, or orphan classifications is associated with altered engagements of protein products of disease-causing genes within PPI networks of proteins encoded by down-stream target genes of retroviral LTR elements. Multiple malfunctions' focal points of these human pathology-driving PPI networks may operate by impacting protein-protein interactions that are significantly enriched in identified herein 1783 multiprotein complexes recorded in the NURSA Human Endogenous Complexome database and 6594 records of virus-host protein-protein interactions reported in the Virus-Host PPI P-HIPSTer 2020 database.

**Insights into mechanisms of potential impacts of human embryo retroviral LTR elements during primate evolution.**

This contribution highlights potential multidimensional impacts of retroviral GRD on emergence of human-specific traits during primate evolution. Genome-wide candidate regulatory targets and high-confidence down-stream target genes of human embryo regulatory LTR elements identified and functionally characterized in this study were cross-referenced with putative down-stream target genes of more than 100 thousands human-specific genomic regulatory elements. These analyses revealed consistent patterns of significant enrichment of common sets of genes comprising down-stream regulatory targets of HSRS and human embryo retroviral LTRs that are related to the 3D topological domains' architecture of linear chromatin fibers as well as regulatory chromatin state signatures. Overall, identification of significantly enriched sets of intersecting genes independently defined as putative down-stream regulatory targets of different families of HSRS and human embryo retroviral LTR elements depicts common phenotypic traits potentially affected by evolutionary and structurally distinct classes of genomic regulatory elements. Future follow-up studies would be required to dissect genetic and molecular mechanisms of connectivity, coopetition, and discordancy effects on down-stream target genes of human embryo regulatory LTRs of distinct retroviral families and human-specific genomic regulatory loci.

Collectively, observations reported in this contribution indicate that two critically important classes of genetic loci appear co-segregated within identified herein human embryo retroviral LTR-governed genomic regulatory units of various architecture: a). Genes essential for the integrity of nervous system development and functions; b). Genes essential for mammalian offspring survival. It seems likely that this striking co-segregation within genomic regulatory units of the GRD governed by retroviral LTRs may reflect co-regulation of these two classes of genes by common sets of retroviral LTR elements. Since knockout of many genes within these retroviral LTR-governed GRNs causes offspring lethality, it is logical to expect that their sustained activity would be essential for offspring survival. Sustained activity of MOS genes within retroviral LTR-governed GRNs could be achieved either by relieving the repressor actions of retroviral LTRs or maintaining the activator effects of retroviral LTRs. Alignments of this genomic regulatory architecture of human embryo retroviral LTR-governed GRNs with control of genes essential for development and functions of nervous system may have contributed to acceleration of central nervous system's dramatic changes during primate evolution.

These findings suggest that during primate evolution functionally essential fertility genes (that is, genes knockout of which causes reduced fertility and/or infertility phenotypes) and genes essential for development and functions of nervous system were placed under regulatory control of human embryo retroviral LTR elements. It follows that biologically meaningful repressor effects on both classes of genes during ontogenesis would cause defects in development and functions of nervous system as well as would result in reduced fertility and/or infertility phenotypes, thus directly affecting the survivability of the offspring lineage and evolutionary fitness of species.

Furthermore, present observations indicate that altered functions of genes essential for mammalian offspring survival (that is, genes knockout of which causes the offspring lethality phenotypes) may represent one of critically important genetic elements contributing to reduced fertility and/or infertility phenotypes. Reported herein observations suggest that during primate evolution functionally essential fertility genes, genes essential for offspring survival, and genes essential for development and functions of nervous system may have been placed under regulatory control of human embryo retroviral LTR elements. This implied genomic regulatory connectivity may constitute a key element of evolutionary concordance linking the requirement of balanced and coherent development and functions of nervous system to the survivability of the offspring lineage and evolutionary fitness of primate species.

**Conclusion: Human embryo retroviral LTR elements function as a pan-primate epigenetic architect contributing to evolution, development, physiological traits, and pathological phenotypes of Modern Humans.**

During millions years of primate evolution, two distinct families of pan-primate endogenous retroviruses, namely HERVL and HERVH, infected primates' germline, colonized host genomes and evolved to contribute to creation of the global retroviral genomic regulatory dominion (GRD) performing vital regulatory and morphogenetic functions during human embryogenesis. Retroviral GRD constitutes of 8839 highly conserved LTR elements linked to 5444 down-stream target genes forged by evolution into a functionally-consonant constellation of 26 genome-wide multimodular genomic regulatory networks (GRNs) each of which is defined by significant enrichment of numerous single gene ontology-specific traits. Locations of GRNs appear scattered across chromosomes to occupy from 5.5% to 15.09% of the human genome. Each GRN harbors from 529 to 1486 human embryo retroviral LTR elements derived from LTR7, MLT2A1, and MLT2A2 sequences that are quantitatively balanced according to their genome-wide abundance. GRNs integrate activities from 199 to 805 down-stream target genes, including transcription factors, chromatin-state remodelers, signal sensing and signal transduction mediators, enzymatic and receptor binding effectors, intracellular complexes and extracellular matrix elements, and cell-cell adhesion molecules. GRN's compositions consist of several hundred to thousands smaller gene ontology enrichment analysis-defined genomic regulatory modules (GRMs), each of which combines from a dozen to hundreds LTRs and down-stream target genes. Overall, this study identifies 69,573 statistically significant retroviral LTR-linked GRMs (Binominal FDR q-value < 0.001), including 27,601 GRMs validated by the single ontology-specific directed acyclic graph (DAG) analyses across 6 gene ontology annotations databases. These observations were corroborated and extended by execution of a comprehensive series of Gene Set Enrichment Analyses (GSEA) of retroviral LTRs down-stream target genes employing more than 70 genomics and proteomics databases, including a large panel of databases developed from single-cell resolution studies of healthy and diseased human's organs and tissues. Genes assigned to distinct GRNs and GRMs appear to operate on individuals' life-span timescale along specific phenotypic avenues selected from a multitude of down-stream gene ontology-defined and signaling pathways-guided frameworks to exert profound effects on patterns of transcription, protein-protein interactions, developmental phenotypes, physiological traits, and pathological conditions of Modern Humans. GO analyses of Mouse phenotype databases and GSEA of the MGI Mammalian Phenotype Level 4 2021 database revealed that down-stream regulatory targets of human embryo retroviral LTRs are enriched for genes making essential contributions to development and functions of all major tissues, organs, and organ systems, including central and peripheral nervous systems, cardio-vascular (circulatory) and lymphatic systems, gastrointestinal (digestive) and urinary (excretory) systems, musculoskeletal and respiratory systems, endocrine and immune systems, as well as integumentary and reproductive systems. These observations were corroborated by numerous developmental defects documented in a single gene KO mouse models. Genes comprising candidate down-stream regulatory targets of human embryo retroviral LTRs are engaged in protein-protein interaction (PPI) networks that have been implicated in pathogenesis of human common and rare disorders (3298 and 2071 significantly enriched records, respectively), in part, by impacting PPIs that are significantly enriched in 1783 multiprotein complexes recorded in the NURSA Human Endogenous Complexome database and 6584 records of virus-host PPIs documented in Virus-Host PPI P-HIPSTer 2020 database. GSEA-guided analytical inference of the preferred cellular targets of human embryo retroviral LTR elements supported by analyses of genes with species-specific expression mapping bias in Human-Chimpanzee hybrids identified Neuronal epithelium, Radial Glia, and Dentate Granule Cells as cell-type-specific marks within a Holy Grail sequence of embryonic and adult neurogenesis. Observations reported in this contribution support the hypothesis that evolution of human embryo retroviral LTR elements created the global GRD consisting of 26 gene ontology enrichment-defined genome-wide GRNs. Decoded herein the hierarchical super-structure of retroviral LTR-associated GRD and GRNs represents an intrinsically integrated developmental compendium of thousands GRMs congregated on specific genotype-phenotypic trait associations. Many highlighted in this contribution GRMs may represent the evolutionary selection units driven by inherent genotype-phenotype associations affecting primate species' fitness and survival by exerting control over mammalian offspring survival genes implicated in reduced fertility and infertility phenotypes. Mechanistically, programmed activation during embryogenesis and ontogenesis of genomic constituents of human embryo retroviral GRD coupled with targeted epigenetic silencing may guide genome-wide heterochromatin patterning within nanodomains and topologically-associated domains during differentiation, thus affecting 3D folding dynamics of linear chromatin fibers and active transcription compartmentalization within interphase chromatin of human cells.

## Methods

*Data source and analytical protocols*

Present analyses were focused on human embryo retroviral LTR loci (see Introduction) the evolutionary origins of which were mapped to genomes of Old World Monkeys (OWM), indicating that these genomic loci could be defined as highly-conserved pan-primate regulatory sequences because they have been present in genomes of primate species for ~30-40 MYA. Three distinct LTR families meet these criteria, namely MLT2A1 (2416 loci), MLT2A2 (3069 loci), and LTR7 (3354 loci). A total of 8839 fixed non-polymorphic sequences of human embryo retroviral LTR elements residing in genomes of Modern Humans (hg38 human reference genome database) were retrieved as described in recent studies (Hashimoto et al., 2021; Carter et al., 2022; Glinsky, 2022) and the number of highly conserved orthologous loci in genomes of sixteen non-human primates (NHP) were determined exactly as previously reported (Glinsky, 2022). Briefly, fixed non-polymorphic retroviral LTR loci residing in the human genome (hg38 human reference genome database) has been considered highly conserved in the genome of non-human primates (NHP) only if the following two requirements are met:

(1) During the direct LiftOver test (https://genome.ucsc.edu/cgi-bin/hgLiftOver ), the human LTR sequence has been mapped in the NHP genome to the single orthologous locus with a threshold of at least 95% sequence identity;

(2) During the reciprocal LiftOver test, the NHP sequence identified in the direct LiftOver test has been remapped with at least 95% sequence identity threshold to the exactly same human orthologous sequence which was queried during the direct LiftOver test.

To ascertain the potential regulatory cross-talks of human embryo retroviral LTRs and human-specific regulatory sequences (HSRS), analyses of a total of 110,339 candidate HSRS were carried-out, including 59,732 HSRS (Glinsky, 2020a); 35,074 neuro-regulatory human-specific SNCs (Kanton et al., 2019; Glinsky, 2020b); 8,297 fixed human-specific regulatory regions defined by chromatin state signatures (Marnetto et al., 2014); 7,236 fixed human-specific regulatory regions devoided of archaic humans' DNA (Schaefer et al., 2021). Detailed catalogues of candidate HSRS including descriptions of diverse origins and regulatory features of different families of HSRS and corresponding references of original contributions are reported in the Supplementary Materials and earlier publications [Barakat et al. 2018; Fuentes et al. 2018; Glinsky 2015; 2016a, b; 2018; 2019; 2020a, b, c, 2021; 2022; Guffanti et al. 2018; Glinsky and Barakat, 2019; McLean et al. 2010; 2011; Pontis et al. 2019; Wang et al. 2014].

To evaluate the potential impacts of human embryo retroviral LTRs on genomic regulatory networks implicated in pluripotency and stemness phenotypes as well as developmental processes, analyses have been performed of 7,598 hESC developmental enhancers defined by chromatin state signatures (Rada-Iglesias et al., 2011); 32,353 and 36,417 functional enhancers operating in Primed and Naïve hESC, respectively (Barakat et al., 2018), as well as 1,619 human-specific hESC functional enhancers (Glinsky and Barakat, 2019). Solely publicly available datasets and resources were used in this contribution. The significance of the differences in the expected and observed numbers of events was calculated using two-tailed Fisher's exact test. Multiple proximity placement enrichment tests were performed for individual families and sub-sets of LTRs and HSRS taking into account the size in bp of corresponding genomic regions, size distributions in human cells of topologically associating domains, distances to putative regulatory targets, bona fide regulatory targets identified in targeted genetic interference and/or epigenetic silencing experiments. Additional details of methodological and analytical approaches are provided in the Supplementary Materials and previously reported contributions [Barakat et al. 2018; Fuentes et al. 2018; Glinsky 2015; 2016a, b; 2018; 2019; 2020a, b, c, 2021; 2022; Guffanti et al. 2018; Glinsky and Barakat, 2019; McLean et al. 2010; 2011; Pontis et al. 2019; Wang et al. 2014].

*Gene set enrichment and genome-wide proximity placement analyses*

Gene set enrichment analyses were carried-out using the Enrichr bioinformatics platform, which enables the interrogation of nearly 200,000 gene sets from more than 100 gene set libraries. The Enrichr API (January 2018 through January 2023 releases) [Chen et al. 2013; Kuleshov et al. 2016; Xie et al. 2021] was used to test genes linked to retroviral LTR elements, HSRS, or other regulatory loci of interest for significant enrichment in numerous functional categories. When technically feasible, larger sets of genes comprising several thousand entries were analyzed. Regulatory connectivity maps between HSRS, retroviral LTRs and coding genes and additional functional enrichment analyses were performed with the Genomic Regions Enrichment of Annotations Tool (GREAT) algorithm [McLean et al. 2010; 2011] at default settings. The reproducibility of the results was validated by implementing two releases of the GREAT algorithm: GREAT version 3.0.0 (02/15/2015 to 08/18/2019) and GREAT version 4.0.4 (08/19/2019) applying default settings at differing maximum extension thresholds as previously reported (Glinsky 2020a, b, c; 2021; 2022). The GREAT algorithm allows investigators to identify and annotate the genome-wide connectivity networks of user-defined distal regulatory loci and their putative target genes. Concurrently, the GREAT algorithm performs functional Gene Ontology (GO) annotations and analyses of statistical enrichment of GO annotations of identified genomic regulatory elements (GREs) and target genes, thus enabling the inference of potential biological impacts of interrogated genomic regulatory networks. The Genomic Regions Enrichment of Annotations Tool (GREAT) algorithm was employed to identify putative down-stream target genes of human embryo retroviral LTRs. Concurrently with the identification of putative regulatory target genes of GREs, the GREAT algorithm performs stringent statistical enrichment analyses of functional annotations of identified down-stream target genes, thus enabling the inference of potential significance of phenotypic impacts of interrogated GRNs. Importantly, the assignment of phenotypic traits as putative statistically valid components of GRN actions entails the assessments of statistical significance of the enrichment of both GREs and down-stream target genes by applying independent statistical tests.

The validity of statistical definitions of genomic regulatory networks (GRNs) and genomic regulatory modules (GRMs) based on the binominal (regulatory elements) and hypergeometrc (target genes) FDR Q values was evaluated using a directed acyclic graph (DAG) test based on the enriched terms from a single ontology-specific table generated by the GREAT algorithm (**Figures 2-3; Table 3; Supplementary Data Sets S1: GRN1-GRN26**). DAG test draws patterns and directions of connections between significantly enriched GO modules based on the experimentally-documented temporal logic of developmental processes and structural/functional relationships between gene ontology enrichment analysis-defined statistically significant terms. A specific DAG test utilizes only a sub-set of statistically significant GRMs from a single gene ontology-specific table generated by the GREAT algorithm by extracting GRMs manifesting connectivity patterns defined by experimentally documented developmental and/or structure/function/activity relationships. These GRMs are deemed valid observations and visualized as a consensus hierarchy network of the ontology-specific DAGs (**Figures 2-3; Tables 3-5; Supplementary Data Sets S1: GRN1-GRN26**). Based on these considerations, the DAG algorithm draws the developmental and structure/function/activity relationships-guided hierarchy of connectivity between statistically significant gene ontology enriched GRMs.

Genome-wide Proximity Placement Analysis (GPPA) of down-stream target genes and distinct genomic features co-localizing with retroviral LTRs and HSRS was carried out as described previously and originally implemented for human-specific transcription factor binding sites [Glinsky et al. 2018; Glinsky, 2015, 2016a, 2016b, 2017, 2018, 2019, 2020a, 2020b, 2020c, 2021; Guffanti et al. 2018].

*Differential GSEA to infer the relative contributions of distinct subsets of retroviral LTR elements and down-stream target genes on phenotypes of interest.*

When technically and analytically feasible, different sets of retroviral LTRs and candidate down-stream target genes defined at several significance levels of statistical metrics and comprising from dozens to several thousand individual genetic loci were analyzed using differential GSEA. This approach was utilized to gain insights into biological effects of retroviral LTRs and down-stream target genes and infer potential mechanisms of phenotype affecting activities. Previously, this approach was successfully implemented for identification and characterization of human-specific regulatory networks governed by human-specific transcription factor-binding sites [Glinsky et al. 2018; Glinsky, 2015, 2016a, 2016b, 2017, 2018, 2019, 2020a, 2020b, 2020c, 2021; Guffanti et al. 2018] and functional enhancer elements [Barakat et al. 2018; Glinsky et al. 2018; Glinsky and Barakat 2019; Glinsky 2015, 2016a, 2016b, 2017, 2018, 2019, 2020a, 2020b, 2020c, 2021]. Differential GSEA approach has been utilized for characterization of phenotypic impacts of 13,824 genes associated with 59,732 human-specific regulatory sequences [Glinsky, 2020a], 8,405 genes associated with 35,074 human-specific neuroregulatory single-nucleotide changes [Glinsky, 2020b], 8,384 genes regulated by stem cell-associated retroviral sequences (SCARS) [Glinsky 2021], as well as human genes and medicinal molecules affecting the susceptibility to SARS-CoV-2 coronavirus [Glinsky, 2020c].

Initial GSEA entail interrogations of each specific set of candidate down-stream target genes using ~70 distinct genomic databases, including comprehensive pathway enrichment Gene Ontology (GO) analyses. Upon completion, these analyses were followed by in-depth interrogations of the identified significantly-enriched gene sets employing selected genomic databases deemed most statistically informative at the initial GSEA. In all reported tables and plots (unless stated otherwise), in addition to the nominal p values and adjusted p values, the Enrichr software calculate the “combined score”, which is a product of the significance estimate and the magnitude of enrichment (combined score c = log(p) * z, where p is the Fisher’s exact test p-value and z is the z-score deviation from the expected rank).

*Statistical Analyses of the Publicly Available Datasets*

All statistical analyses of the publicly available genomic datasets, including error rate estimates, background and technical noise measurements and filtering, feature peak calling, feature selection, assignments of genomic coordinates to the corresponding builds of the reference human genome, and data visualization, were performed exactly as reported in the original publications and associated references linked to the corresponding data visualization tracks (http://genome.ucsc.edu/ ). Additional elements or modifications of statistical analyses are described in the corresponding sections of the Results. Statistical significance of the Pearson correlation coefficients was determined using GraphPad Prism version 6.00 software. Both nominal and Bonferroni adjusted p values were estimated and considered as reported in corresponding sections of the Results. The significance of the differences in the numbers of events between the groups was calculated using two-sided Fisher's exact and Chi-square test, and the significance of the overlap between the events was determined using the hypergeometric distribution test [Tavazoie et al. 1999].

**Supplementary Information is available online.**

Supplementary information includes Supplementary Tables S1-S5, Supplementary Data Sets S1 and S2; Supplemenmtary Figures S1 and S2;  and Supplementary Summaries S1-S6.

**Acknowledgements.** This work was made possible by the open public access policies of major grant funding agencies and international genomic databases and the willingness of many investigators worldwide to share their primary research data. Author would like to thank you Victoria Glinskii for invaluable expert assistance with graphical presentation of the results of this study.

**Author Contributions**

This is a single author contribution. All elements of this work, including the conception of ideas, formulation, and development of concepts, execution of experiments, analysis of data, and writing of the paper, were performed by the author.

**Funding**

In part, this work was supported by OncoScar, LLC.

**Declarations**

**Conflict of interest statement**

Dr. Glinsky is co-founder of the OncoScar, LLC, an early-stage biotechnology company dedicated to exploration of the potential translational utility of stem cell-associated retroviral sequences.

**Data availability statement**

All data supporting the reported observations and required to reproduce the findings are provided in the main body of the paper and Supplementary materials.

## References

Agoglia RM, Sun D, Birey F, Yoon SJ, Miura Y, Sabatini K, Pașca SP, Fraser HB. 2021. Primate cell fusion disentangles gene regulatory divergence in neurodevelopment. Nature. 592:421-427. doi: 10.1038/s41586-021-03343-3.

Albanese, A., Arosio, D., Terreni, M. & Cereseto, A. 2008. HIV-1 pre-integration complexes selectively target decondensed chromatin in the nuclear periphery. PLoS ONE 3, e2413.

Autio MI, Bin Amin T, Perrin A, Wong JY, Foo RS, Prabhakar S. 2021. Transposable elements that have recently been mobile in the human genome. BMC Genomics. 22(1):789. doi: 10.1186/s12864-021-08085-0. PMID: 34732136

Bao W, Kojima KK, Kohany O. 2015. Repbase Update, a database of repetitive elements in eukaryotic genomes. *Mobile DNA* **6**:11. doi:10.1186/s13100-015-0041-9

Barakat TS, Halbritter F, Zhang M, Rendeiro AF, Perenthaler E, Bock C, Chambers I. 2018. Functional dissection of the enhancer repertoire in human embryonic stem cells. Cell Stem Cell. 23: 276-288.

Barbulescu M, Turner G, Seaman MI, Deinard AS, Kidd KK, Lenz J. 1999. Many human endogenous retrovirus K (HERV-K) proviruses are unique to humans. Curr Biol 9:861-868.

Belshaw R, Pereira V, Katzourakis A, Talbot G, Pačes J, Burt A, Tristem M. 2004. Long-term reinfection of the human genome by endogenous retroviruses. Proc Natl Acad Sci 101: 4894–4899. doi:10.1073/pnas.0307800101

Belshaw R, Dawson AL, Woolven-Allen J, Redding J, Burt A, Tristem M. 2005. Genomewide screening reveals high levels of insertional polymorphism in the human endogenous retrovirus family HERV-K(HML2): implications for present-day activity. J Virol 79:12507-12514.

Belshaw R, Watson J, Katzourakis A, Howe A, Woolven-Allen J, Burt A, Tristem M. 2007. Rate of recombinational deletion among human endogenous retroviruses. J Virol 81: 9437–9442. doi:10.1128/JVI.02216-06

Berrens, R.V., Yang, A., Laumer, C.E., Lun, A.T.L., Bieberich, F., Law, C.-T., Lan, G., Imaz, M., Bowness, J.S., Brockdorff, N., et al. 2022. Locus-specific expression of transposable elements in single cells with CELLO-seq. Nat. Biotechnol. 40, 546–554. https://doi.org/10.1038/s41587-021-01093-1.

Bushman, F. et al. Genome-wide analysis of retroviral DNA integration. 2005. Nat. Rev. Microbiol. 3, 848–858.

Camp JG et al. 2015. Human cerebral organoids recapitulate gene expression programs of fetal neocortex development. Proc Natl Acad Sci U S A 112, 15672–15677, doi:10.1073/pnas.1520760112.

Carter T, Singh M, Dumbovic G, Chobirko JD, Rinn JL, Feschotte C. Mosaic cis-regulatory evolution drives transcriptional partitioning of HERVH endogenous retrovirus in the human embryo. Elife. 2022 Feb 18; 11:e76257. doi: 10.7554/eLife.76257.

Cereseto A, Giacca M. 2004. Integration site selection by retroviruses. AIDS Rev 6(1): 13-21.

Chen, EY, et al., Enrichr: interactive and collaborative HTML5 gene list enrichment analysis tool. BMC Bioinformatics, 2013. 14: 128.

Cordaux, R., Batzer, M. 2009. The impact of retrotransposons on human genome evolution. Nat Rev Genet 10, 691–703. https://doi.org/10.1038/nrg2640

Dopkins N, O'Mara MM, Lawrence E, Fei T, Sandoval-Motta S, Nixon DF, Bendall ML. 2022. A field guide to endogenous retrovirus regulatory networks. Mol Cell. 2022 82: 3763-3768. doi: 10.1016/j.molcel.2022.09.011. PMID: 36270247.

Fort A, Hashimoto K, Yamada D, Salimullah M, Keya CA, Saxena A, Bonetti A, Voineagu I, Bertin N, Kratz A, Noro Y, Wong C-H, de Hoon M, Andersson R, Sandelin A, Suzuki H, Wei C-L, Koseki H, Hasegawa Y, Forrest ARR, Carninci P. 2014. Deep transcriptome profiling of mammalian stem cells supports a regulatory role for retrotransposons in pluripotency maintenance. *Nature Genetics* **46**:558–566. 1015 doi:10.1038/ng.2965

Fuentes DR, Swigut T, Wysocka J. Systematic perturbation of retroviral LTRs reveals widespread long-range effects on human gene regulation. Elife. 2018 Aug 2;7:e35989. doi: 10.7554/eLife.35989. PMID: 30070637; PMCID: PMC6158008.

Gallego Romero I et al. 2015. A panel of induced pluripotent stem cells from chimpanzees: a resource for comparative functional genomics. Elife 4, e07103, doi:10.7554/eLife.07103.

Gemmell P, Hein J, Katzourakis A. 2015. Orthologous endogenous retroviruses exhibit directional selection since the chimp-human split. *Retrovirology* **12**. doi:10.1186/s12977-015-0172-6

Gemmell P, Hein J, Katzourakis A. 2019. The Exaptation of HERV-H: Evolutionary Analyses Reveal the Genomic Features of Highly Transcribed Elements. *Front Immunol* **10**. doi:10.3389/fimmu.2019.01339

Glinsky GV. 2009. Human genome connectivity code links disease-associated SNPs, microRNAs and pyknons. Cell Cycle 8(6): 925-30.

Glinsky GV. 2015. Transposable Elements and DNA Methylation Create in Embryonic Stem Cells Human-Specific Regulatory Sequences Associated with Distal Enhancers and Noncoding RNAs. *Genome Biol Evol* **7**:1432–1454. doi:10.1093/gbe/evv081

Glinsky, G.V. 2016a. Mechanistically Distinct Pathways of Divergent Regulatory DNA Creation Contribute to Evolution of Human-Specific Genomic Regulatory Networks Driving Phenotypic Divergence of Homo sapiens. *Genome Biol Evol* 8: 2774-2788.

Glinsky GV. 2016b. Single cell genomics reveals activation signatures of endogenous SCARS networks in aneuploid human embryos and clinically intractable malignant tumors. Cancer Lett. 381: 176-193.

Glinsky G.V. 2017. Human-specific features of pluripotency regulatory networks link NANOG with fetal and adult brain development. BioRxiv. https://www.biorxiv.org/content/10.1101/022913v3 doi: https://doi.org/10.1101/022913

Glinsky G, Durruthy-Durruthy J, Wossidlo M, Grow EJ, Weirather JL, Au KF, Wysocka J, Sebastiano V. Single cell expression analysis of primate-specific retroviruses-derived HPAT lincRNAs in viable human blastocysts identifies embryonic cells co-expressing genetic markers of multiple lineages. Heliyon. 2018 Jun 28;4(6):e00667. doi: 10.1016/j.heliyon.2018.e00667. PMID: 30003161; PMCID: PMC6039856.

Glinsky GV, Barakat TS. 2019. The evolution of Great Apes has shaped the functional enhancers' landscape in human embryonic stem cells. Stem Cell Res 37: 101456.

Glinsky GV. 2020a. A catalogue of 59,732 human-specific regulatory sequences reveals unique to human regulatory patterns associated with virus-interacting proteins, pluripotency and brain development. DNA and Cell Biology 39: 126-143. https://doi.org/10.1089/dna.2019.4988

Glinsky GV. 2020b. Impacts of genomic networks governed by human-specific regulatory sequences and genetic loci harboring fixed human-specific neuro-regulatory single nucleotide mutations on phenotypic traits of Modern Humans. Chromosome Res. 28: 331-354. https://doi.org/10.1007/s10577-020-09639-w

Glinsky GV. 2020c. Tripartite combination of candidate pandemic mitigation agents: Vitamin D, Quercetin, and Estradiol manifest properties of medicinal agents for targeted mitigation of the COVID-19 pandemic defined by genomics-guided tracing of SARS-CoV-2 targets in human cells. Biomedicines. 8: 129. https://doi.org/10.3390/biomedicines8050129

Glinsky GV. 2021. Genomics-Guided Drawing of Molecular and Pathophysiological Components of Malignant Regulatory Signatures Reveals a Pivotal Role in Human Diseases of Stem Cell-Associated Retroviral Sequences and Functionally-Active hESC Enhancers. Frontiers in Oncology. 11: 974. https://doi.org/10.3389/fonc.2021.638363

Göke J, Lu X, Chan Y-S, Ng H-H, Ly L-H, Sachs F, Szczerbinska I. 2015. Dynamic Transcription of Distinct Classes of Endogenous Retroviral Elements Marks Specific Populations of Early Human Embryonic Cells. 1052 *Cell Stem Cell* **16**:135–141. doi:10.1016/j.stem.2015.01.005

Gokhman D, Agoglia RM, Kinnebrew M, Gordon W, Sun D, Bajpai VK, Naqvi S, Chen C, Chan A, Chen C, Petrov DA, Ahituv N, Zhang H, Mishina Y, Wysocka J, Rohatgi R, Fraser HB. Human-chimpanzee fused cells reveal cis-regulatory divergence underlying skeletal evolution. 2021. Nat Genet. 53:467-476. doi: 10.1038/s41588-021-00804-3.

Goodchild NL, Wilkinson DA, Mager DL. 1993. Recent Evolutionary Expansion of a Subfamily of RTVL-H Human Endogenous Retrovirus-like Elements. Virology 196:778–788. doi:10.1006/viro.1993.1535

Grow EJ, Flynn RA, Chavez SL, Bayless NL, Wossidlo M, Wesche DJ, Martin L, Ware CB, Blish CA, Chang HY, Pera RA, Wysocka J. 2015. Intrinsic retroviral reactivation in human preimplantation embryos and pluripotent cells. Nature 522:221–225. DOI: https://doi.org/10.1038/nature14308, PMID: 25896322

Guffanti G, Bartlett A, Klengel T, Klengel C, Hunter R, Glinsky G, Macciardi F. 2018. Novel bioinformatics approach identifies transcriptional profiles of lineage-specific transposable elements at distinct loci in the human dorsolateral prefrontal cortex. Mol Biol Evol. 35: 2435-2453.

Hainmueller T, Bartos M. 2018. Parallel emergence of stable and dynamic memory engrams in the hippocampus. Nature. 558: 292–296. doi:10.1038/s41586-018-0191-2.

Hanke K, Hohn O, Bannert N. 2016. HERV-K (HML-2), a seemingly silent subtenant - but still waters run deep. Apmis 124:67–87. DOI: https://doi.org/10.1111/apm.12475, PMID: 26818263

Hashimoto K, Jouhilahti EM, Töhönen V, Carninci P, Kere J, Katayama S. Embryonic LTR retrotransposons supply promoter modules to somatic tissues. 2021. Genome Res. 2021 31: 1983-1993. doi: 10.1101/gr.275354.121.

He, J., Babarinde, I.A., Sun, L., Xu, S., Chen, R., Shi, J., Wei, Y., Li, Y., Ma, G., Zhuang, Q., et al. 2021. Identifying transposable element expression dynamics and heterogeneity during development at the single-cell level with a processing pipeline scTE. Nat. Commun. 12, 1456. https://doi.org/10.1038/s41467-021-21808-x.

Hendrickson PG, Doráis JA, Grow EJ, Whiddon JL, Lim JW, Wike CL, Weaver BD, Pflueger C, Emery BR, Wilcox AL, et al. 2017. Conserved roles of mouse DUX and human DUX4 in activating cleavage-stage genes and MERVL/HERVL retrotransposons. Nat Genet 49: 925–934. doi:10.1038/ng.3844

Hughes JF, Coffin JM. 2004. Human endogenous retrovirus K solo-LTR formation and insertional polymorphisms: implications for human and viral evolution. Proc Natl Acad Sci 101:1668-1672.

Ito J, Sugimoto R, Nakaoka H, Yamada S, Kimura T, Hayano T, Inoue I. 2017. Systematic identification and characterization of regulatory elements derived from human endogenous retroviruses. *PLOS Genetics* 13(**7)**: e1006883. https://doi.org/10.1371/journal.pgen.1006883

Ito J, Seita Y, Kojima S, Parrish NF, Sasaki K, Sato K. 2022. A hominoid-specific endogenous retrovirus may have rewired the gene regulatory network shared between primordial germ cells and naïve pluripotent cells. PLOS Genetics 18(5): e1009846. https://doi.org/10.1371/journal.pgen.1009846

Izsvák Z, Wang J, Singh M, Mager DL, Hurst LD. 2016. Pluripotency and the endogenous retrovirus HERVH: Conflict or serendipity? *BioEssays* **38**:109–117. doi:10.1002/bies.201500096

Kanton S, Boyle MJ, He Z, Santel M,Weigert A, Sanchís-Calleja F, Guijarro P, Sidow L, Fleck JS, Han D, Qian Z, Heide M, Huttner WB, Khaitovich P, Pääbo S, Treutlein B, Camp JG. 2019. Organoid single-cell genomic atlas uncovers human-specific features of brain development. Nature. 574: 418–422. https://doi.org/10.1038/s41586-019-1654-9

Kelley D, Rinn J. 2012. Transposable elements reveal a stem cell-specific class of long noncoding RNAs. *Genome Biology* **13**:R107. doi:10.1186/gb-2012-13-11-r107.

Kempermann G, Song H, Gage FH. 2015. Neurogenesis in the Adult Hippocampus. Cold Spring Harb Perspect Biol 7:a018812.

Kempermann G, Song H, Gage FH. 2023. Adult neurogenesis in the hippocampus. Hippocampus. 33:269-270. doi: 10.1002/hipo.23525.

Kent D, Marchetti L, Mikulasova A, Russell LJ, Rico D. 2023. Broad H3K4me3 domains: Maintaining cellular identity and their implication in super-enhancer hijacking. BioEssays, 00, e2200239. https://doi.org/10.1002/bies.202200239

Kojima KK. 2018. Human transposable elements in Repbase: genomic footprints from fish to humans. Mobile DNA 9:2. doi:10.1186/s13100-017-0107-y

Kronenberg ZN et al. 2018. High-resolution comparative analysis of great ape genomes. Science 360:eaar6343

Kuhn HG, Toda T, Gage FH. 2018. Adult Hippocampal Neurogenesis: A Coming-of-Age Story. J Neurosci. 38:10401-10410. doi: 10.1523/JNEUROSCI.2144-18.2018.

Kuleshov MV, et al., Enrichr: a comprehensive gene set enrichment analysis web server 2016 update. Nucleic Acids Res, 2016. 44(W1): W90-7.

Kunarso G, Chia N-Y, Jeyakani J, Hwang C, Lu X, Chan Y-S, Ng H-H, Bourque G. 2010. Transposable elements have rewired the core regulatory network of human embryonic stem cells. *Nature Genetics* **42**:631–634. doi:10.1038/ng.600

Lancaster MA et al. 2013. Cerebral organoids model human brain development and microcephaly. Nature 501, 373–379, doi:10.1038/nature12517.

Lerat, E. 2022. Recent bioinformatic progress to identify epigenetic changes associated to transposable elements. Front. Genet. 13, 891194. https://doi.org/10.3389/fgene.2022.891194.

Loewer S, Cabili MN, Guttman M, Loh Y-H, Thomas K, Park IH, Garber M, Curran M, Onder T, Agarwal S, Manos PD, Datta S, Lander ES, Schlaeger TM, Daley GQ, Rinn JL. 2010. Large intergenic non-coding RNA-RoR modulates reprogramming of human induced pluripotent stem cells. *Nat Genet* **42**:1113–1117. 1131 doi:10.1038/ng.710

Lu X, Sachs F, Ramsay L, Jacques P-É, Göke J, Bourque G, Ng H-H. 2014. The retrovirus HERVH is a long noncoding RNA required for human embryonic stem cell identity. *Nature Structural & Molecular Biology* **21**:423–425. doi:10.1038/nsmb.2799

Mager DL, Freeman JD. 1995. HERV-H Endogenous Retroviruses: Presence in the New World Branch but Amplification in the Old World Primate Lineage. Virology 213:395–404. doi:10.1006/viro.1995.0012

Mager DL, Goodchild NL. 1989. Homologous recombination between the LTRs of a human retrovirus-like element causes a 5-kb deletion in two siblings. Am J Hum Genet 45: 848–854.

Marnetto D, Molineris I, Grassi E, Provero P. 2014. Genome-wide identification and characterization of fixed human-specific regulatory regions. Am J Hum Genet 95: 39-48.

Matsumura, Y., Nakaki, R., Inagaki, T., Yoshida, A., Kano, Y., Kimura, H., Tanaka, T., Tsutsumi, S., Nakao, M., Doi, T., et al. 2015. H3K4/H3K9me3 Bivalent Chromatin Domains Targeted by Lineage-Specific DNA Methylation Pauses Adipocyte Differentiation. Mol. Cell 60, 584–596. https://doi.org/10.1016/j.molcel.2015.10.025

McHugh SB, Lopes-dos-Santos V, Gava GP, Hartwich K, Tam, SKE, Bannerman DM, Dupret D. 2022. Adult-born dentate granule cells promote hippocampal population sparsity. Nature Neuroscience. 25: 1481-1491.

McLean, CY, Bristor, D, Hiller, M, Clarke, SL, Schaar, BT, Lowe, CB, Wenger, AM. Bejerano, G. 2010. GREAT improves functional interpretation of cis-regulatory regions. Nat Biotechnol 28: 495-501.

McLean CY, Reno PL, Pollen AA, Bassan AI, Capellini TD, Guenther C, Indjeian VB, Lim X, Menke DB, Schaar BT, Wenger AM, Bejerano G, Kingsley DM. 2011. Human-specific loss of regulatory DNA and the evolution of human-specific traits. Nature 471: 216-9.

Mora-Bermudez F et al. 2016. Differences and similarities between human and chimpanzee neural progenitors during cerebral cortex development. Elife 5, doi:10.7554/eLife.18683.

Muchnik SK, Lorente-Galdos B, Santpere G & Sestan N. 2019. Modeling the Evolution of Human Brain Development Using Organoids. Cell 179, 1250–1253, doi:10.1016/j.cell.2019.10.041.

Ohnuki M, Tanabe K, Sutou K, Teramoto I, Sawamura Y, Narita M, Nakamura Michiko, Tokunaga Y, Nakamura Masahiro, Watanabe A, Yamanaka S, Takahashi K. 2014. Dynamic regulation of human endogenous retroviruses mediates factor-induced reprogramming and differentiation potential. Proc Natl Acad Sci USA **111**: 12426–12431. doi:10.1073/pnas.1413299111

Otani T, Marchetto MC, Gage FH, Simons BD & Livesey FJ. 2016. 2D and 3D Stem Cell Models of Primate Cortical Development Identify Species-Specific Differences in Progenitor Behavior Contributing to Brain Size. Cell Stem Cell 18, 467–480, doi:10.1016/j.stem.2016.03.003

Pasca AM et al. 2015. Functional cortical neurons and astrocytes from human pluripotent stem cells in 3D culture. Nat Methods 12, 671–678, doi:10.1038/nmeth.3415.

Pasca SP. 2018. The rise of three-dimensional human brain cultures. Nature 553, 437–445, doi:10.1038/nature25032.

Pollen AA et al. 2019. Establishing Cerebral Organoids as Models of Human-Specific Brain Evolution. Cell 176, 743–756.e717, doi:10.1016/j.cell.2019.01.017.

Pontis J, Planet E, Offner S, Turelli P, Duc J, Coudray A, Theunissen TW, Jaenisch R, Trono D. 2019. Hominoid-Specific Transposable Elements and KZFPs Facilitate Human Embryonic Genome Activation and Control Transcription in Naive Human ESCs. *Cell Stem Cell* **24**:724-735.e5. doi:10.1016/j.stem.2019.03.012

Prescott SL et al. 2015. Enhancer divergence and cis-regulatory evolution in the human and chimp neural crest. Cell 163, 68–83, doi:10.1016/j.cell.2015.08.036.

Prüfer K et al. 2014. The complete genome sequence of a Neanderthal from the Altai Mountains. *Nature* **505**, 43–49.

Qian X, Song H & Ming GL 2019. Brain organoids: advances, applications and challenges. Development 146, doi:10.1242/dev.166074.

Römer C, Singh M, Hurst LD, Izsvák Z. 2017. How to tame an endogenous retrovirus: HERVH and the evolution of human pluripotency. *Current Opinion in Virology*, Animal models for viral diseases Paleovirology **25**:49–58. doi:10.1016/j.coviro.2017.07.001

Rugg-Gunn, P.J., Cox, B.J., Ralston, A., and Rossant, J. 2010. Distinct histone modifications in stem cell lines and tissue lineages from the early mouse embryo. Proc. Natl. Acad. Sci. USA 107, 10783–10790. https://doi.org/10.1073/pnas.0914507107

Santoni FA, Guerra J, Luban J. 2012. HERV-H RNA is abundant in human embryonic stem cells and a precise marker for pluripotency. *Retrovirology* **9**:111. doi:10.1186/1742-4690-9-111

Schaefer NK, Shapiro B, Green RE. 2021. An ancestral recombination graph of human, Neanderthal, and Denisovan genomes. Sci Adv. 7(29): eabc0776. doi: 10.1126/sciadv.abc0776.

Schröder, A. R. et al. 2002. HIV-1 integration in the human genome favors active genes and local hotspots. Cell 110, 521–529.

Shin W, Lee J, Son SY, Ahn K, Kim HS, Han K. 2013. Human-specific HERV-K insertion causes genomic variations in the human genome. PLoS ONE 8:e60605. DOI: https://doi.org/10.1371/journal.pone.0060605, PMID: 23593260.

Sloan SA et al. 2017. Human Astrocyte Maturation Captured in 3D Cerebral Cortical Spheroids Derived from Pluripotent Stem Cells. Neuron 95, 779–790.e776, doi:10.1016/j.neuron.2017.07.035.

Subramanian RP, Wildschutte JH, Russo C, Coffin JM. 2011. Identification, characterization, and comparative genomic distribution of the HERV-K (HML-2) group of human endogenous retroviruses. Retrovirology 8:90. DOI: https://doi.org/10.1186/1742-4690-8-90, PMID: 22067224

Takahashi K, Nakamura M, Okubo C, Kliesmete Z, Ohnuki M, Narita M, Watanabe A, Ueda M, Takashima Y, Hellmann I, Yamanaka S. 2021. The pluripotent stem cell-specific transcript ESRG is dispensable for human pluripotency. *PLOS Genetics* **17**:e1009587. doi:10.1371/journal.pgen.1009587

Tavazoie, S., Hughes, J.D., Campbell, M.J., Cho, R.J., and Church, GM. 1999. Systematic determination of genetic network architecture. Nat Genet 22, 281–285.

Theunissen TW, Friedli M, He Y, Planet E, O'Neil RC, Markoulaki S, Pontis J, Wang H, Iouranova A, Imbeault M, Duc J, Cohen MA, Wert KJ, Castanon R, Zhang Z, Huang Y, Nery JR, Drotar J, Lungjangwa T, Trono D, Ecker JR, Jaenisch R. 2016. Molecular Criteria for Defining the Naive Human Pluripotent State. *Cell Stem Cell* **19**:502–515. doi:10.1016/j.stem.2016.06.011

Thomas J, Perron H, Feschotte C. 2018. Variation in proviral content among human genomes mediated by LTR recombination. *Mobile DNA* **9**:36. doi:10.1186/s13100-018-0142-3

Thompson PJ, Macfarlan TS, Lorincz MC. 2016. Long terminal repeats: from parasitic elements to building blocks of the transcriptional regulatory repertoire. Mol Cell 62: 766–776. doi:10.1016/j.molcel.2016.03.029

Turner G, Barbulescu M, Su M, Jensen-Seaman MI, Kidd KK, Lenz J. 2001. Insertional polymorphisms of full-length endogenous retroviruses in humans. Curr Biol 11:1531-1535.

Vargiu L, Rodriguez-Tomé P, Sperber GO, Cadeddu M, Grandi N, Blikstad V, Tramontano E, Blomberg J. 2016. Classification and characterization of human endogenous retroviruses; mosaic forms are common. Retrovirology 13:7. doi:10.1186/s12977-015-0232-y

Wang, G. P., Ciuffi, A., Leipzig, J., Berry, C. C., Bushman, F. D. 2007. HIV integration site selection: analysis by massively parallel pyrosequencing reveals association with epigenetic modifications. Genome Res. 17, 1186–1194.

Wang J, Xie G, Singh M, Ghanbarian AT, Raskó T, Szvetnik A, Cai H, Besser D, Prigione A, Fuchs NV, Schumann GG, Chen W, Lorincz MC, Ivics Z, Hurst LD, Izsvák Z. 2014. Primate-specific endogenous retrovirus-driven transcription defines naive-like stem cells. *Nature* **516**:405–409. 1248 doi:10.1038/nature13804

Wildschutte JH, Williams ZH, Montesion M, Subramanian RP, Kidd JM, Coffin JM. 2016. Discovery of unfixed endogenous retrovirus insertions in diverse human populations. Proc Natl Acad Sci USA 113:E2326–E2334. DOI: https://doi.org/10.1073/pnas.1602336113, PMID: 27001843

Xie Z, Bailey A, Kuleshov MV, Clarke DJB., Evangelista JE, Jenkins SL, Lachmann A, Wojciechowicz ML, Kropiwnicki E, Jagodnik KM, Jeon M, & Ma'ayan A. Gene set knowledge discovery with Enrichr. Current Protocols, 1, e90. 2021. doi: 10.1002/cpz1.90

Xu C, Li Q, Efimova O, He L, Tatsumoto S, StepanovaV, Oishi T, Udono T, Yamaguchi K, Shigenobu S, Kakita A, Nawa H, Khaitovich P, Go Y. 2018. Human-specific features of spatial gene expression and regulation in eight brain regions. Genome Res 28:1097–1110. https://doi.org/10.1101/gr.231357.117

Xu R, et al. 2022. Stage-specific H3K9me3 occupancy ensures retrotransposon silencing in human preimplantation embryos. Cell Stem Cell 29: 1051–1066.

Yu H, et al. 2022. Dynamic reprogramming of H3K9me3 at hominoid-specific retrotransposons during human preimplantation development. Cell Stem Cell 29: 1031–1050.

Zhang Y, Li T, Preissl S, Amaral ML, Grinstein JD, Farah EN, Destici E, Qiu Y, Hu R, Lee AY, Chee S, Ma K, Ye Z, Zhu Q, Huang H, Fang R, Yu L, Izpisua Belmonte JC, Wu J, Evans SM, Chi NC, Ren B. 2019. Transcriptionally active HERV-H retrotransposons demarcate topologically associating domains in human pluripotent stem cells. *Nature Genetics* **51**:1380–1388. doi:10.1038/s41588-019-0479-7

**Figure legends.**

**Figure 1.** Evolutionary conservation analysis of human embryo retroviral LTR elements (A; B) and graphical summaries of genomic features defining 26 Genomic Regulatory Networks (GRNs) governed by human embryo retroviral LTRs (C - F).

Conservation profiles reported as percentiles of highly conserved orthologous LTR loci identified in genomes 16 species of non-human primates (NHP) based on sequences identity analyses of 2416 MLT2A1, 3069 MLT2A2, and 3354 LTR7 fixed non-polymorphic LTR sequences in human genome (A). Conservation profiles reported as percentiles of highly conserved orthologous loci of human-specific LTR sequences identified in genomes 10 species of non-human primates (NHP) based on sequences identity analyses of 35 MLT2A1, 45 MLT2A2, and 175 LTR7 human-specific fixed non-polymorphic LTR sequences in human genome (B). Correlation plots (C; D; F) illustrate direct correlations between genome fractions and LTR set coverage values observed for 354 significantly enriched records (Binominal FDR Q value < 0.05) of Gene Ontology annotations. Reported numerical values were documented by the GREAT analyses of 5444 genes representing putative down-stream regulatory targets of 8839 human embryo retroviral LTRs. Correlation plots shown for 227 significantly enriched terms of GO Biological Process database (C), 47 significantly-enriched terms of GO Cellular Component database (D), and 80 significantly enriched terms of GO Molecular Function database (F). Panel (E) shows distribution of numbers of human embryo retroviral LTRs assigned by the GREAT algorithm to each of the 47 significantly enriched terms identified by GO annotations of GO Cellular Component database. Note a sharp drop of the numbers of LTR loci assigned to 13 top-scoring significantly enriched records compared to remaining 34 significant terms. These 13 top-scoring significantly enriched records were defined thereafter as Genomic Regulatory Networks (GRNs) identified by GO annotations of GO Cellular Component database (D). Shaded areas within correlation plots shown in (C; D; F) highlight 26 GRNs representing top-scoring significantly enriched terms identified during the GO annotations of corresponding GO databases. Additional information pertinent to statistical, genomic, molecular, and phenotypic features of 26 GRNs associated with human embryo retroviral LTRs is reported in the text; Tables 2 – 5; and Supplementary Materials (Supplementary Data Set S1: GRN1 – GRN26).

**Figure 2.** Graphical summaries of genomic features (A; B) and Directed Acyclic Graph (DAG) analyses (C – H) of the Genomic Regulatory Network 18 (GRN18: Regulation of Membrane Potential) consisting of 592 human embryo retroviral LTRs and 340 down-stream target genes.

Panel (A; top) reports numerical values of main genomic features assigned to the GRN18: Regulation of Membrane Potential and panel (A; bottom) presents a graphical summary of distribution of numbers of human embryo retroviral LTRs among the significantly enriched terms documented by GO annotations of GO Biological Process database (top-scoring 500 of 1315 significantly-enriched terms are shown, which were identified by the GREAT algorithm at the Binominal FDR Q value < 0.001). Panel (B) shows the correlation plot illustrating direct correlation between genome fractions and LTR set coverage values observed for 1315 significantly enriched records (Binominal FDR Q value < 0.001) of GO annotations of GO Biological Process database. Corresponding numerical values were obtained from the GREAT algorithm-generated tables using as an input 592 human embryo retroviral LTR loci assigned to the GRN18.

Panels (C – H) reports graphical summaries of DAG analyses of the GRN18: Regulation of Membrane Potential. Hierarchical networks were generated by the GREAT algorithm using as an input 592 human embryo retroviral LTR loci assigned to the GRN18 by the GO annotation analyses. Hierarchical networks of local DAG are shown for significantly enriched terms in GO Biological Process annotations (C; 620 significantly-enriched terms); GO Cellular Component annotations (D; 103 significantly-enriched terms); Human phenotype annotations (E; 82 significantly-enriched terms); GO Molecular Function annotations (F; 90 significantly-enriched terms); Mouse Phenotype annotations (G; 268 significantly-enriched terms); Mouse Phenotype Single KO annotations (H; 217 significantly-enriched terms). Complete tabular data and graphical summary reports pertinent to statistical, genomic, molecular, and phenotypic attributes of 26 GRNs and 69,573 retroviral LTR-linked Genomic Regulatory Modules (GRMs) associated with human embryo retroviral LTR elements are presented in the text; Tables 2 – 5; and Supplementary Materials (Supplementary Data Set S1: GRN1 – GRN26).

**Figure 3.** Graphical summaries of Directed Acyclic Graph (DAG) analyses of the Genomic Regulatory Network 5 (GRN5: Somatodendritic Compartment) consisting of 995 human embryo retroviral LTRs and 582 down-stream target genes.

Panels (A – H) reports graphical summaries of DAG analyses of the GRN5: Somatodendritic Compartment. Hierarchical networks were generated by the GREAT algorithm using as an input 995 human embryo retroviral LTR loci assigned by the GO annotation analyses to the GRN5. Hierarchical networks of local DAG are shown for significantly enriched terms in GO Biological Process annotations [A; B; 599 significantly-enriched terms shown in (A) as an array of bars sorted in descending order of the log10(Binominal p values) and reported in (B) as local DAG of a developmental hierarchical network]; Human phenotype annotations [C; D; 62 significantly-enriched terms shown in (C) as an array of bars sorted in descending order of the log10(Binominal p values) and reported in (D) as local DAG of a developmental hierarchical network]; GO Cellular Component annotations (E; 96 significantly-enriched terms reported as local DAG of a developmental hierarchical network); GO Molecular Function annotations (F; 64 significantly-enriched terms reported as local DAG of a developmental hierarchical network); Mouse Phenotype annotations (G; 361 significantly-enriched terms reported as local DAG of a developmental hierarchical network); Mouse Phenotype Single KO annotations (H; 245 significantly-enriched terms reported as local DAG of a developmental hierarchical network). Complete tabular data and graphical summary reports pertinent to statistical, genomic, molecular, and phenotypic attributes of 26 GRNs and 69,573 retroviral LTR-linked Genomic Regulatory Modules (GRMs) associated with human embryo retroviral LTR elements are presented in the text; Tables 2 – 5; and Supplementary Materials (Supplementary Data Set S1: GRN1 – GRN26).

**Figure 4.** Graphical summaries of Directed Acyclic Graph (DAG) analyses of the GSEA-defined networks of protein-protein interactions (PPI) consisting of 88 human embryo retroviral LTRs and 34 down-stream target genes.

Panels (A – I) reports graphical summaries of DAG analyses of the GSEA-defined PPI networks of 34 genes linked with 88 retroviral LTRs. Hierarchical networks were generated by the GREAT algorithm using as an input genomic coordinates of 88 human embryo retroviral LTR loci assigned by the GREAT to the core set of 34 genes of the PPI networks engaged in PPI with 121 transcription factors (Transcription Factors' PPI database) and 124 PPI Hub proteins (PPI Hub Proteins database). Hierarchical networks of local DAG are shown for significantly enriched terms in GO Biological Process annotations [A; B; 142 significantly-enriched terms shown in (A) as an array of bars sorted in descending order of the log10(Binominal p values) and reported in (B) as local DAG of a developmental hierarchical network]; GO Cellular Component annotations [C; D; 6 significantly-enriched terms shown in (C) as an array of bars sorted in descending order of the log10(Binominal p values) and reported in (D) as local DAG of a developmental hierarchical network]; GO Molecular Function annotations [E; F; 4 significantly-enriched terms shown in (E) as an array of bars sorted in descending order of the log10(Binominal p values) and reported in (F) as local DAG of a developmental hierarchical network]; Mouse Phenotype annotations [G; H; 327 significantly-enriched terms shown in (G) as an array of bars sorted in descending order of the log10(Binominal p values) and reported in (H) as local DAG of a developmental hierarchical network]; Mouse Phenotype Single KO annotations [I; J; 69 significantly-enriched terms shown in (I) as an array of bars sorted in descending order of the log10(Binominal p values) and reported in (J) as local DAG of a developmental hierarchical network]. Complete tabular data and graphical summary reports pertinent to statistical, genomic, molecular, and phenotypic attributes of retroviral LTR-linked GSEA-defined Genomic Regulatory Pathways (GRPs) and their constituent Genomic Regulatory Modules (GRMs) associated with human embryo retroviral LTR elements are presented in the text; Tables 5 – 6; and Supplementary Materials (Supplementary Table S1; Supplementary Figure S1; Supplementary Summaries S4-S5; Supplementary Data Set S2).

**Figure 5.** Visualization patterns of GSEA of retroviral LTRs down-stream targets comprising 5444 genes (A; B; E – H) and 34 genes (B; C) vis-a-vis of GSEA of 240 genes (D – H) encoding proteins engaged in PPI networks associated with 34 LTR target genes.

Panel (A) shows a clustergram visualization of GSEA of 5444 genes comprising a cumulative set of down-stream regulatory targets of human embryo retroviral LTRs employing the OMIM Expanded PPI database of human disease-causing genes (Enrichr). Note the uniform pattern of engagement of 34 genes in PPI with protein products of disease-causing genes implicated in thousands common human disorders. Top-scoring 30 of 175 significantly enriched records are shown. Panel (B) reports a scatter plot visualization of GSEA of 5444 genes (left scatter plots) and 34 genes (right scatter plots) employing the Transcription Factors PPI database (top scatter plots) and the Hub Proteins PPI database (bottom scatter plots). Panel (C) reports a clustergram visualization of GSEA of 34 genes employing the Transcription Factors PPI database (left clustergram) and the Hub Proteins PPI database (right clustergram). Top 30 of 121 (Transcription Factors PPI database) and top 30 of 124 (Hub Proteins PPI database) significantly enriched records are shown. Panel (D) reports a clustergram visualization of GSEA of 240 genes employing the Transcription Factors PPI database (left clustergram) and the Hub Proteins PPI database (right clustergram). Top 30 of 268 (Transcription Factors PPI database) and top 30 of 228 (Hub Proteins PPI database) significantly enriched records are shown. Panels (E – G) reports results of GSEA of 5444 genes (left figures) and 240 genes (right figures) employing the DisGeNET database of common human disorders. Panel (E) reports top 10 of 477 (left set of bars; GSEA of 5444 genes) and top 10 of 3298 (right set of bars; GSEA of 240 genes) significantly enriched records. Panel (F) shows a clustergram visualization of top 30 of 477 (left clustergram; GSEA of 5444 genes) and top 30 of 3298 (right clustergram; GSEA of 240 genes) significantly enriched records. Panel (G) reports a scatter plot visualization of 477 (left set of bars; GSEA of 5444 genes) and of 3298 (right set of bars; GSEA of 240 genes) significantly enriched records. Panel (H) a clustergram visualization of top 30 of 1095 (left clustergram; GSEA of 5444 genes) and top 30 of 199 (right clustergram; GSEA of 240 genes) significantly enriched records identified employing the Orphanet Augmented 2021 database of human orphan disorders (Enrichr). Significance threshold levels were set at adjusted p value < 0.05 (Methods).

Complete tabular data and graphical summary reports pertinent to statistical, genomic, molecular, and phenotypic attributes of retroviral LTR-linked GSEA-defined Genomic Regulatory Pathways (GRPs) and their constituent Genomic Regulatory Modules (GRMs) associated with human embryo retroviral LTR elements are presented in the text; Tables 5 – 6; and Supplementary Materials (Supplementary Table S1; Supplementary Figure S1; Supplementary Summaries S4-S5; Supplementary Data Set S2).

**Figure 6.** Visualization patterns of GSEA of retroviral 34 genes comprising down-stream regulatory targets of 88 retroviral LTRs vis-a-vis of GSEA of 240 genes encoding proteins that are engaged in PPI networks

associated with 34 retroviral LTR target genes. Panel (A) reports a scatter plot visualization of GSEA of 240 genes (268 significantly enriched records; left scatter plots) and 34 genes (121 significantly enriched records; right scatter plots) employing the Transcription Factors PPI database. Panel (B) reports a scatter plot visualization of GSEA of 240 genes (228 significantly enriched records; left scatter plots) and 34 genes (124 significantly enriched records; right scatter plots) employing the Hub Proteins PPI database. Panel (C) shows graphical representations of protein-protein interaction networks identified by the Enrichr software (Methods) from GSEA reported in (A) and (B). Panel (D) reports a scatter plot visualization of GSEA of 240 genes (606 significantly enriched records; left scatter plots) and 34 genes (143 significantly enriched records; right scatter plots) employing the CORUM Protein Complexes database. Panel (E) reports a scatter plot visualization of GSEA of 240 genes (62 significantly enriched records; left scatter plots) and 34 genes (22 significantly enriched records; right scatter plots) employing the SILAC Phosphoproteomics database. Panel (F) reports a scatter plot visualization of GSEA of 240 genes (1783 significantly enriched records; left scatter plots) and 34 genes (no significantly enriched records at adjusted p value < 0.05; right scatter plots) employing the NURSA Human Endogenous Complexome database. Panel (G) shows graphical representations of protein-protein interaction networks identified by the Enrichr software (Methods) from GSEA reported in (D - F). Panel (H) shows a clustergram visualization of GSEA of 240 genes that are engaged in PPI networks associated with 34 retroviral LTR target genes. Reported GSEA employed the OMIM Expanded PPI database of human disease-causing genes (Enrichr). Note the uniform pattern of engagement of 51 genes in PPI with protein products of disease-causing genes implicated in pathogenesis of 3298 common human disorders [Figure 5F (right panel); GSEA of the DisGeNET database]. Top-scoring 30 of 181 significantly enriched records are shown. Significance threshold levels were set at adjusted p value < 0.05 (Methods).

Complete tabular data and graphical summary reports pertinent to statistical, genomic, molecular, and phenotypic attributes of retroviral LTR-linked GSEA-defined Genomic Regulatory Pathways (GRPs) and their constituent Genomic Regulatory Modules (GRMs) associated with human embryo retroviral LTR elements are presented in the text; Tables 5 – 6; and Supplementary Materials (Supplementary Table S1; Supplementary Figure S1; Supplementary Summaries S4-S5; Supplementary Data Set S2).

**Figure 7.** Visualization of quantitative features of GSEA-defined genomic regulatory pathways (GRPs) and genomic regulatory modules (GRMs) mediating phenotypic impacts of human embryo regulatory LTR elements. A set of 709 genes comprising common down-stream regulatory targets of three families of human embryo retroviral LTRs (MLT2A1; MLT2A2; LTR7) were selected from 5444 genes comprising a cumulative set of down-stream regulatory targets of 8839 human embryo retroviral LTRs and subjected to GSEA employing a comprehensive panel of genomics and proteomics databases (Methods). GSEA-defined significantly enriched sets of genes assigned to specific GRPs (adjusted p value < 0.05) were cross-referenced to corresponding up-stream regulatory sequences of human embryo retroviral LTR elements to specify for each GRPs defined sets of down-stream target genes and up-stream regulatory retroviral LTR sequences. Each GSEA-defined GRPs were subjected to GO annotations and DAG validation analyses by imputing the corresponding sets retroviral LTRs into the GREAT algorithm software (Methods). Observed numbers of significantly enriched GO annotations' terms (representing constituent GRMs of corresponding GRPs) are visualized as five-leaf flower patterns for corresponding GRPs. Panel (A) reports selected examples of significantly-enriched quantitative features associated with specific GRPs pertinent to GO Biological Process annotations (GRPs of Neurexins & Neuroligins; Neurotransmitter receptors' activity; Receptor tyrosine kinases' signaling; Neuroactive ligands/receptors interactions). For GRPs of PPI networks of 34 genes linked to 88 LTRs, the number of significantly enriched records of the DisGeNET database GSEA is reported. Panel (B) reports selected examples of significantly-enriched quantitative features associated with specific GRPs pertinent to GO Mouse Phenotype annotations (GRPs of Neurotrophins; Neurotransmitter receptors' activity; Receptor tyrosine kinases' signaling; Neuroactive ligands/receptors interactions; PPI networks of 34 genes linked to 88 LTRs). Panel (C) reports numbers of significantly-enriched quantitative features associated with the GRP

Neurotrophins (Neurotrophic factors) in respective GO annotations' categories; Panel (D) reports numbers of significantly-enriched quantitative features associated with the GRP RET/GDNF signaling in respective GO annotations' categories; Panel (E) reports numbers of significantly-enriched quantitative features associated with the GRP Neurexins & Neuroligins in respective GO annotations' categories; Panel (F) reports numbers of significantly-enriched quantitative features associated with the GRP Neurotransmitter receptors' activity in respective GO annotations' categories; Panel (G) reports numbers of significantly-enriched quantitative features associated with the GRP Tyrosine kinase receptors in respective GO annotations' categories; Panel (H) reports numbers of significantly-enriched quantitative features associated with the GRP Neuroactive ligands/receptors interactions in respective GO annotations' categories; Panel (I) reports numbers of significantly-enriched quantitative features associated with the GRP Human diseases' PPI networks of 78 genes linked to 88 LTRs in respective GO annotations' categories; Panel (J) reports numbers of significantly-enriched quantitative features associated with the hypothetical GSEA-defined Genomic Regulatory Network (GRN) comprising of 255 genes linked with 966 retroviral LTRs in respective GO annotations' categories (see text for details).

Graphical summary reports of DAG validation analyses for corresponding GRPs are presented in Figure 8 – 13. Complete tabular data and graphical summary reports pertinent to statistical, genomic, molecular, and phenotypic attributes of retroviral LTR-linked GSEA-defined Genomic Regulatory Pathways (GRPs) and their constituent Genomic Regulatory Modules (GRMs) associated with human embryo retroviral LTR elements are presented in the text; Tables 5 – 6; and Supplementary Materials (Supplementary Table S1; Supplementary Figure S1; Supplementary Summaries S4-S5; Supplementary Data Set S2).

**Figure 8.** Graphical summaries of Directed Acyclic Graph (DAG) analyses of the GSEA-defined *RET/GNF* genomic regulatory pathway consisting of 115 human embryo retroviral LTRs and 28 down-stream target genes.

Panels (A – H) reports graphical summaries of DAG analyses of the GSEA-defined *RET/GNF* pathway comprising of 28 genes linked with 115 retroviral LTRs. Hierarchical networks were generated by the GREAT algorithm using as an input genomic coordinates of 115 human embryo retroviral LTR loci assigned by the GREAT to the set of 28 genes of the *RET/GNF* genomic regulatory pathway identified by GSEA of 709 genes comprising common down-stream regulatory targets of three families of human embryo retroviral LTRs (MLT2A1; MLT2A2; LTR7) (see Fig. 7 legend and text for details). Hierarchical networks of local DAG are shown for significantly enriched terms in GO Biological Process annotations [A; B; 181 significantly-enriched terms shown in (A) as an array of bars sorted in descending order of the log10(Binominal p values) and reported in (B) as local DAG of a developmental hierarchical network]; Human Phenotype annotations [C; D; 8 significantly-enriched terms shown in (C) as an array of bars sorted in descending order of the log10(Binominal p values) and reported in (D) as local DAG of a developmental hierarchical network]; Mouse Phenotype annotations [E; F; 191 significantly-enriched terms shown in (E) as an array of bars sorted in descending order of the log10(Binominal p values) and reported in (F) as local DAG of a developmental hierarchical network]; Mouse Phenotype Single KO annotations [G; H; 1077 significantly-enriched terms shown in (G) as an array of bars sorted in descending order of the log10(Binominal p values) and reported in (H) as local DAG of a developmental hierarchical network]. Complete tabular data and graphical summary reports pertinent to statistical, genomic, molecular, and phenotypic attributes of retroviral LTR-linked GSEA-defined Genomic Regulatory Pathways (GRPs) and their constituent Genomic Regulatory Modules (GRMs) associated with human embryo retroviral LTR elements are presented in the text; Tables 5 – 6; and Supplementary Materials (Supplementary Table S1; Supplementary Figure S1; Supplementary Summaries S4-S5; Supplementary Data Set S2).

**Figure 9.** Graphical summaries of Directed Acyclic Graph (DAG) analyses of the GSEA-defined Neurexins & Neuroligins genomic regulatory pathway consisting of 120 human embryo retroviral LTRs and 30 down-stream target genes.

Panels (A – F) reports graphical summaries of DAG analyses of the GSEA-defined Neurexins & Neuroligins pathway comprising of 30 genes linked with 120 retroviral LTRs. Hierarchical networks were generated by the GREAT algorithm using as an input genomic coordinates of 120 human embryo retroviral LTR loci assigned by the GREAT to the set of 30 genes of the Neurexins & Neuroligins genomic regulatory pathway identified by GSEA of 709 genes comprising common down-stream regulatory targets of three families of human embryo retroviral LTRs (MLT2A1; MLT2A2; LTR7) (see Fig. 7 legend and text for details). Hierarchical networks of local DAG are shown for significantly enriched terms in GO Biological Process annotations [A; B; 61 significantly-enriched terms shown in (A) as an array of bars sorted in descending order of the log10(Binominal p values) and reported in (B) as local DAG of a developmental hierarchical network]; GO Cellular Component annotations [C; D; 19 significantly-enriched terms shown in (C) as an array of bars sorted in descending order of the log10(Binominal p values) and reported in (D) as local DAG of a developmental hierarchical network]; GO Molecular Function annotations [E; F; 4 significantly-enriched terms shown in (E) as an array of bars sorted in descending order of the log10(Binominal p values) and reported in (F) as local DAG of a developmental hierarchical network]. Complete tabular data and graphical summary reports pertinent to statistical, genomic, molecular, and phenotypic attributes of retroviral LTR-linked GSEA-defined Genomic Regulatory Pathways (GRPs) and their constituent Genomic Regulatory Modules (GRMs) associated with human embryo retroviral LTR elements are presented in the text; Tables 5 – 6; and Supplementary Materials (Supplementary Table S1; Supplementary Figure S1; Supplementary Summaries S4-S5; Supplementary Data Set S2).

**Figure 10.** Graphical summaries of Directed Acyclic Graph (DAG) analyses of the GSEA-defined Neurotrophins (Neurotrophic Factors) genomic regulatory pathway consisting of 8 human embryo retroviral LTRs and 12 down-stream target genes.

Panels (A – H) reports graphical summaries of DAG analyses of the GSEA-defined Neurotrophins (Neurotrophic Factors) pathway comprising of 12 genes linked with 8 retroviral LTRs. Hierarchical networks were generated by the GREAT algorithm using as an input genomic coordinates of 8 human embryo retroviral LTR loci assigned by the GREAT to the set of 12 genes of the Neurotrophins (Neurotrophic Factors) genomic regulatory pathway identified by GSEA of 709 genes comprising common down-stream regulatory targets of three families of human embryo retroviral LTRs (MLT2A1; MLT2A2; LTR7) (see Fig. 7 legend and text for details). Hierarchical networks of local DAG are shown for significantly enriched terms in GO Biological Process annotations [A; B; 12 significantly-enriched terms shown in (A) as an array of bars sorted in descending order of the log10(Binominal p values) and reported in (B) as local DAG of a developmental hierarchical network]; Human Phenotype annotations [C; D; 30 significantly-enriched terms shown in (C) as an array of bars sorted in descending order of the log10(Binominal p values) and reported in (D) as local DAG of a developmental hierarchical network]; Mouse Phenotype annotations [E; F; 77 significantly-enriched terms shown in (E) as an array of bars sorted in descending order of the log10(Binominal p values) and reported in (F) as local DAG of a developmental hierarchical network]; Mouse Phenotype Single KO annotations [G; H; 70 significantly-enriched terms shown in (G) as an array of bars sorted in descending order of the log10(Binominal p values) and reported in (H) as local DAG of a developmental hierarchical network]. Complete tabular data and graphical summary reports pertinent to statistical, genomic, molecular, and phenotypic attributes of retroviral LTR-linked GSEA-defined Genomic Regulatory Pathways (GRPs) and their constituent Genomic Regulatory Modules (GRMs) associated with human embryo retroviral LTR elements are presented in the text; Tables 5 – 6; and Supplementary Materials (Supplementary Table S1; Supplementary Figure S1; Supplementary Summaries S4-S5; Supplementary Data Set S2).

**Figure 11.** Graphical summaries of Directed Acyclic Graph (DAG) analyses of the GSEA-defined Neurotransmitters Receptors Activity genomic regulatory pathway consisting of 137 human embryo retroviral LTRs and 30 down-stream target genes.

Panels (A – J) reports graphical summaries of DAG analyses of the GSEA-defined Neurotransmitters Receptors Activity pathway comprising of 30 genes linked with 137 retroviral LTRs. Hierarchical networks were generated by the GREAT algorithm using as an input genomic coordinates of 137 human embryo retroviral LTR loci assigned by the GREAT to the set of 30 genes of the Neurotransmitters Receptors Activity genomic regulatory pathway identified by GSEA of 709 genes comprising common down-stream regulatory targets of three families of human embryo retroviral LTRs (MLT2A1; MLT2A2; LTR7) (see Fig. 7 legend and text for details). Hierarchical networks of local DAG are shown for significantly enriched terms in GO Biological Process annotations [A; B; 34 significantly-enriched terms shown in (A) as an array of bars sorted in descending order of the log10(Binominal p values) and reported in (B) as local DAG of a developmental hierarchical network]; GO Cellular Component annotations [C; D; 29 significantly-enriched terms shown in (C) as an array of bars sorted in descending order of the log10(Binominal p values) and reported in (D) as local DAG of a developmental hierarchical network]; GO Molecular Function annotations [E; F; 38 significantly-enriched terms shown in (E) as an array of bars sorted in descending order of the log10(Binominal p values) and reported in (F) as local DAG of a developmental hierarchical network]; Mouse Phenotype annotations [G; H; 32 significantly-enriched terms shown in (G) as an array of bars sorted in descending order of the log10(Binominal p values) and reported in (H) as local DAG of a developmental hierarchical network]; Mouse Phenotype Single KO annotations [I; J; 20 significantly-enriched terms shown in (I) as an array of bars sorted in descending order of the log10(Binominal p values) and reported in (J) as local DAG of a developmental hierarchical network]. Complete tabular data and graphical summary reports pertinent to statistical, genomic, molecular, and phenotypic attributes of retroviral LTR-linked GSEA-defined Genomic Regulatory Pathways (GRPs) and their constituent Genomic Regulatory Modules (GRMs) associated with human embryo retroviral LTR elements are presented in the text; Tables 5 – 6; and Supplementary Materials (Supplementary Table S1; Supplementary Figure S1; Supplementary Summaries S4-S5; Supplementary Data Set S2).

**Figure 12.** Graphical summaries of Directed Acyclic Graph (DAG) analyses of the GSEA-defined Neuroactive ligands/receptors interactions genomic regulatory pathway consisting of 305 human embryo retroviral LTRs and 84 down-stream target genes.

Panels (A – J) reports graphical summaries of DAG analyses of the GSEA-defined Neuroactive ligands/receptors interactions pathway comprising of 84 genes linked with 305 retroviral LTRs. Hierarchical networks were generated by the GREAT algorithm using as an input genomic coordinates of 305 human embryo retroviral LTR loci assigned by the GREAT to the set of 84 genes of the Neuroactive ligands/receptors interactions genomic regulatory pathway identified by GSEA of 709 genes comprising common down-stream regulatory targets of three families of human embryo retroviral LTRs (MLT2A1; MLT2A2; LTR7) (see Fig. 7 legend and text for details). Hierarchical networks of local DAG are shown for significantly enriched terms in GO Biological Process annotations [A; B; 85 significantly-enriched terms shown in (A) as an array of bars sorted in descending order of the log10(Binominal p values) and reported in (B) as local DAG of a developmental hierarchical network]; GO Cellular Component annotations [C; D; 30 significantly-enriched terms shown in (C) as an array of bars sorted in descending order of the log10(Binominal p values) and reported in (D) as local DAG of a developmental hierarchical network]; GO Molecular Function annotations [E; F; 43 significantly-enriched terms shown in (E) as an array of bars sorted in descending order of the log10(Binominal p values) and reported in (F) as local DAG of a developmental hierarchical network]; Mouse Phenotype annotations [G; H; 85 significantly-enriched terms shown in (G) as an array of bars sorted in descending order of the log10(Binominal p values) and reported in (H) as local DAG of a developmental hierarchical network]; Mouse Phenotype Single KO annotations [I; J; 80 significantly-enriched terms shown in (I) as an array of bars sorted in descending order of the log10(Binominal p values) and reported in (J) as local DAG of a developmental hierarchical network]. Complete tabular data and graphical summary reports pertinent to statistical, genomic, molecular, and phenotypic attributes of retroviral LTR-linked GSEA-defined Genomic Regulatory Pathways (GRPs) and their constituent Genomic Regulatory Modules (GRMs) associated with human embryo retroviral LTR elements are presented in the text; Tables 5 – 6; and Supplementary Materials (Supplementary Table S1; Supplementary Figure S1; Supplementary Summaries S4-S5; Supplementary Data Set S2).

**Figure 13.** Graphical summaries of Directed Acyclic Graph (DAG) analyses of the GSEA-defined Tyrosine Kinase Receptors Activity genomic regulatory pathway consisting of 241 human embryo retroviral LTRs and 81 down-stream target genes.

Panels (A – J) reports graphical summaries of DAG analyses of the GSEA-defined Tyrosine Kinase Receptors Activity pathway comprising of 81 genes linked with 241 retroviral LTRs. Hierarchical networks were generated by the GREAT algorithm using as an input genomic coordinates of 241 human embryo retroviral LTR loci assigned by the GREAT to the set of 81 genes of the Tyrosine Kinase Receptors Activity genomic regulatory pathway identified by GSEA of 709 genes comprising common down-stream regulatory targets of three families of human embryo retroviral LTRs (MLT2A1; MLT2A2; LTR7) (see Fig. 7 legend and text for details). Hierarchical networks of local DAG are shown for significantly enriched terms in GO Biological Process annotations [A; B; 172 significantly-enriched terms shown in (A) as an array of bars sorted in descending order of the log10(Binominal p values) and reported in (B) as local DAG of a developmental hierarchical network]; GO Cellular Component annotations [C; D; 8 significantly-enriched terms shown in (C) as an array of bars sorted in descending order of the log10(Binominal p values) and reported in (D) as local DAG of a developmental hierarchical network]; GO Molecular Function annotations [E; F; 12 significantly-enriched terms shown in (E) as an array of bars sorted in descending order of the log10(Binominal p values) and reported in (F) as local DAG of a developmental hierarchical network]; Mouse Phenotype annotations [G; H; 69 significantly-enriched terms shown in (G) as an array of bars sorted in descending order of the log10(Binominal p values) and reported in (H) as local DAG of a developmental hierarchical network]; Mouse Phenotype Single KO annotations [I; J; 24 significantly-enriched terms shown in (I) as an array of bars sorted in descending order of the log10(Binominal p values) and reported in (J) as local DAG of a developmental hierarchical network]. Complete tabular data and graphical summary reports pertinent to statistical, genomic, molecular, and phenotypic attributes of retroviral LTR-linked GSEA-defined Genomic Regulatory Pathways (GRPs) and their constituent Genomic Regulatory Modules (GRMs) associated with human embryo retroviral LTR elements are presented in the text; Tables 5 – 6; and Supplementary Materials (Supplementary Table S1; Supplementary Figure S1; Supplementary Summaries S4-S5; Supplementary Data Set S2).

**Figure 14.** Graphical summaries of Directed Acyclic Graph (DAG) analyses of the GSEA-defined composite Genomic Regulatory Network (GRN) consisting of 966 human embryo retroviral LTRs and 255 down-stream target genes.

Panels (A – J) reports graphical summaries of DAG analyses of the GSEA-defined composite GRN comprising of 255 genes linked with 966 retroviral LTRs. Hierarchical networks were generated by the GREAT algorithm using as an input genomic coordinates of 966 human embryo retroviral LTR loci assigned by the GREAT to the set of 255 genes of the composite GRN assembled from GRPs (Figs. 8 – 13) identified by GSEA of 709 genes comprising common down-stream regulatory targets of three families of human embryo retroviral LTRs (MLT2A1; MLT2A2; LTR7) (see Fig. 7 legend and text for details). Hierarchical networks of local DAG are shown for significantly enriched terms in GO Biological Process annotations [A; B; 287 significantly-enriched terms shown in (A) as an array of bars sorted in descending order of the log10(Binominal p values) and reported in (B) as local DAG of a developmental hierarchical network]; GO Cellular Component annotations [C; D; 38 significantly-enriched terms shown in (C) as an array of bars sorted in descending order of the log10(Binominal p values) and reported in (D) as local DAG of a developmental hierarchical network]; GO Molecular Function annotations [E; F; 20 significantly-enriched terms shown in (E) as an array of bars sorted in descending order of the log10(Binominal p values) and reported in (F) as local DAG of a developmental hierarchical network]; Mouse Phenotype annotations [G; H; 264 significantly-enriched terms shown in (G) as an array of bars sorted in descending order of the log10 (Binominal p values) and reported in (H) as local DAG of a developmental hierarchical network]; Mouse Phenotype Single KO annotations [I; J; 136 significantly-enriched terms shown in (I) as an array of bars sorted in descending order of the log10(Binominal p values) and reported in (J) as local DAG of a developmental hierarchical network]. Complete tabular data and graphical summary reports pertinent to statistical, genomic, molecular, and phenotypic attributes of retroviral LTR-linked GSEA-defined Genomic Regulatory Pathways (GRPs) and their constituent Genomic Regulatory Modules (GRMs) associated with human embryo retroviral LTR elements are presented in the text; Tables 5 – 6; and Supplementary Materials (Supplementary Table S1; Supplementary Figure S1; Supplementary Summaries S4-S5; Supplementary Data Set S2).

**Figure 15.** Regulatory cross-talks of 8,839 human embryo retroviral LTR elements and 110,339 human specific regulatory sequences (HSRS) representing different families of HSRS.

Comparative analyses of enrichment and abundance patterns of genes comprising candidate down-stream regulatory targets of human embryo retroviral LTR elements. Genes identified in this contribution as candidate down-stream regulatory targets of human embryo retroviral LTR elements were compared with previously reported gene sets expression of which is governed by different categories of genomic regulatory loci, including various families of primate-specific and human-specific regulatory sequences (HSRS). Panels A; C; E; G report graphical summaries of the enrichment analyses, while Panels B; D; F; H report graphical summaries of the analyses of abundance profiles. Results of the analyses of 3971 down-stream target genes associated with MLT2A1 & MLT2A2 loci and 2955 genes associated with LTR7 elements are shown in the panels (A; B) and (C; D), respectively. Panels (E; F) report results of the analyses of a cumulative set of 5444 genes comprising a cumulative set of candidate down-stream regulatory targets of LTR7, MLT2A1, and MLT2A2 human embryo regulatory LTRs. Panels (G; H) show results of the analyses of a common set of 709 down-stream regulatory target genes shared by LTR7, MLT2A1, and MLT2A2 human embryo regulatory LTR elements. Please refer to text for additional information and interpretations of reported observations. Detailed reports of the quantitative metrics pertinent to these analyses are presented in the Table 12; Supplementary Table S4 (Excel spreadsheets st1 – st5).

**Figure 16.** Identification and characterization of down-stream target genes of human embryo retroviral LTR elements that are significantly enriched among genes with species-specific expression mapping bias in Human-Chimpanzee hybrids of induced pluripotent stem cells (iPSC) and brain organoids (cortical spheroids, CS). Genes manifesting species-specific expression mapping bias in Human-Chimpanzee hybrids were analyzed to identify down-stream target genes of human embryo retroviral LTR elements with species-specific expression bias in Human-Chimpanzee hybrids (Table 13; Supplementary Summary S6). Significance of the enrichment of human embryo retroviral LTR targets among genes with species-specific expression bias in Human-Chimpanzee hybrids were ascertained for three different categories of genes, including 374 genes with species-specific expression bias in hybrid iPSC (Panels A – B; Table 13), 291 genes with species-specific expression bias in hybrid CS (Panels C –D; Table 13), and 183 genes manifesting species allele-specific expression profiles in human-chimpanzee hybrid cells (Table 13). Panels (A – D) show species-specific expression profiles in hybrid iPSC (A; B) and hybrid CS (C; D) of 374 genes (A) and 95 retroviral LTR target genes (B) as well as 291 genes (C) and 63 retroviral LTR target genes (D). In panels (A – D) genes were sorted in descending orders of human alleles' expression values (blue bars) to highlight species-specific differences of gene expression profiles in Human-Chimpanzee hybrids. Panel (E) reports the visualization summary of GSEA-guided identification of 53 retroviral LTR target genes constituting genetic markers of Dentate Granule Cells (Table 14; Supplementary Table S5; Supplementary Summary S6). A total of 177 human embryo retroviral LTR target genes with species-specific expression mapping bias in Human-Chimpanzee hybrids were identified and subjected to GSEA employing a panel of genomics and proteomics database. GSEA of 177 genes employing the ARCHS4 Human Tissues database identified the Dentate Granule Cells as the top-scoring significantly enriched record (Panel E; Table 14; Supplementary Table S5; Supplementary Summary S6). Panels (F – G) reports significantly enriched terms of the Mouse Phenotype Single KO database annotations of the 223 retroviral LTRs linked to 53 down-stream target genes comprising genetic markers of the Dentate Granule Cells. Panels (F; G) report 6 significantly-enriched terms shown in (F) as an array of bars sorted in descending order of the log10 (Binominal p values) and reported in (G) as local DAG of a developmental hierarchical network (see text for details). Panel (H) reports enrichment/depletion patterns of human embryo retroviral LTR regulatory targets among genes comprising 24 differential gene expression modules of primates' brain regions delineated by comparative analyses of eight brain regions of humans, chimpanzees, gorillas, a gibbon, and macaques (see text for details). Red bars marked with stars denote 3 modules manifesting most significant enrichment (modules 22; 16; 15), while yellow bars denote 7 modules, including module 7 with chimpanzee-specific expression, exhibiting most apparent relative depletion of human embryo retroviral LTR target genes.

**Table 1.** Numbers of human embryo retroviral LTR loci and their putative down-stream regulatory target genes in human genome.

| Human embryo regulatory LTR families | Number of loci in human genome | Number of putative regulatory target genes |
|---|---|---|
| **MLT2A1** | 2416 | 2450 |
| **MLT2A2** | 3069 | 2701 |
| **LTR7** | 3354 | 2955 |
| **MLT2A1 & MLTA2 (all loci)** | 5485 | 3971 |
| **LTR7 & MLT2A1 & MLT2A2 (all loci)** | 8839 | 5444 |
| **Human embryo regulatory LTR families** | **Number of loci in human genome** | **Number of common regulatory target genes** |
| **MLT2A1 & MLTA2 loci common regulatory target genes** | 5485 | 1180 |
| **MLTA1 & LTR7 loci common regulatory target genes** | 5770 | 1044 |
| **MLTA2 & LTR7 loci common regulatory target genes** | 6423 | 1149 |
| **LTR7 & MLT2A1 & MLT2A2 loci common regulatory target genes** | 8839 | 709 |

**Table 2.**Top-ranked biological processes preferentially affected by 8,839 human embryo retroviral LTRs identified by proximity placement-guided Genomic Regulatory Network (GRN) analysis* of Gene Ontology (GO) Biological Process database.

| GO ID | Description | Binominal Rank | Binominal P value | Binominal Bonferroni P value | Binominal FDR Q value | LTR Fold Enrichment | Observed LTRs | Gene Fold Enrichment | Observed Genes |
|---|---|---|---|---|---|---|---|---|---|
| **GO:0098742** | cell-cell adhesion via plasma-membrane adhesion molecules | 1 | 2.05E-19 | 2.65E-15 | 2.65E-15 | 1.42 | 677 | 1.46 | 95 |
| **GO:0007156** | homophilic cell adhesion via plasma membrane adhesion molecules | 2 | 1.79E-16 | 2.31E-12 | 1.15E-12 | 1.47 | 481 | 1.38 | 62 |
| **GO:0051965** | positive regulation of synapse assembly | 3 | 1.12E-12 | 1.45E-08 | 4.83E-09 | 1.48 | 355 | 2.12 | 40 |
| **GO:0006408** | snRNA export from nucleus | 4 | 1.03E-10 | 1.33E-06 | 3.33E-07 | 14.46 | 12 | 1.72 | 1 |
| **GO:0051963** | regulation of synapse assembly | 5 | 4.35E-10 | 5.62E-06 | 1.12E-06 | 1.38 | 392 | 2.05 | 51 |
| **GO:0051030** | snRNA transport | 6 | 8.87E-10 | 1.15E-05 | 1.91E-06 | 11.92 | 12 | 1.15 | 1 |
| **GO:0044691** | tooth eruption | 7 | 3.88E-09 | 5.02E-05 | 7.17E-06 | 3.74 | 29 | 2.30 | 2 |
| **GO:0050803** | regulation of synapse structure or activity | 8 | 1.21E-08 | 1.56E-04 | 1.95E-05 | 1.31 | 448 | 1.80 | 68 |
| **GO:0098609** | cell-cell adhesion | 9 | 2.21E-08 | 2.86E-04 | 3.18E-05 | 1.17 | 1089 | 1.29 | 238 |
| **GO:0050807** | regulation of synapse organization | 10 | 2.87E-08 | 3.71E-04 | 3.71E-05 | 1.30 | 440 | 1.82 | 66 |
| **GO:0098904** | regulation of AV node cell action potential | 11 | 8.99E-08 | 1.16E-03 | 1.06E-04 | 3.22 | 29 | 3.45 | 3 |
| **GO:0086072** | AV node cell-bundle of His cell adhesion involved in cell communication | 12 | 1.47E-07 | 1.90E-03 | 1.58E-04 | 3.89 | 22 | 3.45 | 1 |
| **GO:0007210** | serotonin receptor signaling pathway | 13 | 1.60E-07 | 2.07E-03 | 1.59E-04 | 2.21 | 54 | 3.45 | 7 |
| **GO:1904292** | regulation of ERAD pathway | 14 | 1.78E-07 | 2.30E-03 | 1.64E-04 | 2.13 | 58 | 0.89 | 7 |
| **GO:0046068** | cGMP metabolic process | 15 | 3.43E-07 | 4.43E-03 | 2.95E-04 | 1.78 | 90 | 1.56 | 14 |
| **GO:0051096** | positive regulation of helicase activity | 16 | 3.66E-07 | 4.72E-03 | 2.95E-04 | 3.16 | 27 | 1.15 | 2 |
| **GO:0007416** | synapse assembly | 17 | 3.77E-07 | 4.86E-03 | 2.86E-04 | 1.36 | 285 | 1.57 | 35 |
| **GO:0000710** | meiotic mismatch repair | 18 | 3.77E-07 | 4.87E-03 | 2.70E-04 | 8.73 | 10 | 1.72 | 1 |
| **GO:0034497** | protein localization to pre-autophagosomal structure | 19 | 4.80E-07 | 6.20E-03 | 3.27E-04 | 2.56 | 37 | 1.15 | 4 |
| **GO:0072378** | blood coagulation, fibrin clot formation | 20 | 4.90E-07 | 6.34E-03 | 3.17E-04 | 2.29 | 46 | 1.46 | 11 |
| **GO:0097104** | postsynaptic membrane assembly | 21 | 5.18E-07 | 6.69E-03 | 3.19E-04 | 1.79 | 86 | 2.07 | 6 |
| **GO:0099560** | synaptic membrane adhesion | 22 | 5.29E-07 | 6.83E-03 | 3.10E-04 | 2.05 | 59 | 3.45 | 3 |
| **GO:0006182** | cGMP biosynthetic process | 23 | 5.44E-07 | 7.03E-03 | 3.06E-04 | 1.99 | 63 | 1.29 | 9 |
| **GO:0007215** | glutamate receptor signaling pathway | 24 | 7.16E-07 | 9.25E-03 | 3.85E-04 | 1.48 | 173 | 2.12 | 27 |
| **GO:0098792** | xenophagy | 25 | 7.41E-07 | 9.57E-03 | 3.83E-04 | 5.79 | 13 | 0.86 | 1 |
| **GO:0046629** | gamma-delta T cell activation | 26 | 7.53E-07 | 9.72E-03 | 3.74E-04 | 3.30 | 24 | 1.15 | 2 |
| **GO:1904293** | negative regulation of ERAD pathway | 27 | 8.32E-07 | 1.07E-02 | 3.98E-04 | 2.43 | 39 | 0.80 | 3 |
| **GO:0052652** | cyclic purine nucleotide metabolic process | 28 | 1.19E-06 | 1.53E-02 | 5.47E-04 | 1.89 | 68 | 1.08 | 10 |
| **GO:1900371** | regulation of purine nucleotide biosynthetic process | 29 | 1.19E-06 | 1.54E-02 | 5.30E-04 | 1.36 | 260 | 1.47 | 67 |
| **GO:0006538** | glutamate catabolic process | 30 | 1.22E-06 | 1.58E-02 | 5.26E-04 | 2.28 | 43 | 1.72 | 4 |
| **GO:0009190** | cyclic nucleotide biosynthetic process | 31 | 1.41E-06 | 1.82E-02 | 5.86E-04 | 1.88 | 68 | 1.05 | 10 |
| **GO:0051962** | positive regulation of nervous system development | 32 | 1.47E-06 | 1.90E-02 | 5.95E-04 | 1.16 | 983 | 1.50 | 204 |
| **GO:0030808** | regulation of nucleotide biosynthetic process | 33 | 1.48E-06 | 1.91E-02 | 5.78E-04 | 1.35 | 260 | 1.46 | 67 |
| **GO:0097105** | presynaptic membrane assembly | 34 | 1.95E-06 | 2.52E-02 | 7.42E-04 | 1.79 | 77 | 1.92 | 5 |
| **GO:0051095** | regulation of helicase activity | 35 | 2.02E-06 | 2.61E-02 | 7.47E-04 | 2.88 | 27 | 0.77 | 2 |
| **GO:0034332** | adherens junction organization | 36 | 2.45E-06 | 3.17E-02 | 8.81E-04 | 1.39 | 216 | 1.39 | 29 |
| **GO:0007158** | neuron cell-cell adhesion | 37 | 2.60E-06 | 3.36E-02 | 9.09E-04 | 1.61 | 106 | 2.30 | 10 |
| **GO:0045409** | negative regulation of interleukin-6 biosynthetic process | 38 | 2.87E-06 | 3.70E-02 | 9.75E-04 | 4.43 | 15 | 2.07 | 3 |
| **GO:0010838** | positive regulation of keratinocyte proliferation | 39 | 3.44E-06 | 4.44E-02 | 1.14E-03 | 2.11 | 47 | 2.30 | 6 |
| **GO:0097090** | presynaptic membrane organization | 40 | 3.60E-06 | 4.65E-02 | 1.16E-03 | 1.76 | 77 | 1.72 | 5 |
| **GO:0007157** | heterophilic cell-cell adhesion via plasma membrane cell adhesion molecules | 41 | 4.17E-06 | 5.38E-02 | 1.31E-03 | 1.38 | 208 | 1.93 | 28 |
| **GO:0061739** | protein lipidation involved in autophagosome assembly | 42 | 4.22E-06 | 5.45E-02 | 1.30E-03 | 4.58 | 14 | 1.72 | 2 |
| **GO:0070295** | renal water absorption | 43 | 4.45E-06 | 5.75E-02 | 1.34E-03 | 3.25 | 21 | 0.86 | 1 |
| **GO:0045656** | negative regulation of monocyte differentiation | 44 | 4.62E-06 | 5.97E-02 | 1.36E-03 | 2.64 | 29 | 2.46 | 5 |
| **GO:0009065** | glutamine family amino acid catabolic process | 45 | 5.01E-06 | 6.47E-02 | 1.44E-03 | 1.92 | 57 | 1.38 | 10 |
| **GO:0045659** | negative regulation of | 46 | 5.89E-06 | 7.61E-02 | 1.65E-03 | 3.57 | 18 | 3.45 | 2 |
| | neutrophil differentiation | | | | | | | | |
| **GO:0003097** | renal water transport | 47 | 6.18E-06 | 7.99E-02 | 1.70E-03 | 3.18 | 21 | 0.69 | 1 |
| **GO:0016074** | snoRNA metabolic process | 48 | 6.30E-06 | 8.14E-02 | 1.69E-03 | 2.98 | 23 | 1.72 | 7 |
| **GO:1902567** | negative regulation of eosinophil activation | 49 | 6.38E-06 | 8.25E-02 | 1.68E-03 | 10.53 | 7 | 3.45 | 2 |
| **GO:0045658** | regulation of neutrophil differentiation | 50 | 6.72E-06 | 8.69E-02 | 1.74E-03 | 3.54 | 18 | 2.30 | 2 |
| **GO:0006536** | glutamate metabolic process | 51 | 7.24E-06 | 9.35E-02 | 1.83E-03 | 1.83 | 63 | 1.19 | 10 |
| **GO:0030853** | negative regulation of granulocyte differentiation | 52 | 7.30E-06 | 9.43E-02 | 1.81E-03 | 3.14 | 21 | 1.29 | 3 |
| **GO:0071362** | cellular response to ether | 53 | 7.34E-06 | 9.48E-02 | 1.79E-03 | 2.80 | 25 | 2.07 | 3 |
| **GO:1904637** | cellular response to ionomycin | 54 | 7.82E-06 | 1.01E-01 | 1.87E-03 | 2.94 | 23 | 1.72 | 2 |

Legend: *Proximity-placement-guided Genomic Regulatory Network (GRN) analyses were performed employing the GREAT algorithm software as described in Methods.

**Table 3.** Quantitative profiles of retroviral LTR elements, down-stream target genes, and genomic regulatory modules (GRMs) * within 26 gene ontology enrichment analysis-defined genomic regulatory networks (GRNs) ** of the genomic dominion of human embryo retroviral LTRs.

| Gene ontology ID | Genomic regulatory networks | Number of LTRs | Number of genes | GO Biological Process | GO Cellular Component | GO Molecular Function | Human Phenotype | Mouse Phenotype Single KO | Mouse Phenotype | Number of GRNs & GRMs |
|---|---|---|---|---|---|---|---|---|---|---|
| **NA** | **Genomic dominion of human embryo retroviral LTRs** | 8839 | 5444 | 38 | 20 | 15 | 7 | 6 | 9 | 95 |
| **GO:0051960** | **Regulation of nervous system development** | 1424 | 716 | 1676 | 110 | 132 | 261 | 1028 | 1413 | 4620 |
| **GO:0051962** | **Positive regulation of nervous system development** | 983 | 443 | 1390 | 103 | 115 | 178 | 887 | 1121 | 3794 |
| **GO:0098609** | **Cell-cell adhesion** | 1089 | 492 | 787 | 77 | 51 | 135 | 491 | 707 | 2231 |
| **GO:0098742** | **Cell-cell adhesion via plasma-membrane adhesion molecules** | 677 | 199 | 375 | 46 | 44 | 52 | 299 | 326 | 1142 |
| **GO:0099537** | **Trans-synaptic signaling** | 616 | 344 | 919 | 112 | 146 | 142 | 599 | 781 | 2699 |
| **GO:0007268** | **Chemical synaptic transmission** | 604 | 332 | 888 | 110 | 145 | 142 | 564 | 725 | 2574 |
| **GO:0050804** | **Modulation of synaptic transmission** | 529 | 298 | 1253 | 118 | 161 | 154 | 625 | 768 | 3079 |
| **GO:0044708** | **Single organism behavior** | 674 | 372 | 1336 | 91 | 111 | 196 | 718 | 843 | 3295 |
| **GO:0042391** | **Regulation of membrane potential** | 592 | 340 | 1315 | 145 | 203 | 387 | 732 | 822 | 3604 |
| **GO:1901137** | **Carbohydrate derivative biosynthetic process** | 678 | 448 | 526 | 41 | 148 | 70 | 218 | 249 | 1252 |
| **GO:0043005** | **Neuron projection** | 1486 | 805 | 1526 | 188 | 200 | 163 | 868 | 981 | 3926 |
| **GO:0098590** | **Plasma membrane region** | 1310 | 788 | 1016 | 193 | 247 | 75 | 642 | 833 | 3006 |
| **GO:0044456** | **Synapse part** | 1089 | 604 | 951 | 167 | 196 | 136 | 611 | 732 | 2793 |
| **GO:0009986** | **Cell surface** | 921 | 560 | 1965 | 142 | 172 | 137 | 1048 | 1164 | 4628 |
| **GO:0098794** | **Postsynapse** | 826 | 406 | 842 | 159 | 170 | 148 | 588 | 727 | 2634 |
| **GO:0044297** | **Cell body** | 737 | 409 | 1644 | 204 | 198 | 163 | 881 | 977 | 4067 |
| **GO:0097060** | **Synaptic membrane** | 679 | 305 | 664 | 133 | 169 | 138 | 572 | 680 | 2356 |
| **GO:0045211** | **Postsynaptic membrane** | 577 | 240 | 515 | 116 | 148 | 136 | 543 | 646 | 2104 |
| **GO:0036477** | **Somatodendritic compartment** | 995 | 582 | 1640 | 180 | 208 | 246 | 898 | 999 | 4171 |
| **GO:0043025** | **Neuronal cell body** | 645 | 359 | 1480 | 174 | 176 | 191 | 855 | 874 | 3750 |
| **GO:0030424** | **Axon** | 651 | 335 | 1029 | 152 | 130 | 143 | 465 | 562 | 2481 |
| **GO:0031012** | **Extracellular matrix** | 674 | 395 | 259 | 57 | 97 | 309 | 402 | 405 | 1529 |
| **GO:0005578** | **Proteinaceous extracellular matrix** | 554 | 311 | 356 | 40 | 87 | 232 | 229 | 245 | 1189 |
| **GO:0005509** | **Calcium ion binding** | 979 | 462 | 397 | 26 | 98 | 43 | 192 | 212 | 968 |
| **GO:0008233** | **Peptidase activity** | 574 | 400 | 303 | 30 | 83 | 71 | 175 | 204 | 849 |
| **GO:0070011** | **Peptidase activity, acting on L-amino acid peptides** | 547 | 384 | 306 | 29 | 83 | 47 | 162 | 197 | 824 |
| **NA** | **Cumulative, within 26 genomic regulatory networks** | 5237 | 3054 | 1105 | 121 | 190 | 71 | 424 | 486 | 2397 |
| **NA** | **Cumulative, excluding 26 genomic regulatory networks** | 3602 | 2685 | 27 | 12 | 23 | 17 | 26 | 24 | 129 |

Legend: *For each GRN, reported numbers of LTRs, genes, and GRMs were determined at the statistical significance threshold of the Binominal FDR Q-Value < 0.001 applied to 6 independent gene ontology-specific enrichment analyses employing the GREAT algorithm (Supplementary Materials); **, A total of 26 gene ontology enrichment analyses-defined GRNs associated with human embryo retroviral LTRs were identified at the statistical significance threshold of the Binominal FDR Q-value < 0.05 (Supplementary Table S1). GRNs contain from 529 to 1486 retroviral LTR elements and from 199 to 805 down-stream target genes.

**Table 4.** Quantitative profiles of retroviral LTR elements and down-stream target genes within 26 gene ontology enrichment analysis-defined genomic regulatory networks (GRNs) of the genomic dominion of human embryo retroviral LTRs.

| Gene ontology ID | Genomic regulatory networks | Number, LTR7 | Number, MLT2A1 | Number, MLT2A2 | Number of linked genes | Percent, LTR7 | Percent, MLT2A1 | Percent, MLT2A2 |
|---|---|---|---|---|---|---|---|---|
| NA | Genomic dominion of human embryo retroviral LTRs | 3354 | 2416 | 3069 | 5444 | 100.0 | 100.0 | 100.0 |
| GO:0051960 | Regulation of nervous system development | 500 | 409 | 515 | 716 | 14.9 | 16.9 | 16.8 |
| GO:0051962 | Positive regulation of nervous system development | 333 | 290 | 360 | 443 | 9.9 | 12.0 | 11.7 |
| GO:0098609 | Cell-cell adhesion | 385 | 301 | 403 | 492 | 11.5 | 12.5 | 13.1 |
| GO:0098742 | Cell-cell adhesion via plasma-membrane adhesion molecules | 235 | 186 | 256 | 199 | 7.0 | 7.7 | 8.3 |
| GO:0099537 | Trans-synaptic signaling | 215 | 161 | 240 | 344 | 6.4 | 6.7 | 7.8 |
| GO:0007268 | Chemical synaptic transmission | 211 | 158 | 235 | 332 | 6.3 | 6.5 | 7.7 |
| GO:0050804 | Modulation of synaptic transmission | 174 | 133 | 222 | 298 | 5.2 | 5.5 | 7.2 |
| GO:0044708 | Single organism behavior | 230 | 185 | 259 | 372 | 6.9 | 7.7 | 8.4 |
| GO:0042391 | Regulation of membrane potential | 217 | 151 | 224 | 340 | 6.5 | 6.3 | 7.3 |
| GO:1901137 | Carbohydrate derivative biosynthetic process | 261 | 175 | 242 | 448 | 7.8 | 7.2 | 7.9 |
| GO:0043005 | Neuron projection | 542 | 411 | 533 | 805 | 16.2 | 17.0 | 17.4 |
| GO:0098590 | Plasma membrane region | 519 | 337 | 454 | 788 | 15.5 | 13.9 | 14.8 |
| GO:0044456 | Synapse part | 373 | 293 | 423 | 604 | 11.1 | 12.1 | 13.8 |
| GO:0009986 | Cell surface | 370 | 239 | 312 | 560 | 11.0 | 9.9 | 10.2 |
| GO:0098794 | Postsynapse | 276 | 229 | 321 | 406 | 8.2 | 9.5 | 10.5 |
| GO:0044297 | Cell body | 275 | 204 | 258 | 409 | 8.2 | 8.4 | 8.4 |
| GO:0097060 | Synaptic membrane | 232 | 182 | 265 | 305 | 6.9 | 7.5 | 8.6 |
| GO:0045211 | Postsynaptic membrane | 195 | 156 | 226 | 240 | 5.8 | 6.5 | 7.4 |
| GO:0036477 | Somatodendritic compartment | 346 | 281 | 368 | 582 | 10.3 | 11.6 | 12.0 |
| GO:0043025 | Neuronal cell body | 225 | 188 | 232 | 359 | 6.7 | 7.8 | 7.6 |
| GO:0030424 | Axon | 246 | 174 | 231 | 335 | 7.3 | 7.2 | 7.5 |
| GO:0031012 | Extracellular matrix | 244 | 199 | 231 | 395 | 7.3 | 8.2 | 7.5 |
| GO:0005578 | Proteinaceous extracellular matrix | 191 | 176 | 187 | 311 | 5.7 | 7.3 | 6.1 |
| GO:0005509 | Calcium ion binding | 354 | 228 | 397 | 462 | 10.6 | 9.4 | 12.9 |
| GO:0008233 | Peptidase activity | 240 | 158 | 176 | 400 | 7.2 | 6.5 | 5.7 |
| GO:0070011 | Peptidase activity, acting on L-amino acid peptides | 226 | 152 | 169 | 384 | 6.7 | 6.3 | 5.5 |
| NA | Cumulative, within 26 genomic regulatory networks | 1952 | 1393 | 1892 | 3054 | 58.2 | 57.7 | 61.6 |
| NA | Cumulative, excluding 26 genomic regulatory networks | 1402 | 1023 | 1177 | 2685 | 41.8 | 42.3 | 38.4 |

Legend: *For each GRN, reported numbers of LTRs and genes were determined at the statistical significance threshold of the Binominal FDR Q-Value < 0.001 applied to 6 independent gene ontology-specific enrichment analyses employing the GREAT algorithm (Supplementary Materials); **, A total of 26 gene ontology enrichment analyses-defined GRNs associated with human embryo retroviral LTRs were identified at the statistical significance threshold of the Binominal FDR Q-value < 0.05 (Supplementary Table S1). GRNs contain from 529 to 1486 retroviral LTR elements and from 199 to 805 down-stream target genes.

**Table 5.** Quantitative profiles of a directed acyclic graph (DAG) * test-validated genomic regulatory modules (GRMs) ** within 26 gene ontology enrichment analysis-defined genomic regulatory networks (GRNs) *** of the genomic dominion of human embryo retroviral LTRs.

| Gene ontology ID | Genomic regulatory networks | Number of LTRs | GRN linked genes | GO Biological Process | GO Cellular Component | GO Molecular Function | Human Phenotype | Mouse Phenotype Single KO | Mouse Phenotype | Number of GRNs & GRMs |
|---|---|---|---|---|---|---|---|---|---|---|
| **NA** | **Genomic dominion of human embryo retroviral LTRs** | 8839 | NA | 1 | 0 | 0 | 1 | 0 | 0 | 2 |
| **GO:0051960** | **Regulation of nervous system development** | 1424 | 333 | 972 | 46 | 31 | 50 | 521 | 838 | 2458 |
| **GO:0051962** | **Positive regulation of nervous system development** | 983 | 204 | 800 | 39 | 19 | 27 | 401 | 598 | 1884 |
| **GO:0098609** | **Cell-cell adhesion** | 1089 | 238 | 470 | 64 | 14 | 52 | 230 | 369 | 1199 |
| **GO:0098742** | **Cell-cell adhesion via plasma-membrane adhesion molecules** | 677 | 95 | 69 | 34 | 2 | 0 | 1 | 1 | 107 |
| **GO:0099537** | **Trans-synaptic signaling** | 616 | 151 | 482 | 64 | 102 | 6 | 322 | 435 | 1411 |
| **GO:0007268** | **Chemical synaptic transmission** | 604 | 146 | 440 | 60 | 96 | 7 | 253 | 356 | 1212 |
| **GO:0050804** | **Modulation of synaptic transmission** | 529 | 130 | 746 | 71 | 42 | 18 | 279 | 357 | 1513 |
| **GO:0044708** | **Single organism behavior** | 674 | 165 | 638 | 59 | 19 | 10 | 344 | 467 | 1537 |
| **GO:0042391** | **Regulation of membrane potential** | 592 | 153 | 620 | 103 | 90 | 82 | 217 | 268 | 1380 |
| **GO:1901137** | **Carbohydrate derivative biosynthetic process** | 678 | 209 | 215 | 26 | 52 | 0 | 0 | 0 | 293 |
| **GO:0043005** | **Neuron projection** | 1486 | 386 | 606 | 107 | 58 | 28 | 335 | 419 | 1553 |
| **GO:0098590** | **Plasma membrane region** | 1310 | 371 | 433 | 153 | 110 | 0 | 266 | 345 | 1307 |
| **GO:0044456** | **Synapse part** | 1089 | 270 | 468 | 118 | 83 | 4 | 238 | 320 | 1231 |
| **GO:0009986** | **Cell surface** | 921 | 267 | 854 | 96 | 43 | 3 | 394 | 523 | 1913 |
| **GO:0098794** | **Postsynapse** | 826 | 179 | 361 | 104 | 68 | 0 | 229 | 279 | 1041 |
| **GO:0044297** | **Cell body** | 737 | 188 | 536 | 81 | 50 | 32 | 190 | 250 | 1139 |
| **GO:0097060** | **Synaptic membrane** | 679 | 132 | 209 | 83 | 64 | 0 | 162 | 220 | 738 |
| **GO:0045211** | **Postsynaptic membrane** | 577 | 102 | 154 | 72 | 62 | 0 | 146 | 167 | 601 |
| **GO:0036477** | **Somatodendritic compartment** | 995 | 286 | 599 | 96 | 64 | 62 | 245 | 361 | 1427 |
| **GO:0043025** | **Neuronal cell body** | 645 | 165 | 464 | 71 | 45 | 51 | 170 | 204 | 1005 |
| **GO:0030424** | **Axon** | 651 | 158 | 516 | 93 | 50 | 0 | 161 | 208 | 1028 |
| **GO:0031012** | **Extracellular matrix** | 674 | 187 | 211 | 31 | 43 | 178 | 146 | 163 | 772 |
| **GO:0005578** | **Proteinaceous extracellular matrix** | 554 | 145 | 136 | 29 | 41 | 119 | 86 | 96 | 507 |
| **GO:0005509** | **Calcium ion binding** | 979 | 236 | 101 | 7 | 36 | 0 | 9 | 5 | 158 |
| **GO:0008233** | **Peptidase activity** | 574 | 199 | 50 | 6 | 38 | 0 | 0 | 0 | 94 |
| **GO:0070011** | **Peptidase activity, acting on L-amino acid peptides** | 547 | 191 | 49 | 5 | 39 | 0 | 0 | 0 | 93 |
| **NA** | **Cumulative, within 26 genomic regulatory networks** | 5237 | ND | 173 | 21 | 37 | 8 | 68 | 68 | 375 |
| **NA** | **Cumulative, excluding 26 genomic regulatory networks** | 3602 | 0 | 0 | 0 | 1 | 1 | 0 | 0 | 2 |

Legend: *, A directed acyclic graph (DAG) test was carried out on the enriched terms from a single ontology-specific table generated by the GREAT algorithm; **, For each GRN, reported numbers of GRMs were determined using a directed acyclic graph (DAG) test applied to 6 independent gene ontology-specific enrichment analyses employing the GREAT algorithm (Supplementary Materials); ***, A total of 26 gene ontology enrichment analyses-defined GRNs associated with human embryo retroviral LTRs were identified at the statistical significance threshold of the Binominal FDR Q-value < 0.05 (Supplementary Table S1). GRNs contain from 529 to 1486 retroviral LTR elements and from 199 to 805 down-stream target genes.

**Table 6.** Genes regulating Mammalian Offspring's Survival (MOS) phenotypes are intrinsic components of twenty six Genomic Regulatory Networks (GRNs) governed by human embryo retroviral LTR elements.

| GRN ID | GO Category | GO ID | GRN GO Name | Number of MOS phenotypes* | Top scoring MOS phenotype | GO ID | Binominal Rank | Binominal FDR Q-value |
|---|---|---|---|---|---|---|---|---|
| **GRN1** | GO Cellular Component | GO:0043005 | **neuron projection** | 5 | **postnatal lethality, complete penetrance** | MP:0011085 | 167 | 5.56E-16 |
| **GRN2** | GO Biological Process | GO:0051960 | **regulation of nervous system development** | 12 | **neonatal lethality, complete penetrance** | MP:0011087 | 33 | 1.35E-33 |
| **GRN3** | GO Cellular Component | GO:0044456 | **synapse part** | 6 | **postnatal lethality, incomplete penetrance** | MP:0011086 | 245 | 1.18E-10 |
| **GRN4** | GO Biological Process | GO:0098609 | **cell-cell adhesion** | 4 | **lethality at weaning, complete penetrance** | MP:0011083 | 41 | 4.63E-13 |
| **GRN5** | GO Cellular Component | GO:0036477 | **somatodendritic compartment** | 5 | **lethality at weaning** | MP:0008569 | 180 | 3.16E-13 |
| **GRN6** | GO Biological Process | GO:0051962 | **positive regulation of nervous system development** | 9 | **neonatal lethality, complete penetrance** | MP:0011087 | 110 | 7.28E-20 |
| **GRN7** | GO Molecular Function | GO:0005509 | **calcium ion binding** | 3 | **embryonic lethality between implantation and placentation, complete penetrance** | MP:0011095 | 329 | 2.01E-02 |
| **GRN8** | GO Cellular Component | GO:0098794 | **postsynapse** | 4 | **lethality at weaning, complete penetrance** | MP:0011083 | 150 | 9.09E-13 |
| **GRN9** | GO Cellular Component | GO:0097060 | **synaptic membrane** | 4 | **lethality at weaning, complete penetrance** | MP:0011083 | 183 | 2.98E-10 |
| **GRN10** | GO Biological Process | GO:1901137 | **carbohydrate derivative biosynthetic process** | 6 | **embryonic lethality between implantation and somite formation, incomplete penetrance** | MP:0011106 | 40 | 1.44E-11 |
| **GRN11** | GO Biological Process | GO:0098742 | **cell-cell adhesion via plasma-membrane adhesion molecules** | 5 | **neonatal lethality, complete penetrance** | MP:0011087 | 106 | 1.45E-08 |
| **GRN12** | GO Biological Process | GO:0044708 | **single-organism behavior** | 6 | **postnatal lethality, incomplete penetrance** | MP:0011086 | 65 | 9.98E-25 |
| **GRN13** | GO Cellular Component | GO:0031012 | **extracellular matrix** | 5 | **lethality throughout fetal growth and development, incomplete penetrance** | MP:0011109 | 14 | 2.45E-14 |
| **GRN14** | GO Cellular Component | GO:0030424 | **axon** | 5 | **postnatal lethality, complete penetrance** | MP:0011085 | 178 | 4.09E-10 |
| **GRN15** | GO Cellular Component | GO:0043025 | **neuronal cell body** | 3 | **embryonic lethality between implantation and placentation, complete penetrance** | MP:0011095 | 341 | 3.66E-07 |
| **GRN16** | GO Biological Process | GO:0099537 | **trans-synaptic signaling** | 8 | **lethality at weaning** | MP:0008569 | 104 | 9.72E-24 |
| **GRN17** | GO Biological Process | GO:0007268 | **chemical synaptic transmission** | 8 | **lethality at weaning** | MP:0008569 | 94 | 3.79E-24 |
| **GRN18** | GO Biological Process | GO:0042391 | **regulation of membrane potential** | 7 | **postnatal lethality, incomplete penetrance** | MP:0011086 | 129 | 3.50E-16 |
| **GRN19** | GO Cellular Component | GO:0045211 | **postsynaptic membrane** | 3 | **lethality at weaning, complete penetrance** | MP:0011083 | 143 | 6.43E-12 |
| **GRN20** | GO Molecular Function | GO:0008233 | **peptidase activity** | 2 | **embryonic lethality during organogenesis, complete penetrance** | MP:0011098 | 304 | 2.26E-02 |
| **GRN21** | GO Cellular Component | GO:0005578 | **proteinaceous extracellular matrix** | 5 | **lethality throughout fetal growth and development, incomplete penetrance** | MP:0011109 | 6 | 1.90E-17 |
| **GRN22** | GO Molecular Function | GO:0070011 | **peptidase activity, acting on L-amino acid peptides** | 2 | **postnatal lethality, complete penetrance** | MP:0011085 | 236 | 1.06E-02 |
| **GRN23** | GO Biological Process | GO:0050804 | **modulation of synaptic transmission** | 6 | **preweaning lethality, incomplete penetrance** | MP:0011110 | 194 | 6.09E-11 |
| **GRN24** | GO Cellular Component | GO:0044297 | **cell body** | 2 | **embryonic lethality between implantation and placentation, complete penetrance** | MP:0011095 | 72 | 5.69E-16 |
| **GRN25** | GO Cellular Component | GO:0009986 | **cell surface** | 7 | **neonatal lethality, complete penetrance** | MP:0011087 | 22 | 1.97E-21 |
| **GRN26** | GO Cellular Component | GO:0098590 | **plasma membrane region** | 3 | **postnatal lethality, incomplete penetrance** | MP:0011086 | 352 | 2.31E-07 |

Legend: *, number of MOS phenotypes among top 500 enriched GO terms.

**Table 7.** Gene Set Enrichment Analyses (GSEA) of 5444 genes comprising putative down-stream targets of human embryo regulatory LTRs revealed their global multifaceted impacts on physiological traits, developmental phenotypes, and pathological conditions of Modern Humans.

| Classification category | Number of significantly enriched records* |
|---|---|
| ***Common & rare human diseases*** | |
| **DisGeNET** | 477 |
| **Orphanet Augmented 2021** | 1095 |
| **Rare Diseases AutoRIF ARCHS4 Predictions** | 140 |
| **Rare Diseases GeneRIF ARCHS4 Predictions** | 97 |
| **Rare Diseases GeneRIF Gene Lists** | 96 |
| **Rare Diseases AutoRIF Gene Lists** | 610 |
| **Disease Perturbations from GEO down** | 128 |
| **Disease Perturbations from GEO up** | 116 |
| **Diabetes Perturbations GEO 2022** | 217 |
| **MAGMA Drugs and Diseases** | 82 |
| **HDSigDB Human 2021 (Huntington disease)** | 1004 |
| **HDSigDB Mouse 2021 (Huntington disease)** | 677 |
| **MSigDB Oncogenic Signatures** | 49 |
| **Jensen database of human diseases** | 79 |
| **RNA-Seq Disease Gene and Drug Signatures from GEO** | 323 |
| ***Developmental & Regulatory Pathways*** | |
| **WikiPathway 2021 Human** | 33 |
| **KEGG 2021 Human** | 17 |
| **BioPlanet 2019** | 48 |
| **Panther 2016** | 1 |
| **Reactome 2022** | 26 |
| **Transcription Factor PPIs** | 1 |
| **ARCHS4 TF Co-expression** | 284 |
| **ARCHS4 Kinases Co-expression** | 106 |
| **TF Perturbations Followed by Expression** | 532 |
| **Gene Perturbations from GEO up** | 342 |
| **Gene Perturbations from GEO down** | 252 |
| **Endogenous Ligand Perturbations from GEO up** | 43 |
| **Endogenous Ligand Perturbations from GEO down** | 46 |
| **lncHUB lncRNA Co-Expression** | 220 |
| **Kinase Perturbations from GEO down** | 17 |
| **Kinase Perturbations from GEO up** | 27 |
| ***Gene Ontology*** | |
| **GO Cellular Component 2021** | 17 |
| **GO Molecular Function 2021** | 12 |
| **GO Biological Process 2021** | 177 |
| **MGI Mammalian Phenotype Level 4 2021** | 332 |
| **Human Phenotype Ontology** | 27 |
| ***Single cell genomics & epigenetics*** | |
| **PanglaoDB Augmented 2021** | 87 |
| **CellMarker Augmented 2021** | 255 |
| **Azimuth Cell Types 2021** | 101 |
| **HuBMAP ASCTplusB Augmented 2022** | 191 |
| ***Human tissues*** | |
| **GTEx Tissue Expression Up** | 513 |
| **GTEx Tissue Expression Down** | 163 |
| **Human Gene Atlas** | 15 |
| **ARCHS4 Tissues** | 48 |
| **Allen Brain Atlas up** | 787 |
| **Allen Brain Atlas down** | 245 |
| ***RNAseq and Proteomics*** | |
| **RNAseq Automatic GEO Signatures Human Up** | 1494 |
| **RNAseq Automatic GEO Signatures Human Down** | 1571 |
| **ProteomicsDB 2020** | 6 |
| **CCLE Proteomics 2020** | 49 |
| ***Aging phenotype signatures*** | |
| **GTEx Aging Signatures 2021** | 13 |
| **Aging Perturbations from GEO up** | 10 |
| **Aging Perturbations from GEO down** | 30 |
| ***Genotype-phenotype associations*** | |
| **GWAS Catalog 2019** | 271 |
| **OMIM Expanded** | 175 |
| **ClinVar 2019** | 1 |
| **PhenGenI Association 2021** | 150 |
| **PheWeb 2019** | 862 |
| **dbGaP** | 144 |
| ***Host-pathogen interactions*** | |
| **COVID-19 Related Gene Sets 2021** | 138 |
| **Virus Perturbations from GEO up** | 55 |
| **Virus Perturbations from GEO down** | 55 |
| **Microbe Perturbations from GEO up** | 18 |
| **Microbe Perturbations from GEO down** | 13 |

Legend: * Numbers of significantly enriched records identified by GSEA during analyses of corresponding genomics and proteomics databases to define entries at the statistical significance threshold of adjusted P value (FDR q value) < 0.05.

**Table 8.** Quantitative profiles of a directed acyclic graph (DAG)* test-validated genomic regulatory modules (GRMs)** within GSEA-defined genomic regulatory panels (GRPs)*** of the genomic dominion of human embryo retroviral LTRs.

| Genomic regulatory panels (GRPs) | Number of LTRs | Number of linked genes | GO Biological Process | GO Cellular Component | GO Molecular Function | Human Phenotype | Mouse Phenotype Single KO | Mouse Phenotype | Number of GRMs | DAG Visualization Report |
|---|---|---|---|---|---|---|---|---|---|---|
| **Neurotrophins (neurotrophic factors)** | 8 | 12 | 12 | 0 | 2 | 30 | 70 | 77 | 191 | Figure 10 |
| **Neurotrophins (expanded)** | 29 | 147 | 4 | 3 | 8 | 0 | 70 | 59 | 144 | Supplementary Data Sets S2 |
| **Neurexins and Neuroligins R-HSA-6794361** | 120 | 30 | 61 | 19 | 4 | 0 | 0 | 0 | 84 | Figure 9 |
| **Neurotransmitter receptor activity** | 137 | 30 | 34 | 29 | 38 | 0 | 20 | 32 | 153 | Figure 11 |
| **Signaling by Receptor Tyrosine Kinases R-HSA-9006934** | 241 | 81 | 172 | 8 | 12 | 0 | 24 | 69 | 285 | Figure 13 |
| **RET-GDNF pathway** | 115 | 28 | 181 | 0 | 1 | 8 | 107 | 191 | 488 | Figure 8 |
| **Neuroactive ligand-receptor interactions** | 305 | 84 | 85 | 30 | 43 | 0 | 80 | 85 | 323 | Figure 12 |
| **GSEA-defined GRN of retroviral LTRs** | 966 | 255 | 287 | 38 | 20 | 2 | 136 | 264 | 747 | Figure 14 |
| **GSEA-defined GRN of retroviral LTRs (all significant GRMs)#** | 966 | 255 | 1359 | 96 | 132 | 178 | 955 | 1165 | 3885 | Supplementary Data Sets S2 |

Legend: *, A directed acyclic graph (DAG) test was carried out on the enriched terms from a single ontology-specific table generated by the GREAT algorithm;
**, For each GRP, reported numbers of GRMs were determined using a directed acyclic graph (DAG) test applied to 6 independent gene ontology-specific enrichment analyses employing the GREAT algorithm (see Supplementary Data Sets S2 for complete reports);
***, GSEA-defined GRPs associated with human embryo retroviral LTRs were identified using the Enrichr Bioinformatics Platform at the statistical significance threshold of the adjusted P-value < 0.05 (Supplementary Data Sets S2);
#, reported numbers of GRMs were identified by the GREAT algorithm at the statistical significance threshold of the Binominal FDR Q-value < 0.001 (Supplementary Data Sets S2).

**Table 9.** Enrichment of retroviral LTR elements and genes associated with Mammalian Offspring Survival phenotypes within the retroviral LTR-governed GRM of reduced fertility phenotype.

| Phenotype | Binominal Rank | Binominal Raw P-Value | Binominal FDR Q-Value | Binominal Fold Enrichment | Binominal Observed LTR Hits | Binominal LTR Set Coverage | Hypergeometric Rank | Hypergeometric FDR Q-Value | Hypergeometric Fold Enrichment | Hypergeometric Observed Gene Hits | Hypergeometric Total Genes | Hypergeometric Gene Set Coverage |
|---|---|---|---|---|---|---|---|---|---|---|---|---|
| **reduced fertility** | 1 | 0 | 0 | 11.34 | 311 | 1.00 | 1 | 1.08E-62 | 13.04 | 77 | 652 | 0.45 |
| **postnatal lethality** | 14 | 8.35E-84 | 5.71E-81 | 3.62 | 218 | 0.70 | 20 | 2.46E-15 | 4.01 | 50 | 1378 | 0.29 |
| **lethality during fetal growth through weaning** | 29 | 7.68E-68 | 2.53E-65 | 2.51 | 248 | 0.80 | 21 | 3.00E-15 | 3.09 | 65 | 2325 | 0.38 |
| **postnatal lethality, incomplete penetrance** | 48 | 5.59E-49 | 1.11E-46 | 3.83 | 141 | 0.45 | 29 | 5.90E-14 | 5.07 | 37 | 806 | 0.22 |
| **infertility** | 104 | 7.29E-34 | 6.70E-32 | 3.61 | 109 | 0.35 | 115 | 1.65E-07 | 3.59 | 29 | 893 | 0.17 |
| **male infertility** | 113 | 2.04E-32 | 1.73E-30 | 4.29 | 89 | 0.29 | 154 | 5.62E-06 | 3.77 | 22 | 644 | 0.13 |
| **premature death** | 116 | 8.52E-32 | 7.02E-30 | 2.61 | 147 | 0.47 | 74 | 1.05E-08 | 3.04 | 41 | 1491 | 0.24 |
| **perinatal lethality** | 141 | 3.62E-30 | 2.46E-28 | 2.41 | 155 | 0.50 | 39 | 9.61E-12 | 3.60 | 44 | 1349 | 0.26 |
| **postnatal lethality, complete penetrance** | 157 | 1.63E-28 | 9.94E-27 | 3.30 | 103 | 0.33 | 128 | 6.13E-07 | 3.97 | 24 | 667 | 0.14 |
| **preweaning lethality, incomplete penetrance** | 282 | 5.62E-20 | 1.91E-18 | 3.60 | 67 | 0.22 | 268 | 0.000239 | 3.73 | 16 | 474 | 0.09 |
| **neonatal lethality,** | 288 | 1.06E-19 | 3.53E-18 | 3.51 | 68 | 0.22 | 137 | 2.03E-06 | 4.92 | 18 | 404 | 0.11 |
| **incomplete penetrance** | | | | | | | | | | | | |
| **lethality at weaning, complete penetrance** | 310 | 6.32E-19 | 1.95E-17 | 7.88 | 32 | 0.10 | 833 | 0.038637 | 6.50 | 4 | 68 | 0.02 |
| **neonatal lethality** | 319 | 1.67E-18 | 5.01E-17 | 2.30 | 112 | 0.36 | 58 | 5.92E-10 | 3.87 | 35 | 998 | 0.21 |
| **perinatal lethality, incomplete penetrance** | 356 | 6.07E-17 | 1.63E-15 | 3.49 | 59 | 0.19 | 185 | 2.14E-05 | 4.95 | 15 | 335 | 0.09 |
| **neonatal lethality, complete penetrance** | 359 | 7.04E-17 | 1.88E-15 | 2.52 | 90 | 0.29 | 67 | 3.47E-09 | 4.49 | 28 | 689 | 0.16 |
| **lethality at weaning** | 399 | 1.62E-15 | 3.87E-14 | 5.73 | 33 | 0.11 | 812 | 0.035547 | 5.11 | 5 | 108 | 0.03 |
| **abnormal cell death** | 436 | 2.42E-14 | 5.31E-13 | 2.07 | 107 | 0.34 | 291 | 0.000436 | 2.27 | 31 | 1510 | 0.18 |
| **prenatal lethality, incomplete penetrance** | 492 | 6.78E-13 | 1.32E-11 | 2.99 | 54 | 0.17 | 572 | 0.011505 | 3.01 | 12 | 440 | 0.07 |
| **increased cell death** | 565 | 2.24E-11 | 3.79E-10 | 2.25 | 74 | 0.24 | 375 | 0.001637 | 2.58 | 21 | 898 | 0.12 |
| **prenatal lethality, complete penetrance** | 926 | 2.07E-07 | 2.14E-06 | 2.38 | 42 | 0.14 | 837 | 0.038723 | 2.78 | 10 | 397 | 0.06 |
| **female infertility** | 1090 | 2.70E-06 | 2.37E-05 | 2.49 | 32 | 0.10 | 531 | 0.008139 | 3.37 | 11 | 360 | 0.06 |

**Table 10.** High-confidence LTR-regulated genes comprise a significant majority of 5444 genes and 709 genes defined as cumulative and common sets of putative down-stream regulatory targets for LTR7 & MLT2A1 & MLT2A2 human embryo retroviral LTR loci.

| Classification category | Number of genes | Percent |
|---|---|---|
| **Common set of LTR7 & MLT2A1 & MLT2A2 target genes** | 709 | 100.00 |
| **LTR5_Hs regulated genes** | 195 | 27.50 |
| **LTR7Y/B regulated genes** | 202 | 28.49 |
| **HERVH lncRNA regulated genes** | 253 | 35.68 |
| **SVA regulated genes** | 96 | 13.54 |
| **LTR5_Hs/SVA regulated genes** | 158 | 22.28 |
| **All LTR-regulated common target genes** | 473 | 66.71 |
| **Cumulative set of LTR7 & MLT2A1 & MLT2A2 target genes** | 5444 | 100.00 |
| **LTR5_Hs regulated genes** | 1344 | 24.69 |
| **LTR7Y/B regulated genes** | 1873 | 34.40 |
| **HERVH lncRNA regulated genes** | 1876 | 34.46 |
| **SVA regulated genes** | 841 | 15.45 |
| **LTR5_Hs/SVA regulated genes** | 937 | 17.21 |
| **All LTR-regulated target genes** | 3659 | 67.21 |

**Table 11.** GSEA of phenotypic impacts of common sets of LTR7, MLT2A1, and MLT2A2 down-stream target genes defined by proximity placement analyses (709 genes) and by genetic/epigenetic silencing of retroviral LTR loci (473 genes).

| Classification category | Proximity placement defined genes* | Genetic/epigenetic silencing defined genes* |
|---|---|---|
| **Genomics & proteomics databases** | **709 genes** | **473 genes** |
| **DisGeNET database of human diseases** | 52 | 72 |
| **Orphanet Augmented 2021** | 697 | 732 |
| **Rare Diseases AutoRIF ARCHS4 Predictions** | 63 | 65 |
| **Rare Diseases GeneRIF ARCHS4 Predictions** | 41 | 43 |
| **Rare Diseases GeneRIF Gene Lists** | 8 | 13 |
| **Rare Diseases AutoRIF Gene Lists** | 16 | 25 |
| **WikiPathway 2021 Human** | 5 | 11 |
| **KEGG 2021 Human** | 2 | 1 |
| **BioPlanet 2019** | 6 | 6 |
| **GO Cellular Component 2021** | 17 | 18 |
| **GO Molecular Function 2021** | 17 | 13 |
| **GO Biological Process 2021** | 67 | 55 |
| **MGI Mammalian Phenotype Level 4 2021** | 28 | 57 |
| **PanglaoDB Augmented 2021** | 66 | 72 |
| **CellMarker Augmented 2021** | 159 | 187 |
| **Azimuth Cell Types 2021** | 75 | 89 |
| **GTEx Tissue Expression Up** | 431 | 555 |
| **GTEx Tissue Expression Down** | 147 | 221 |
| **Human Gene Atlas** | 5 | 5 |
| **ARCHS4 Human Tissues** | 43 | 52 |
| **Allen Brain Atlas up** | 336 | 312 |
| **Allen Brain Atlas down** | 165 | 159 |
| **Disease Perturbations from GEO down** | 55 | 85 |
| **Disease Perturbations from GEO up** | 22 | 31 |
| **RNAseq Automatic GEO Signatures Human Up** | 507 | 821 |
| **RNAseq Automatic GEO Signatures Human Down** | 541 | 856 |
| **Transcription Factor PPIs** | 1 | 1 |
| **Reactome 2022** | 13 | 9 |
| **ARCHS4 Kinases Co-expression** | 61 | 68 |
| **Panther 2016** | 2 | 2 |
| **NCI-Nature 2016** | 1 | 4 |
| **HuBMAP ASCTplusB augmented 2022** | 94 | 91 |
| **ARCHS4 TF Co-expression** | 204 | 190 |
| **ProteomicsDB 2020** | 0 | 0 |

Legend: * Numbers of significantly enriched records identified by GSEA of 709 genes and 473 genes during analyses of corresponding genomics and proteomics databases to define entries at the statistical significance threshold of adjusted P value (FDR q value) < 0.05.

**Table 12.** Enrichment of genes linked to distinct families of Human-Specific Regulatory Sequences (HSRS) among putative down-stream target genes of distinct families of human embryo retroviral LTR elements.

| Classification category | Observed | Expected | Enrichment/Depletion | P value* |
|---|---|---|---|---|
| **<u>Genes associated with LTR7 regulatory elements</u>** | 2957 | 2957 | 1.00 | NA |
| **Genes linked with Naïve & Primed hESC functional enhancers** | 1703 | 1822 | 0.93 | 0.001763 |
| **Genes linked with 1619 human-specific Naïve & Primed hESC functional enhancers** | 522 | 203 | **2.57** | 1.67E-37 |
| **Genes linked with TE loci expressed in human adult DLPFC** | 2262 | 2094 | 1.08 | 8.085E-07 |
| **Genes linked with 4690 human-specific TE loci expressed in human adult DLPFC** | 1328 | 661 | **2.01** | 4.04E-76 |
| **Genes associated with 59,732 HSRS** | 2615 | 2080 | 1.26 | 7.732E-68 |
| **Genes associated with 12,262 created de novo HSRS** | 1390 | 685 | **2.03** | 3.274E-83 |
| **Genes associated with HSRS (except genes linked with de novo HSRS)** | 1225 | 1395 | 0.88 | 9.634E-06 |
| **<u>Genes linked with MLT2A1 & MLT2A2</u>** | 3971 | 3971 | 1.00 | NA |
| **Genes linked with Naïve & Primed hESC functional enhancers** | 2121 | 2447 | 0.87 | 1.54E-13 |
| **Genes linked with 1619 human-specific Naïve & Primed hESC functional enhancers** | 655 | 272 | **2.40** | 8.30E-41 |
| **Genes linked with TE loci expressed in human adult DLPFC** | 2059 | 2811 | 0.73 | 1.30E-67 |
| **Genes linked with 4690 human-specific TE loci expressed in human adult DLPFC** | 1903 | 887 | **2.15** | 5.15E-117 |
| **Genes associated with 59,732 HSRS** | 3525 | 2794 | 1.26 | 1.66E-94 |
| **Genes associated with 12,262 created de novo HSRS** | 1910 | 920 | **2.08** | 1.12E-120 |
| **Genes associated with HSRS (except genes linked with de novo HSRS)** | 1615 | 1874 | 0.86 | 5.35E-09 |
| **<u>Genes linked with LTR7 & MLT2A1 & MLT2A2</u>** | 5444 | 5444 | 1.00 | NA |
| **Genes linked with Naïve & Primed hESC functional enhancers** | 3040 | 3354 | 0.91 | 1.10E-09 |
| **Genes linked with 1619 human-specific Naïve & Primed hESC functional enhancers** | 814 | 373 | **2.18** | 3.26E-43 |
| **Genes linked with TE loci expressed in human adult DLPFC** | 4147 | 3854 | 1.08 | 2.24E-10 |
| **Genes linked with 4690 human-specific TE loci expressed in human adult DLPFC** | 2306 | 1216 | **1.90** | 7.80E-119 |
| **Genes associated with 59,732 HSRS** | 4782 | 3830 | 1.25 | 5.12E-114 |
| **Genes associated with 12,262 created de novo HSRS** | 2361 | 1261 | **1.87** | 5.66E-112 |
| **Genes associated with HSRS (except genes linked with de novo HSRS)** | 2421 | 2569 | 0.94 | 0.0046903 |

Legend: *, 2-tail p values were estimated using the Fisher's exact test.

**Table 13.** Enrichment of human embryo retroviral LTR down-stream target genes among genes with species-specific expression mapping bias in hybrids of Human and Chimpanzee cells.

| Classification category | Number of genes | Percent |
|---|---|---|
| **Genes with expression mapping bias in hybrid induced Pluripotent Stem (iPS) Cells of Human and Chimpanzee** | 374 | 100 |
| **Genes regulated by human embryo retroviral LTR elements** | 95 | 25.40 |
| **Expected by chance** | 44 | |
| **Enrichment** | 2.2 | |
| **P value*** | 9.58E-14 | |
| **Genes with expression mapping bias in hybrid Cortical Spheroids (Brain Organoids) of Human and Chimpanzee** | 291 | 100 |
| **Genes regulated by human embryo retroviral LTR elements** | 63 | 21.65 |
| **Expected by chance** | 34 | |
| **Enrichment** | 1.9 | |
| **P value*** | 5.06E-07 | |
| **Genes manifesting allele-specific expression in hybrids of Human and Chimpanzee cells** | 183 | 100 |
| **Genes regulated by human embryo retroviral LTR elements** | 56 | 30.60 |
| **Expected by chance** | 21 | |
| **Enrichment** | 2.7 | |
| **P value*** | 4.36E-12 | |

Legend: * p values were estimated using the hypergeometric distribution test.

**Table 14.** Summary of significantly enriched phenotypic traits and morphological elements associated with 177 down-stream target genes of human embryo retroviral LTRs manifesting species-specific expression bias in human-chimpanzee hybrids.

| **KEGG 2021 Human Database** | Human embryo retroviral LTR target genes | Adjusted P-value | Genes manifesting species-specific expression bias | Adjusted P-value | Percent, LTR regulated | Relative enrichment* |
|---|---|---|---|---|---|---|
| Neuroactive ligand-receptor interactions | 14 | 4.50E-04 | 24 | 0.116711 | 58 | 2.2 |
| **HuBMAP ASCTplusB augmented 2022 Database** | Human embryo retroviral LTR target genes | Adjusted P-value | Genes manifesting species-specific expression bias | Adjusted P-value | Percent, LTR regulated | Relative enrichment* |
| Inh L1 SST DEFB108B - Brain | 8 | 8.29E-04 | 8 | 0.387543 | 100 | 3.8 |
| Inh L1-2 VIP PTGER3 - Brain | 8 | 8.29E-04 | 9 | 0.207386 | 89 | 3.4 |
| Inh L1-3 SST FAM20A - Brain | 7 | 0.004064 | 11 | 0.129625 | 64 | 2.4 |
| Inh L1-2 PVALB CDK20 - Brain | 7 | 0.004064 | 13 | 0.028948 | 54 | 2.1 |
| Inh L1-2 VIP SCML4 - Brain | 6 | 0.02756 | 10 | 0.129625 | 60 | 2.3 |
| **ARCHS4 Human Tissues Database** | Human embryo retroviral LTR target genes | Adjusted P-value | Genes manifesting species-specific expression bias | Adjusted P-value | Percent, LTR regulated | Relative enrichment* |
| DENTATE GRANULE CELLS | 53 | 3.62E-09 | 119 | 6.25E-05 | 45 | **1.7** |
| PREFRONTAL CORTEX | 52 | 6.03E-09 | 118 | 7.90E-05 | 44 | **1.7** |
| CINGULATE GYRUS | 51 | 1.30E-08 | 109 | 0.001818 | 47 | **1.8** |
| CEREBELLUM | 48 | 1.40E-07 | 98 | 0.062407 | 49 | **1.9** |
| CEREBRAL CORTEX | 48 | 1.40E-07 | 116 | 1.74E-04 | 41 | 1.6 |
| SPINAL CORD | 48 | 1.40E-07 | 115 | 2.04E-04 | 42 | 1.6 |
| SPINAL CORD (BULK) | 48 | 1.40E-07 | 115 | 2.04E-04 | 42 | 1.6 |
| SUPERIOR FRONTAL GYRUS | 48 | 1.40E-07 | 136 | 5.62E-09 | 35 | 1.4 |
| BRAIN (BULK) | 47 | 3.62E-07 | 125 | 3.29E-06 | 38 | 1.4 |
| DORSAL STRIATUM | 45 | 2.51E-06 | 112 | 7.31E-04 | 40 | 1.5 |
| MOTOR NEURON | 41 | 9.55E-05 | 111 | 9.16E-04 | 37 | 1.4 |
| MIDBRAIN | 38 | 0.001047 | 97 | 0.076015 | 39 | 1.5 |
| SMALL INTESTINE (BULK TISSUE) | 37 | 0.002074 | 76 | 0.999994 | 49 | **1.9** |
| FETAL BRAIN | 36 | 0.003502 | 109 | 0.001818 | 33 | 1.3 |
| OLIGODENDROCYTE | 36 | 0.003502 | 100 | 0.039555 | 36 | 1.4 |
| TESTIS (BULK TISSUE) | 36 | 0.003502 | 111 | 9.16E-04 | 32 | 1.2 |
| ALPHA CELLS | 35 | 0.005636 | 101 | 0.032484 | 35 | 1.3 |
| FETAL BRAIN CORTEX | 35 | 0.005636 | 96 | 0.092408 | 36 | 1.4 |
| RETINA | 35 | 0.005636 | 99 | 0.05124 | 35 | 1.4 |
| SENSORY NEURON | 35 | 0.005636 | 92 | 0.229934 | 38 | 1.5 |
| NEURONAL EPITHELIUM | 33 | 0.018769 | 79 | 0.999994 | 42 | 1.6 |
| OOCYTE | 33 | 0.018769 | 100 | 0.039555 | 33 | 1.3 |
| HUMAN ZYGOTE | 32 | 0.032657 | 91 | 0.268966 | 35 | 1.4 |
| SPERM | 25 | 0.514091 | 102 | 0.024863 | 25 | 0.9 |

Legend: *Relative enrichment values were estimated by dividing observed values by expected values in corresponding category based on 26% representation of all retroviral LTR down-stream target genes among all genes manifesting species-specific expression bias in human-chimpanzee hybrids.

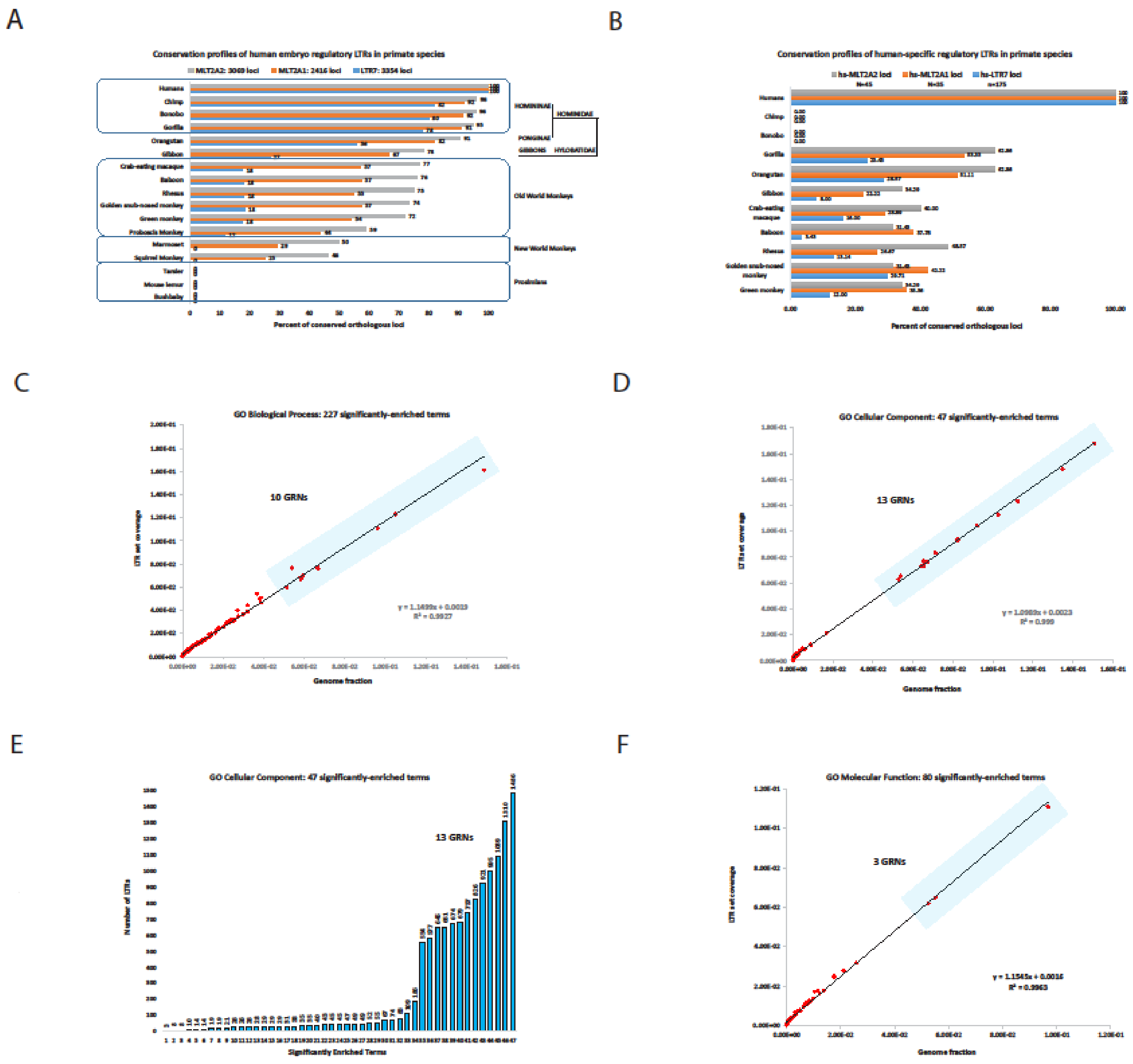
A
Conservation profiles of human embryo regulatory LTRs in primate species
MLT2A2: 3069 loci
MLT2A1: 2416 loci
LTR7: 3354 loci
Humans
Chimp
Bonobo
Gorilla
Orangutan
Gibbon
Crab-eating macaque
Baboon
Rhesus
Golden snub-nosed monkey
Green monkey
Proboscis Monkey
Marmoset
Squirrel Monkey
Tarsier
Mouse lemur
Bushbaby
HOMININAE
HOMINIDAE
PONGINAE
GIBBONS
HYLOBATIDAE
Old World Monkeys
New World Monkeys
Prosimians
Percent of conserved orthologous loci
B
Conservation profiles of human-specific regulatory LTRs in primate species
hs-MLT2A2 loci
hs-MLT2A1 loci
hs-LTR7 loci
Percent of conserved orthologous loci
C
GO Biological Process: 227 significantly-enriched terms
10 GRNs
Genome fraction
D
GO Cellular Component: 47 significantly-enriched terms
13 GRNs
Genome fraction
E
GO Cellular Component: 47 significantly-enriched terms
13 GRNs
Significantly Enriched Terms
F
GO Molecular Function: 80 significantly-enriched terms
3 GRNs
Genome fraction

**Figure 1.**

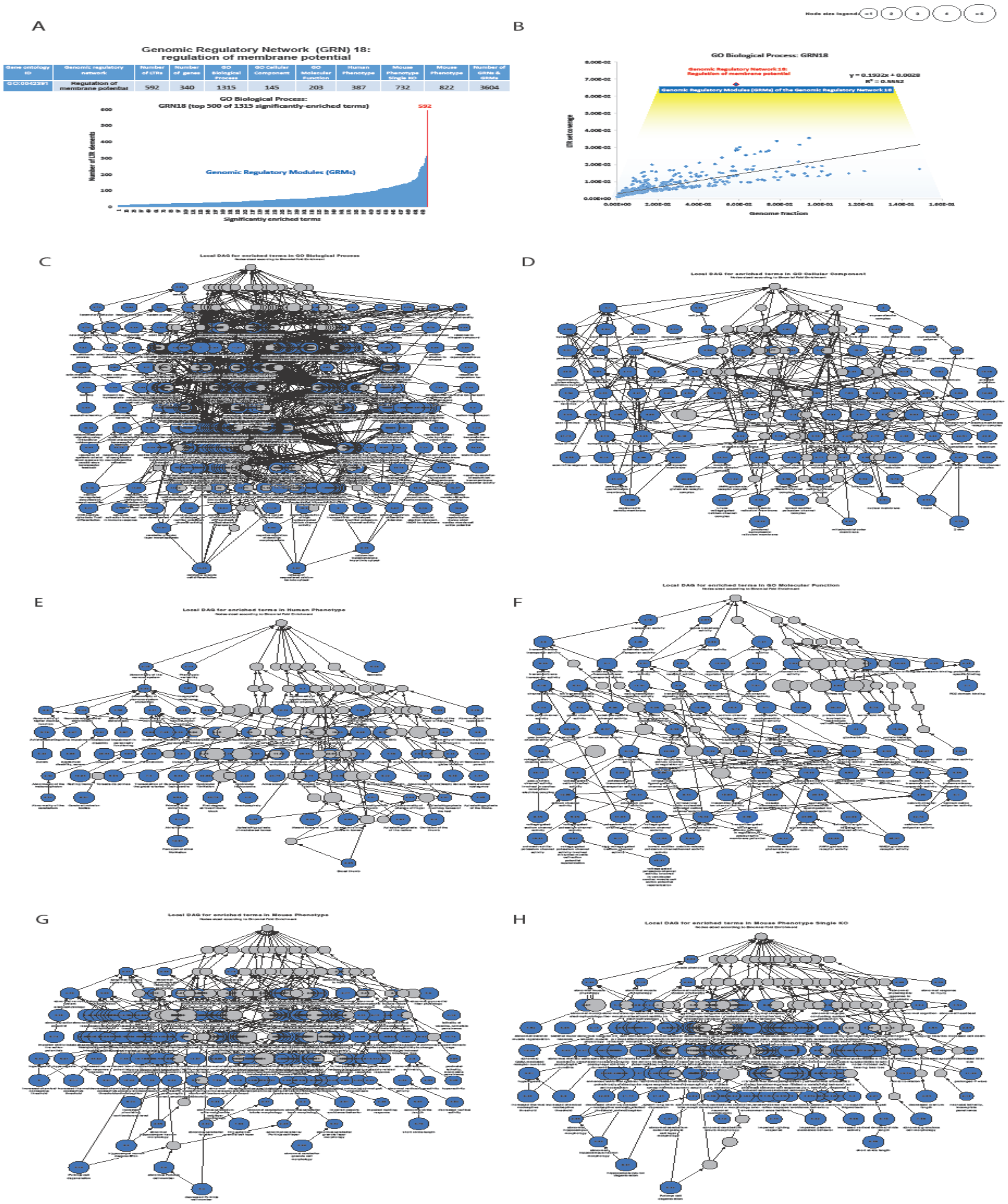
A
B
Genomic Regulatory Network (GRN) 18:
regulation of membrane potential
Gene ontology ID
Genomic regulatory network
Number of LTRs
Number of genes
GO Biological Process
GO Cellular Component
GO Molecular Function
Human Phenotype
Mouse Phenotype Single KO
Mouse Phenotype
Number of GRNs & GRMs
GO:0042391
Regulation of membrane potential
592
340
1315
145
203
387
732
822
3604
GO Biological Process:
GRN18 (top 500 of 1315 significantly-enriched terms)
592
Number of LTR elements
Genomic Regulatory Modules (GRMs)
Significantly enriched terms
GO Biological Process: GRN18
Genomic Regulatory Network 18:
Regulation of membrane potential
y = 0.1932x + 0.0028
R² = 0.5552
Genomic Regulatory Modules (GRMs) of the Genomic Regulatory Network 18
Genome fraction
C
D
E
F
G
H
Local DAG for enriched terms in GO Biological Process
Local DAG for enriched terms in GO Cellular Component
Local DAG for enriched terms in Human Phenotype
Local DAG for enriched terms in GO Molecular Function
Local DAG for enriched terms in Mouse Phenotype
Local DAG for enriched terms in Mouse Phenotype Single KO

**Figure 2.**

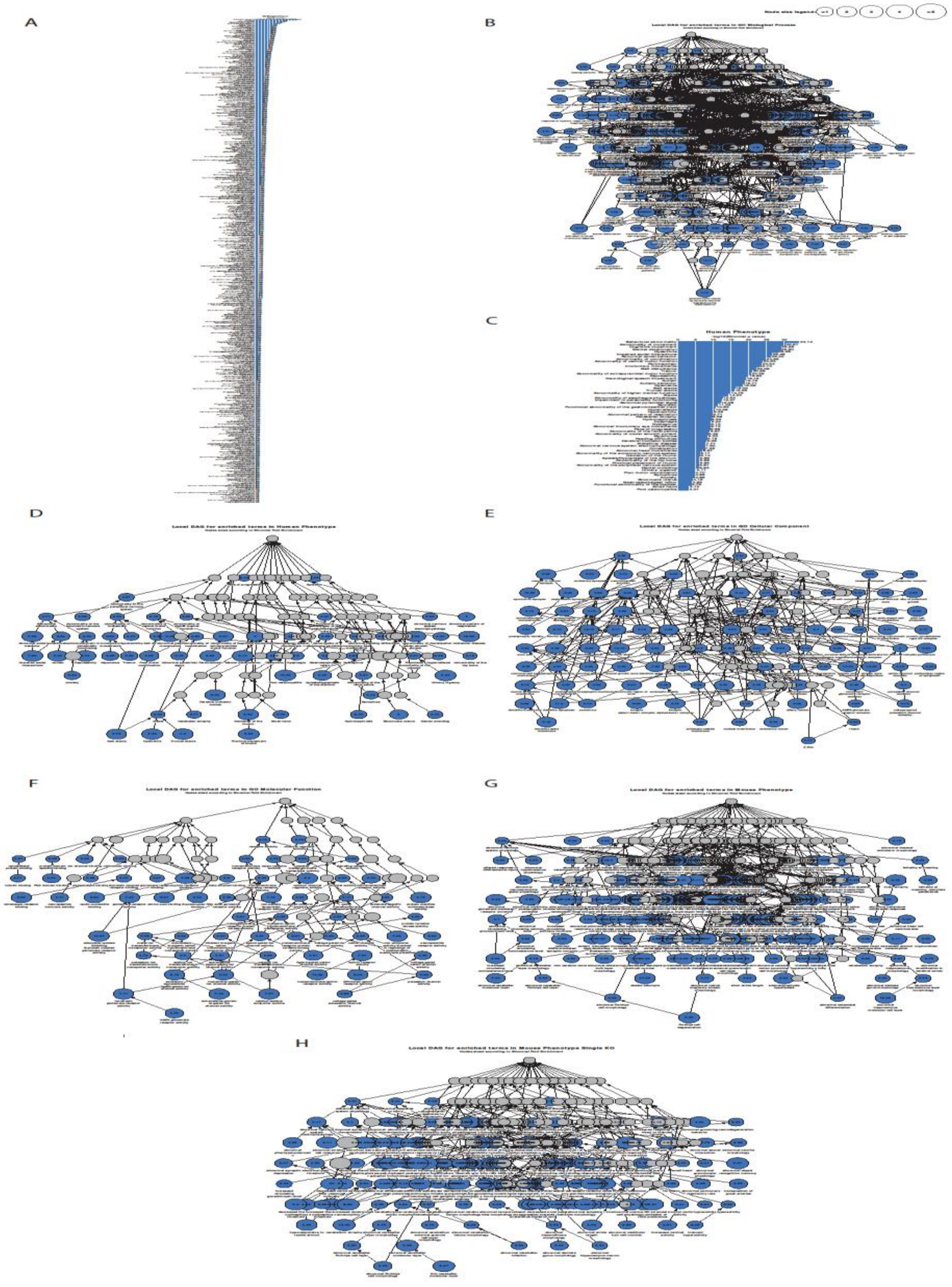

**Figure 3.**

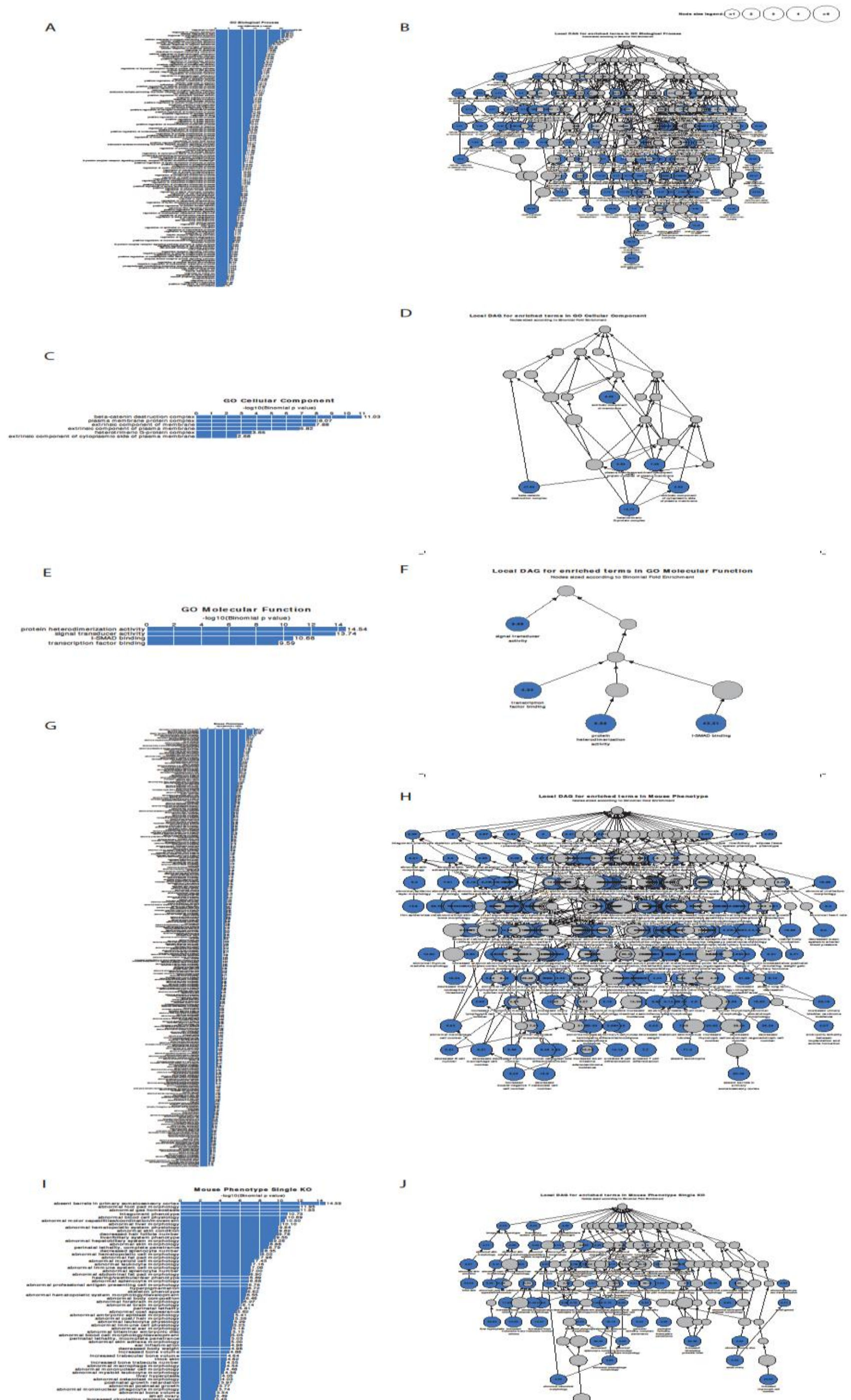
A
B
Local DAG for enriched terms in GO Cellular Component
D
C
GO Cellular Component
-log10(Binomial p value)
beta-catenin destruction complex 11.03
plasma membrane protein complex 8.07
extrinsic component of membrane 7.88
extrinsic component of plasma membrane 6.82
heterotrimeric G-protein complex 3.64
extrinsic component of cytoplasmic side of plasma membrane 2.68
E
GO Molecular Function
-log10(Binomial p value)
protein heterodimerization activity 14.54
signal transducer activity 13.74
I-SMAD binding 10.68
transcription factor binding 9.59
F
Local DAG for enriched terms in GO Molecular Function
Nodes sized according to Binomial Fold Enrichment
signal transducer activity
transcription factor binding
protein heterodimerization activity
I-SMAD binding
G
H
Local DAG for enriched terms in Mouse Phenotype
I
Mouse Phenotype Single KO
-log10(Binomial p value)
J

**Figure 4.**

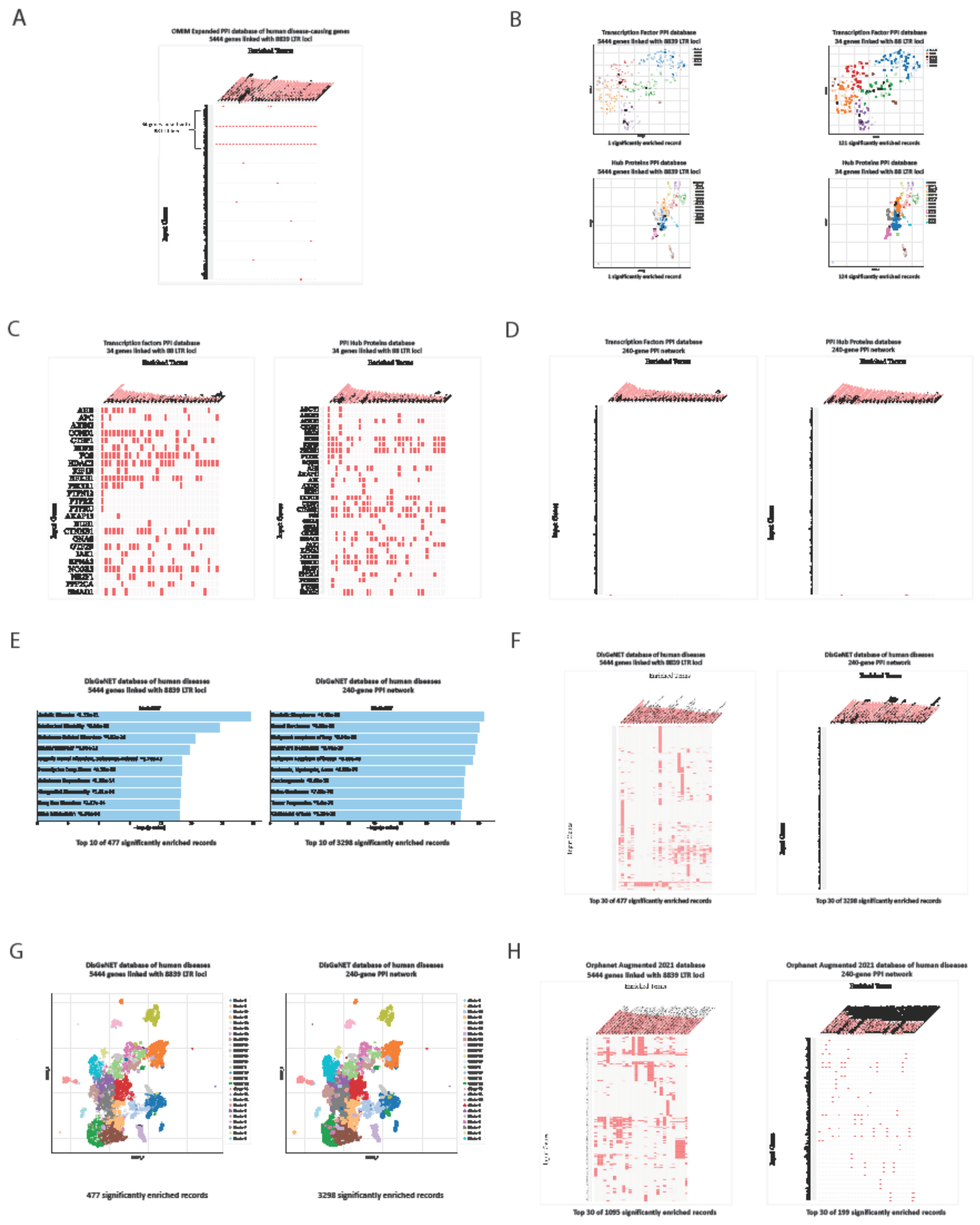
A
OMIM Expanded PPI database of human disease-causing genes
5444 genes linked with 8839 LTR loci
B
Transcription Factor PPI database
5444 genes linked with 8839 LTR loci
1 significantly enriched record
Transcription Factor PPI database
34 genes linked with 88 LTR loci
121 significantly enriched records
Hub Proteins PPI database
5444 genes linked with 8839 LTR loci
1 significantly enriched record
Hub Proteins PPI database
34 genes linked with 88 LTR loci
124 significantly enriched records
C
Transcription factors PPI database
34 genes linked with 88 LTR loci
PPI Hub Proteins database
34 genes linked with 88 LTR loci
D
Transcription Factors PPI database
240-gene PPI network
PPI Hub Proteins database
240-gene PPI network
E
DisGeNET database of human diseases
5444 genes linked with 8839 LTR loci
Top 10 of 477 significantly enriched records
DisGeNET database of human diseases
240-gene PPI network
Top 10 of 3298 significantly enriched records
F
DisGeNET database of human diseases
5444 genes linked with 8839 LTR loci
Top 30 of 477 significantly enriched records
DisGeNET database of human diseases
240-gene PPI network
Top 30 of 3298 significantly enriched records
G
DisGeNET database of human diseases
5444 genes linked with 8839 LTR loci
477 significantly enriched records
DisGeNET database of human diseases
240-gene PPI network
3298 significantly enriched records
H
Orphanet Augmented 2021 database
5444 genes linked with 8839 LTR loci
Top 30 of 1095 significantly enriched records
Orphanet Augmented 2021 database of human diseases
240-gene PPI network
Top 30 of 199 significantly enriched records

**Figure 5.**

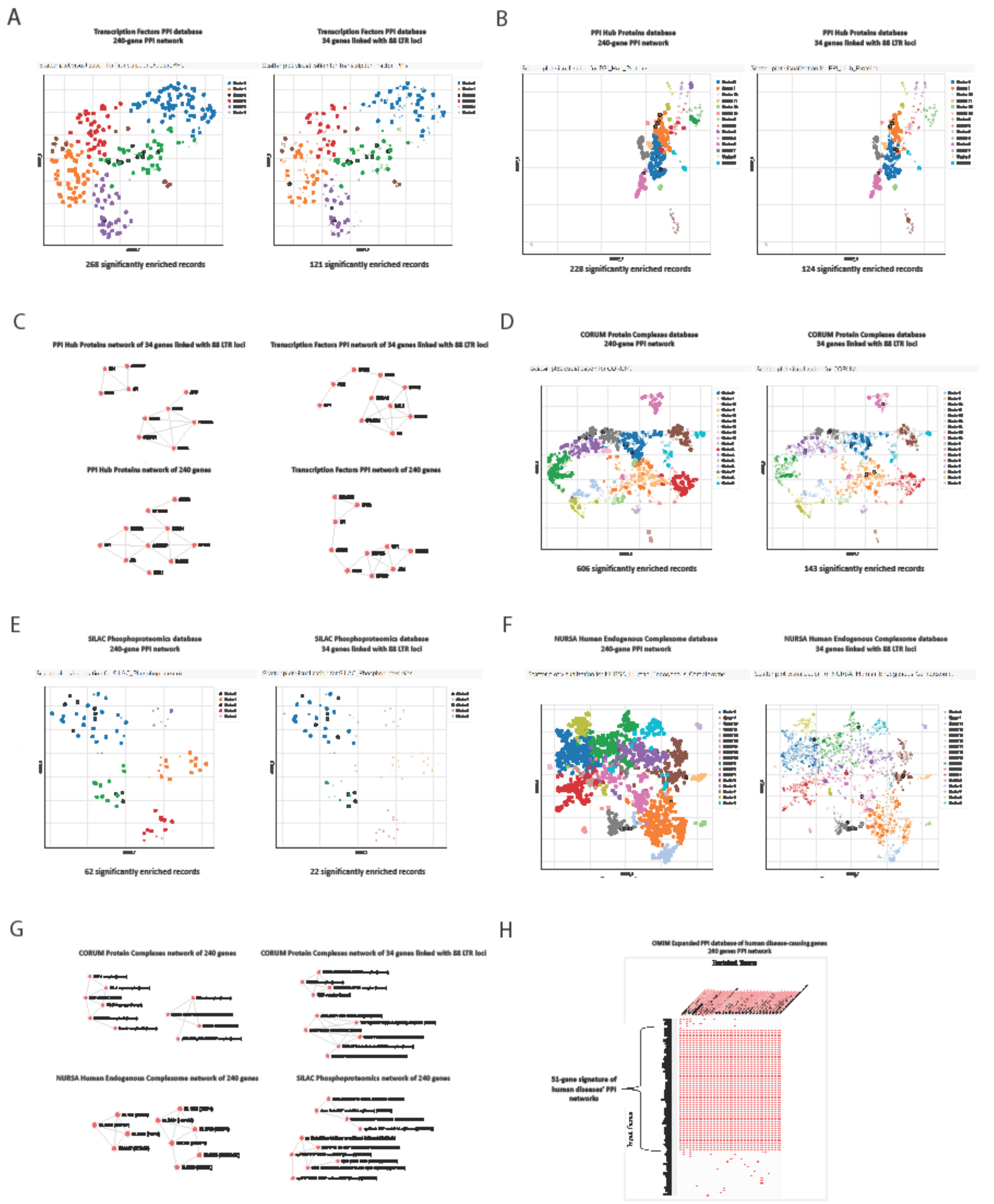
A
Transcription Factors PPI database
240-gene PPI network
268 significantly enriched records
Transcription Factors PPI database
34 genes linked with 88 LTR loci
121 significantly enriched records
B
PPI Hub Proteins database
240-gene PPI network
228 significantly enriched records
PPI Hub Proteins database
34 genes linked with 88 LTR loci
124 significantly enriched records
C
PPI Hub Proteins network of 34 genes linked with 88 LTR loci
Transcription Factors PPI network of 34 genes linked with 88 LTR loci
PPI Hub Proteins network of 240 genes
Transcription Factors PPI network of 240 genes
D
CORUM Protein Complexes database
240-gene PPI network
606 significantly enriched records
CORUM Protein Complexes database
34 genes linked with 88 LTR loci
143 significantly enriched records
E
SILAC Phosphoproteomics database
240-gene PPI network
62 significantly enriched records
SILAC Phosphoproteomics database
34 genes linked with 88 LTR loci
22 significantly enriched records
F
NURSA Human Endogenous Complexome database
240-gene PPI network
NURSA Human Endogenous Complexome database
34 genes linked with 88 LTR loci
G
CORUM Protein Complexes network of 240 genes
CORUM Protein Complexes network of 34 genes linked with 88 LTR loci
NURSA Human Endogenous Complexome network of 240 genes
SILAC Phosphoproteomics network of 240 genes
H
OMIM Expanded PPI database of human disease-causing genes
240 genes PPI network
51-gene signature of human diseases' PPI networks

**Figure 6.**

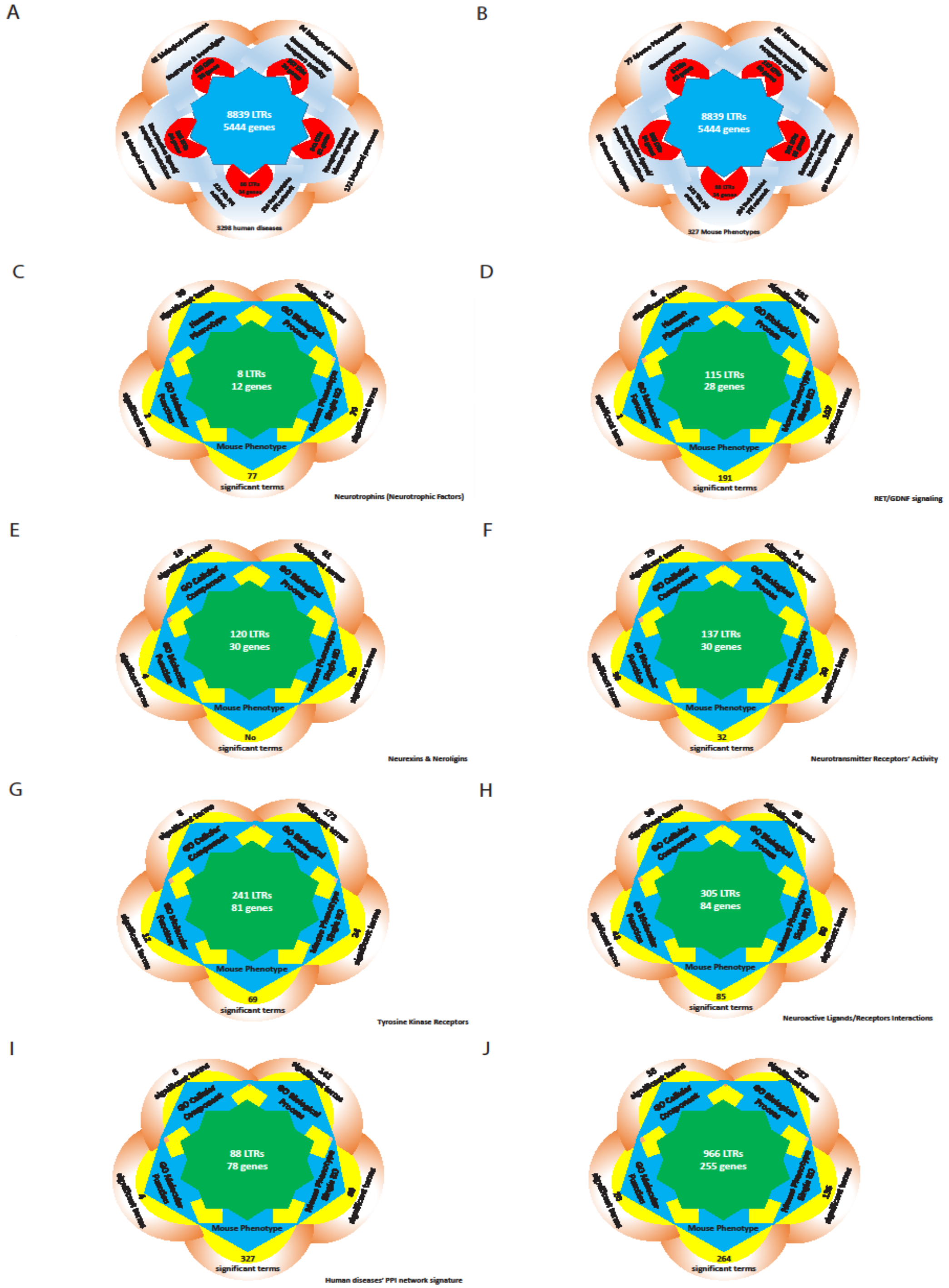

**Figure 7.**

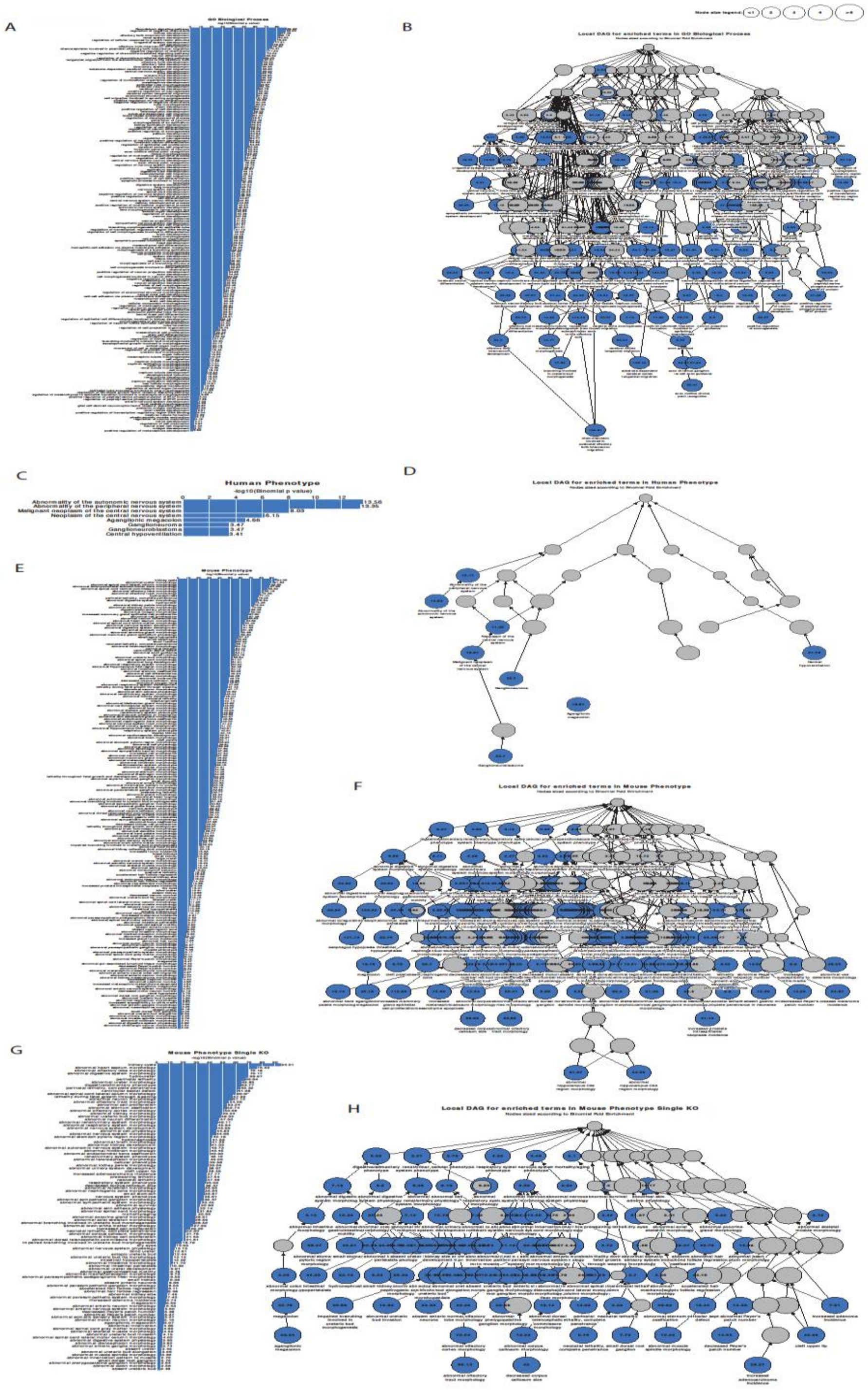

A
GO Biological Process
B
Local DAG for enriched terms in GO Biological Process
C
Human Phenotype
-log10(Binomial p value)
0
2
4
6
8
10
12
Abnormality of the autonomic nervous system
Abnormality of the peripheral nervous system
Malignant neoplasm of the central nervous system
Neoplasm of the central nervous system
Aganglionic megacolon
Ganglioneuroma
Ganglioneuroblastoma
Central hypoventilation
13.56
13.35
8.03
6.15
4.66
3.47
3.47
3.41
D
Local DAG for enriched terms in Human Phenotype
E
Mouse Phenotype
F
Local DAG for enriched terms in Mouse Phenotype
G
Mouse Phenotype Single KO
H
Local DAG for enriched terms in Mouse Phenotype Single KO

**Figure 8.**

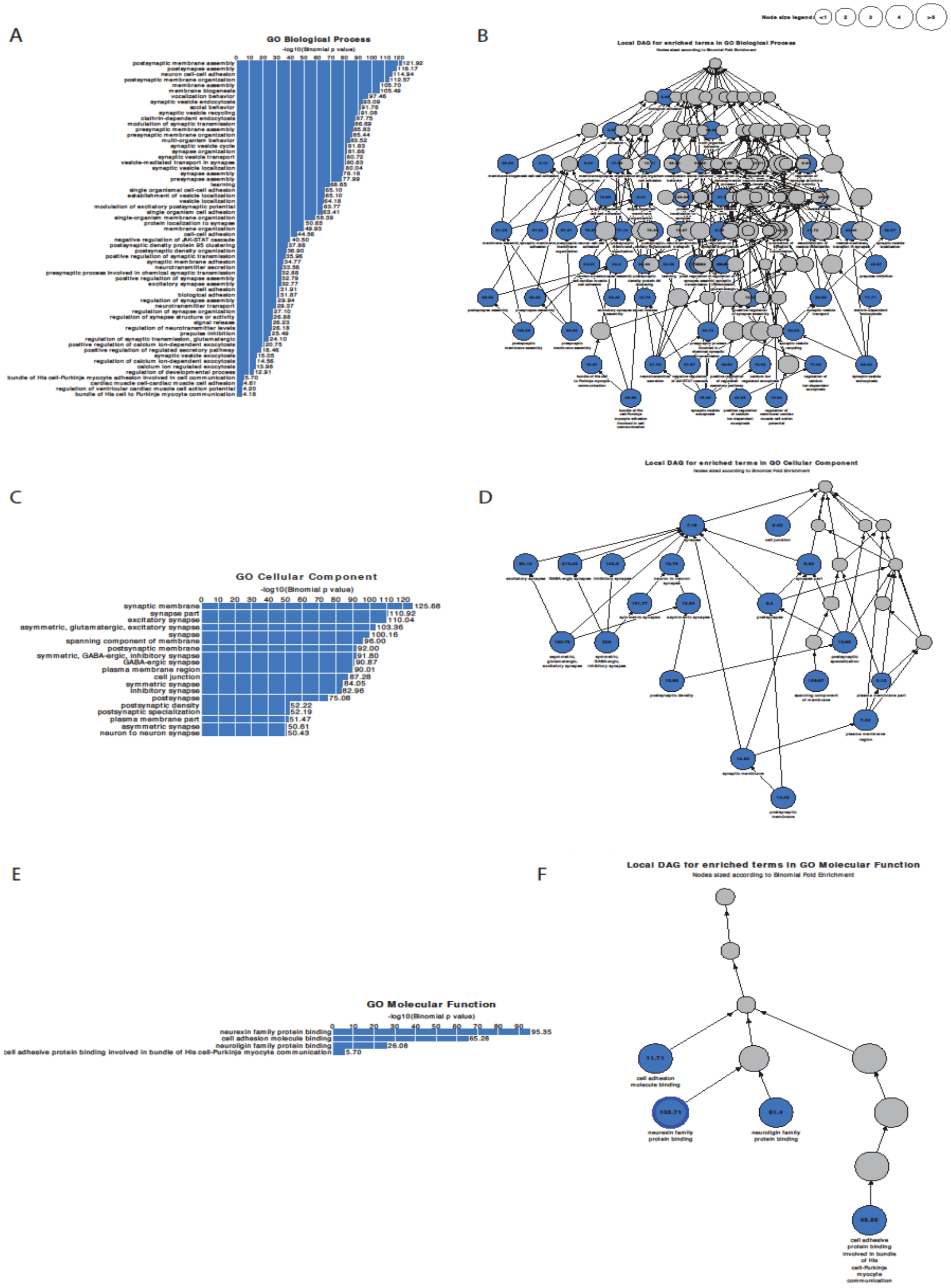

A
GO Biological Process
-log10(Binomial p value)
0 10 20 30 40 50 60 70 80 90 100 110 120
postsynaptic membrane assembly 121.92
postsynapse assembly 118.17
neuron cell-cell adhesion 114.94
postsynaptic membrane organization 112.57
membrane assembly 105.70
membrane biogenesis 105.49
vocalization behavior 97.46
synaptic vesicle endocytosis 93.09
social behavior 91.76
synaptic vesicle recycling 91.08
clathrin-dependent endocytosis 87.75
modulation of synaptic transmission 86.69
presynaptic membrane assembly 85.83
presynaptic membrane organization 85.44
multi-organism behavior 83.52
synaptic vesicle cycle 81.83
synapse organization 81.66
synaptic vesicle transport 80.72
vesicle-mediated transport in synapse 80.63
synaptic vesicle localization 80.04
synapse assembly 78.18
presynapse assembly 77.99
learning 68.65
single organismal cell-cell adhesion 65.10
establishment of vesicle localization 65.10
vesicle localization 64.18
modulation of excitatory postsynaptic potential 63.77
single organism cell adhesion 63.41
single-organism membrane organization 58.39
protein localization to synapse 50.65
membrane organization 49.93
cell-cell adhesion 44.56
negative regulation of JAK-STAT cascade 40.50
postsynaptic density protein 95 clustering 37.88
postsynaptic density organization 36.90
positive regulation of synaptic transmission 35.96
synaptic membrane adhesion 34.77
neurotransmitter secretion 33.56
presynaptic process involved in chemical synaptic transmission 32.88
positive regulation of synapse assembly 32.79
excitatory synapse assembly 32.77
cell adhesion 31.91
biological adhesion 31.87
regulation of synapse assembly 29.94
neurotransmitter transport 29.37
regulation of synapse organization 27.10
regulation of synapse structure or activity 26.88
signal release 26.23
regulation of neurotransmitter levels 26.18
prepulse inhibition 25.49
regulation of synaptic transmission, glutamatergic 24.10
positive regulation of calcium ion-dependent exocytosis 20.75
positive regulation of regulated secretory pathway 18.46
synaptic vesicle exocytosis 15.05
regulation of calcium ion-dependent exocytosis 14.56
calcium ion regulated exocytosis 13.98
regulation of developmental process 12.91
bundle of His cell-Purkinje myocyte adhesion involved in cell communication 5.70
cardiac muscle cell-cardiac muscle cell adhesion 4.61
regulation of ventricular cardiac muscle cell action potential 4.20
bundle of His cell to Purkinje myocyte communication 4.18
B
Local DAG for enriched terms in GO Biological Process
C
GO Cellular Component
-log10(Binomial p value)
0 10 20 30 40 50 60 70 80 90 100 110 120
synaptic membrane 125.68
synapse part 110.92
excitatory synapse 110.04
asymmetric, glutamatergic, excitatory synapse 103.36
synapse 100.16
spanning component of membrane 96.00
postsynaptic membrane 92.00
symmetric, GABA-ergic, inhibitory synapse 91.80
GABA-ergic synapse 90.87
plasma membrane region 90.01
cell junction 87.28
symmetric synapse 84.05
inhibitory synapse 82.96
postsynapse 75.08
postsynaptic density 52.22
postsynaptic specialization 52.19
plasma membrane part 51.47
asymmetric synapse 50.61
neuron to neuron synapse 50.43
D
Local DAG for enriched terms in GO Cellular Component
Nodes sized according to Binomial Fold Enrichment
E
GO Molecular Function
-log10(Binomial p value)
0 10 20 30 40 50 60 70 80 90
neurexin family protein binding 95.35
cell adhesion molecule binding 65.28
neuroligin family protein binding 26.08
cell adhesive protein binding involved in bundle of His cell-Purkinje myocyte communication 5.70
F
Local DAG for enriched terms in GO Molecular Function
Nodes sized according to Binomial Fold Enrichment
cell adhesion molecule binding
neurexin family protein binding
neuroligin family protein binding
cell adhesive protein binding involved in bundle of His cell-Purkinje myocyte communication

**Figure 9.**

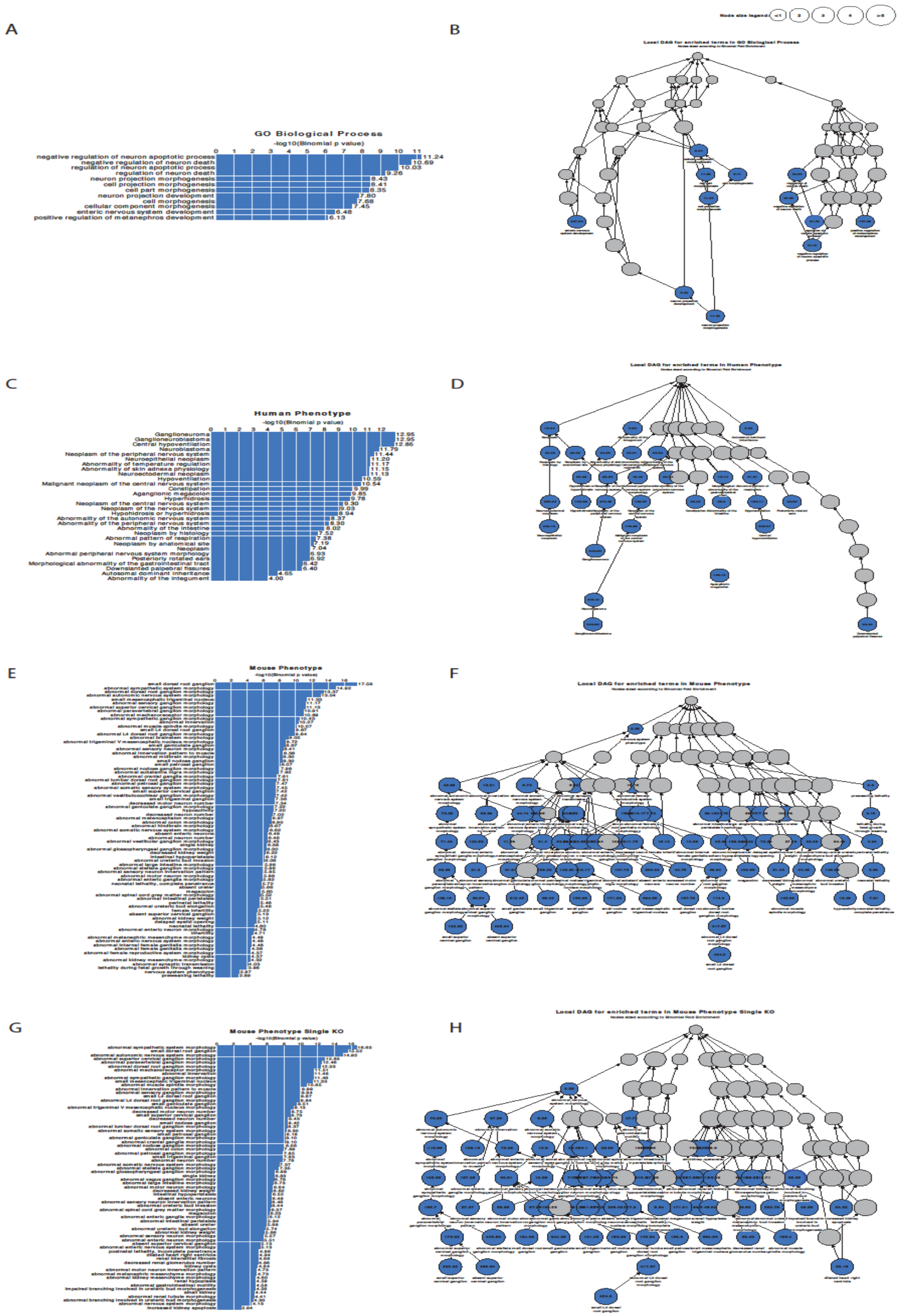
A
B
GO Biological Process
-log10(Binomial p value)
0 1 2 3 4 5 6 7 8 9 10 11
negative regulation of neuron apoptotic process 11.24
negative regulation of neuron death 10.69
regulation of neuron apoptotic process 10.03
regulation of neuron death 9.26
neuron projection morphogenesis 8.43
cell projection morphogenesis 8.41
cell part morphogenesis 8.35
neuron projection development 7.80
cell morphogenesis 7.68
cellular component morphogenesis 7.45
enteric nervous system development 6.48
positive regulation of metanephros development 6.13
Local DAG for enriched terms in GO Biological Process
C
D
Human Phenotype
-log10(Binomial p value)
0 1 2 3 4 5 6 7 8 9 10 11 12
Ganglioneuroma 12.95
Ganglioneuroblastoma 12.95
Central hypoventilation 12.86
Neuroblastoma 11.79
Neoplasm of the peripheral nervous system 11.44
Neuroepithelial neoplasm 11.20
Abnormality of temperature regulation 11.17
Abnormality of skin adnexa physiology 11.15
Neuroectodermal neoplasm 11.13
Hypoventilation 10.59
Malignant neoplasm of the central nervous system 10.54
Constipation 9.99
Aganglionic megacolon 9.85
Hyperhidrosis 9.78
Neoplasm of the central nervous system 9.30
Neoplasm of the nervous system 9.03
Hypohidrosis or hyperhidrosis 8.94
Abnormality of the autonomic nervous system 8.37
Abnormality of the peripheral nervous system 8.30
Abnormality of the intestine 8.02
Neoplasm by histology 7.52
Abnormal pattern of respiration 7.38
Neoplasm by anatomical site 7.19
Neoplasm 7.04
Abnormal peripheral nervous system morphology 6.93
Posteriorly rotated ears 6.92
Morphological abnormality of the gastrointestinal tract 6.42
Downslanted palpebral fissures 6.40
Autosomal dominant inheritance 4.65
Abnormality of the integument 4.00
Local DAG for enriched terms in Human Phenotype
E
F
Mouse Phenotype
Local DAG for enriched terms in Mouse Phenotype
G
H
Mouse Phenotype Single KO
Local DAG for enriched terms in Mouse Phenotype Single KO

**Figure 10.**

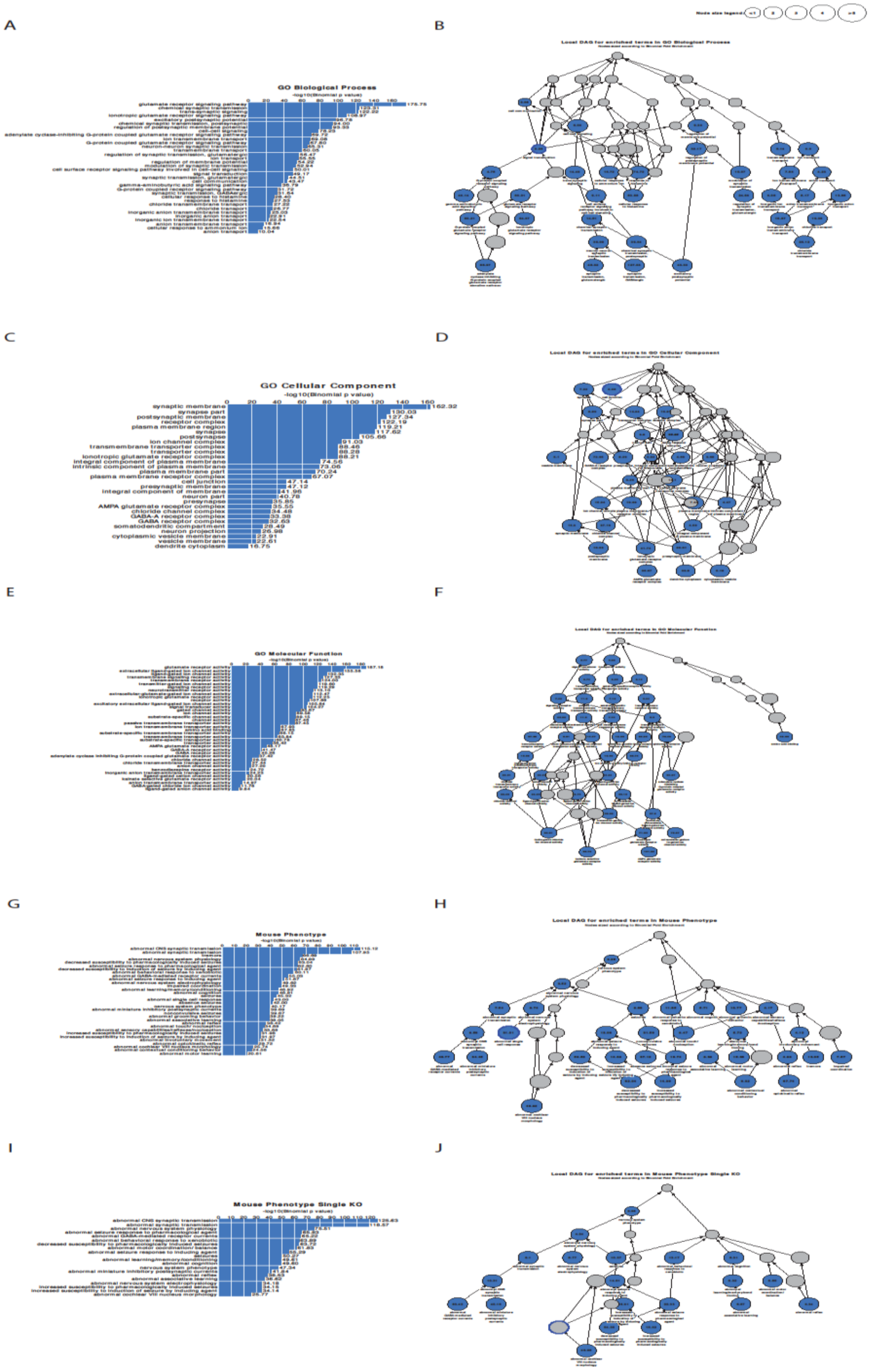

**Figure 11.**

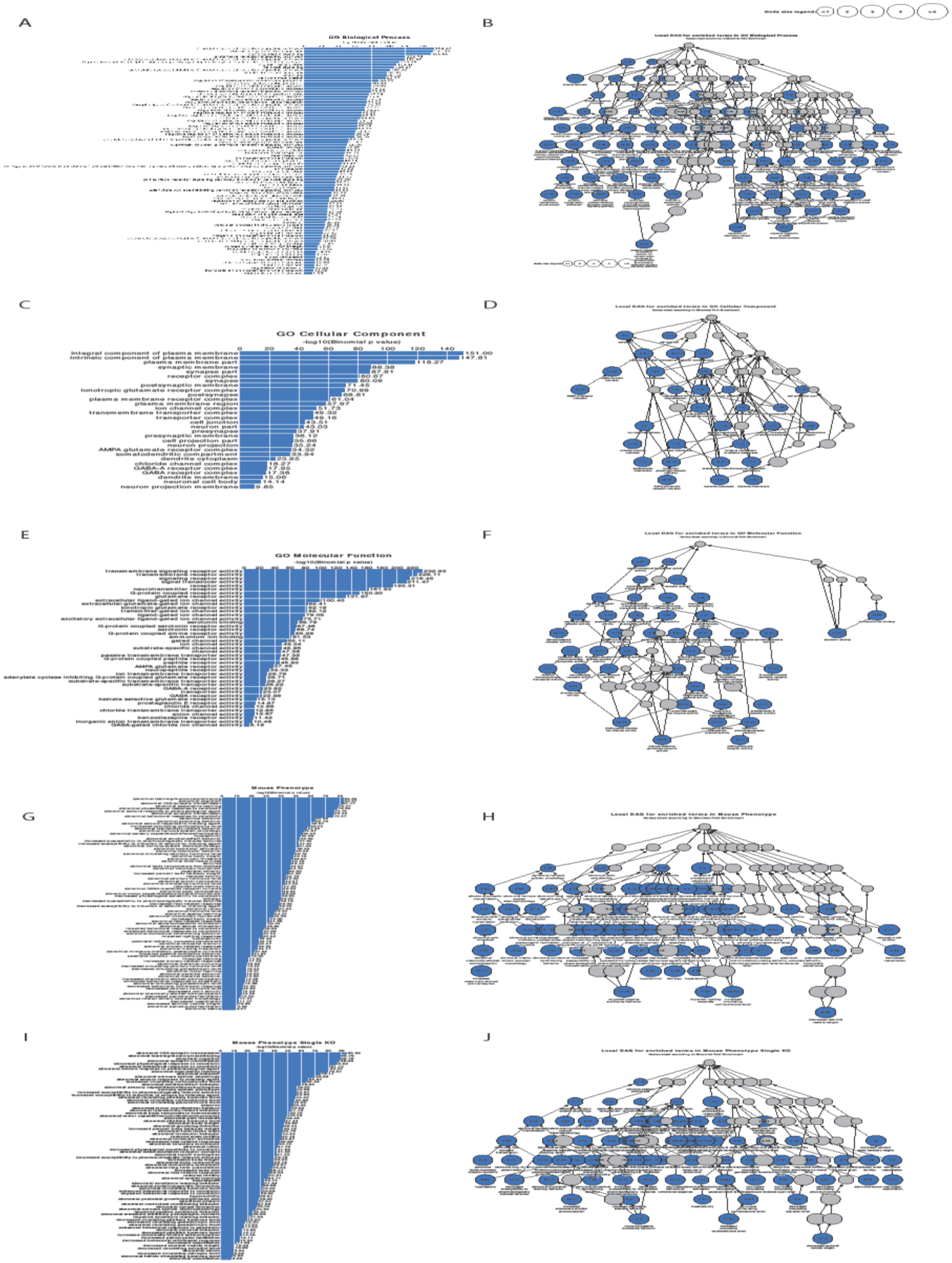

A
GO Biological Process
B
Local DAG for enriched terms in GO Biological Process
C
GO Cellular Component
-log10(Binomial p value)
integral component of plasma membrane
intrinsic component of plasma membrane
plasma membrane part
synaptic membrane
synapse part
receptor complex
synapse
postsynaptic membrane
ionotropic glutamate receptor complex
postsynapse
plasma membrane receptor complex
plasma membrane region
ion channel complex
transmembrane transporter complex
transporter complex
cell junction
neuron part
presynapse
presynaptic membrane
cell projection part
neuron projection
AMPA glutamate receptor complex
somatodendritic compartment
dendrite cytoplasm
chloride channel complex
GABA-A receptor complex
GABA receptor complex
dendrite membrane
neuronal cell body
neuron projection membrane
D
Local DAG for enriched terms in GO Cellular Component
E
GO Molecular Function
F
Local DAG for enriched terms in GO Molecular Function
G
Mouse Phenotype
H
Local DAG for enriched terms in Mouse Phenotype
I
Mouse Phenotype Single KO
J
Local DAG for enriched terms in Mouse Phenotype Single KO

**Figure 12.**

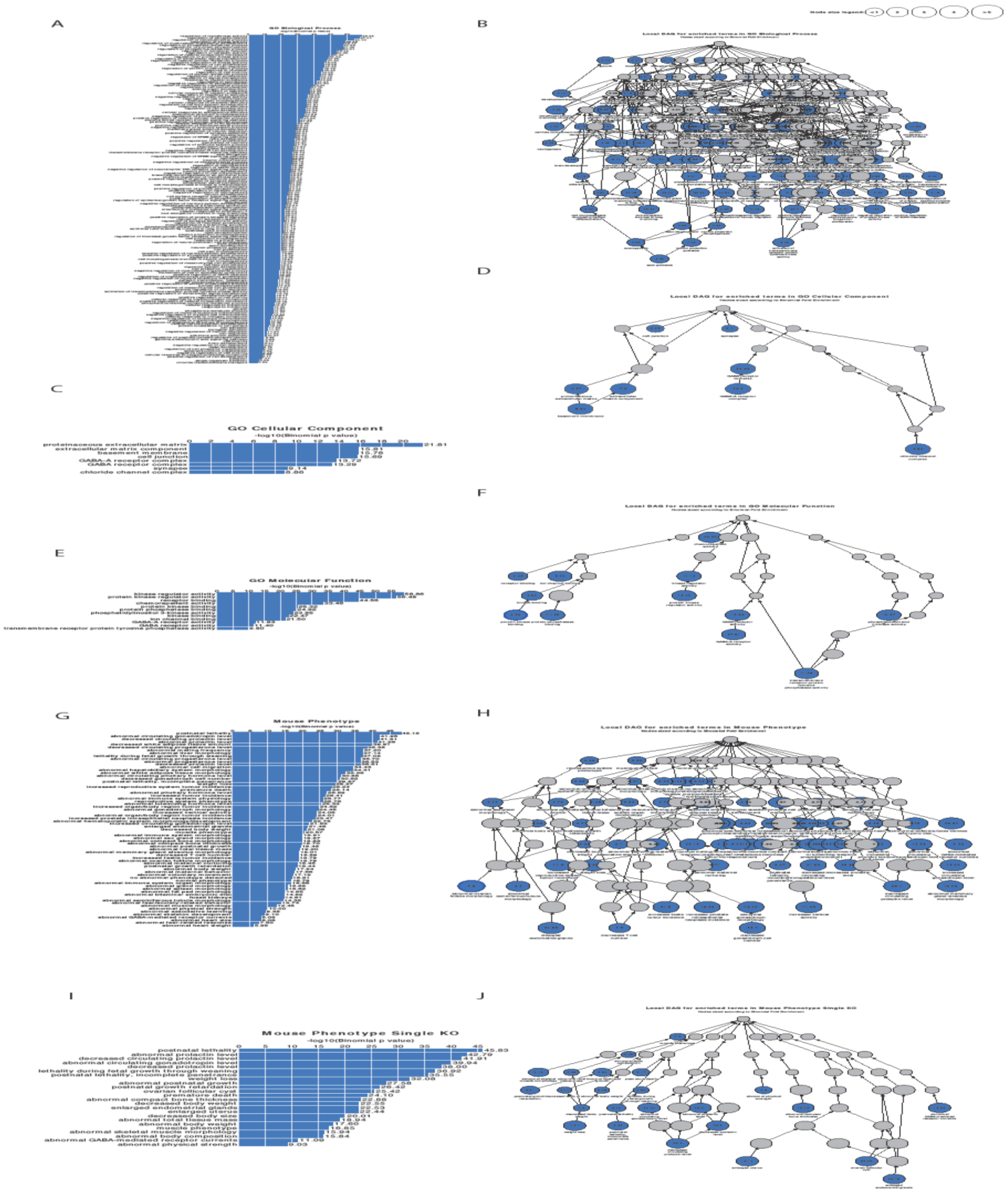

**Figure 13.**

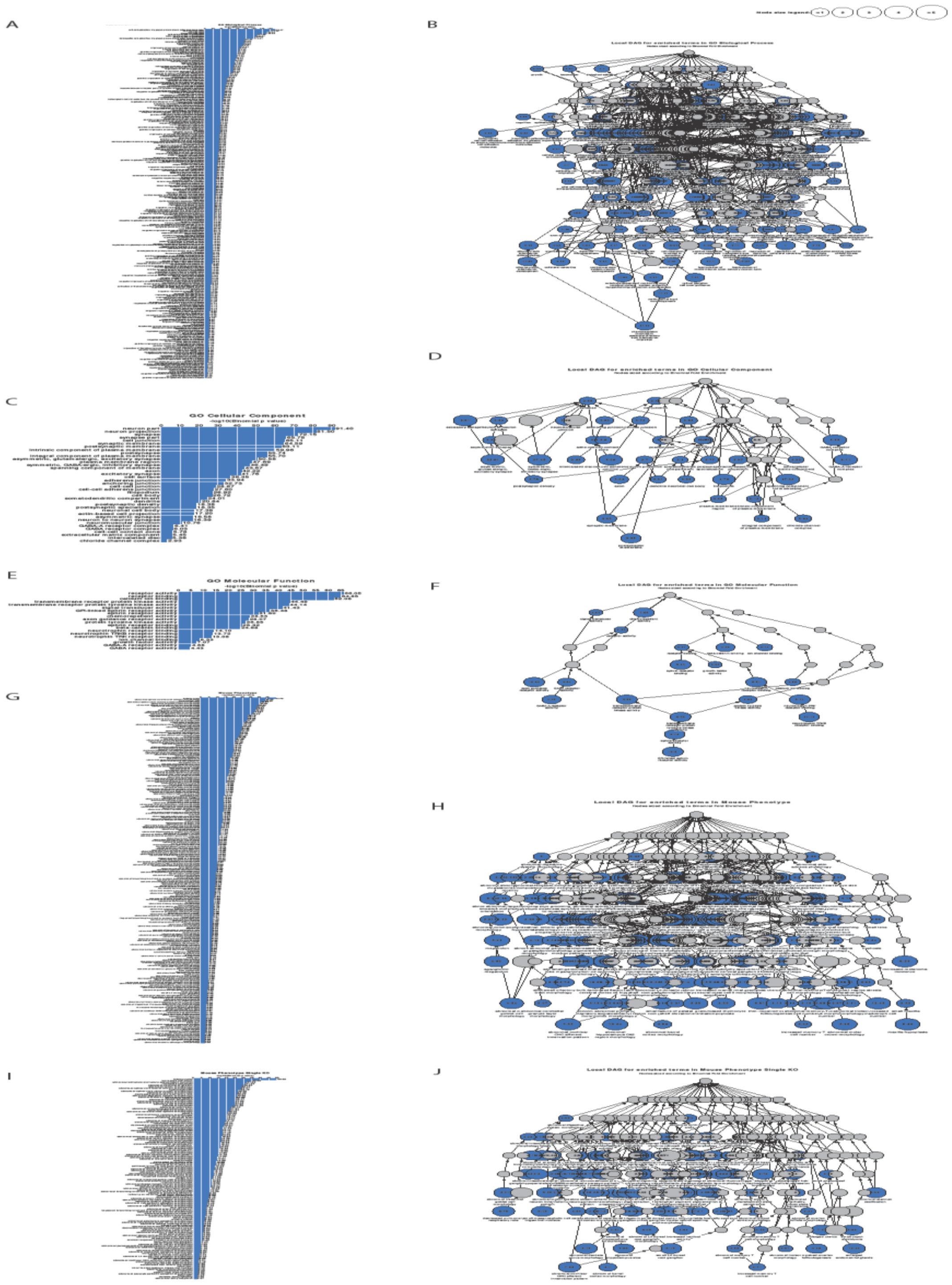

**Figure 14.**

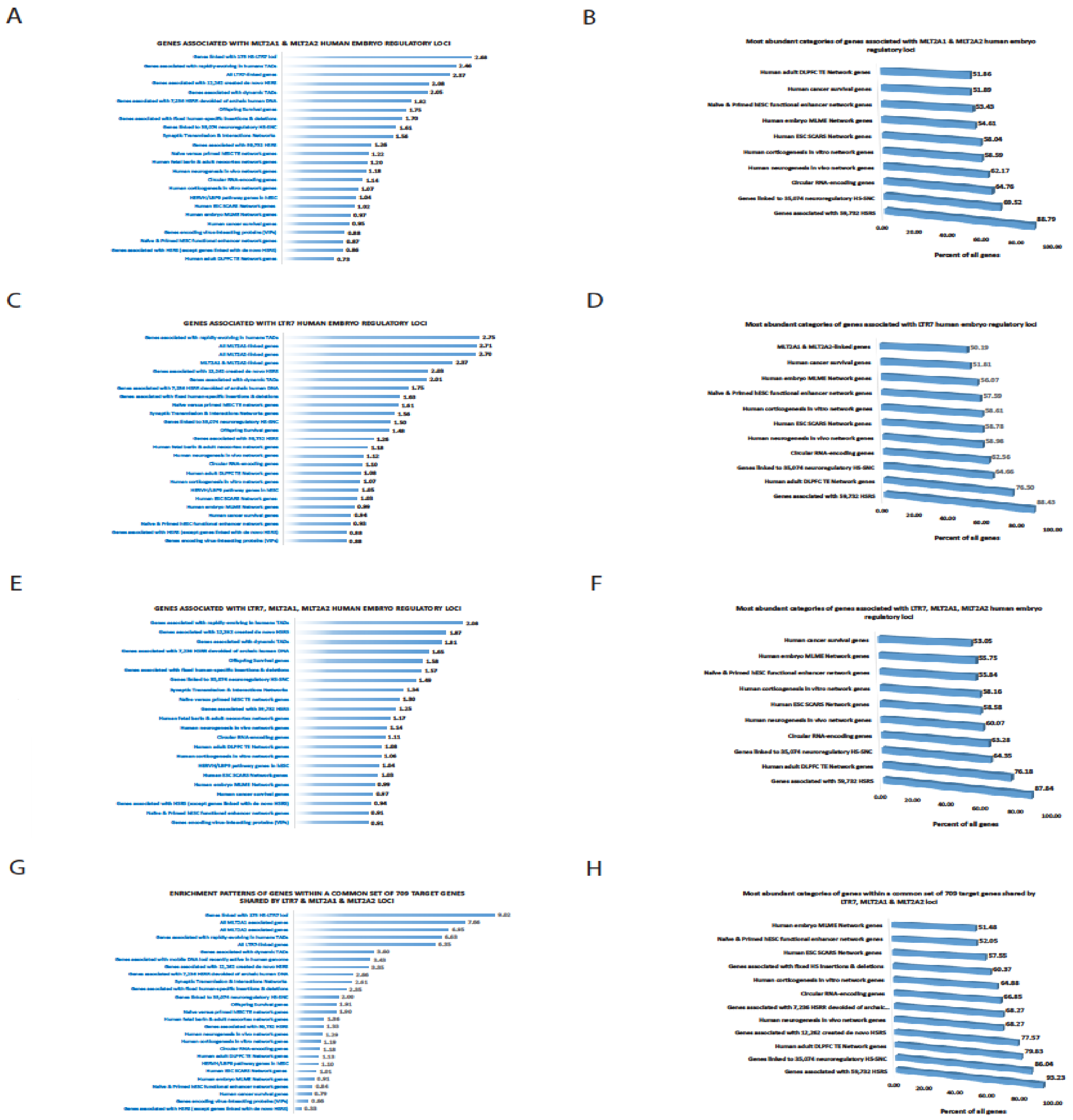
A
GENES ASSOCIATED WITH MLT2A1 & MLT2A2 HUMAN EMBRYO REGULATORY LOCI
B
Most abundant categories of genes associated with MLT2A1 & MLT2A2 human embryo regulatory loci
Percent of all genes
C
GENES ASSOCIATED WITH LTR7 HUMAN EMBRYO REGULATORY LOCI
D
Most abundant categories of genes associated with LTR7 human embryo regulatory loci
Percent of all genes
E
GENES ASSOCIATED WITH LTR7, MLT2A1, MLT2A2 HUMAN EMBRYO REGULATORY LOCI
F
Most abundant categories of genes associated with LTR7, MLT2A1, MLT2A2 human embryo regulatory loci
Percent of all genes
G
ENRICHMENT PATTERNS OF GENES WITHIN A COMMON SET OF 709 TARGET GENES SHARED BY LTR7 & MLT2A1 & MLT2A2 LOCI
H
Most abundant categories of genes within a common set of 709 target genes shared by LTR7, MLT2A1 & MLT2A2 loci
Percent of all genes

**Figure 15.**

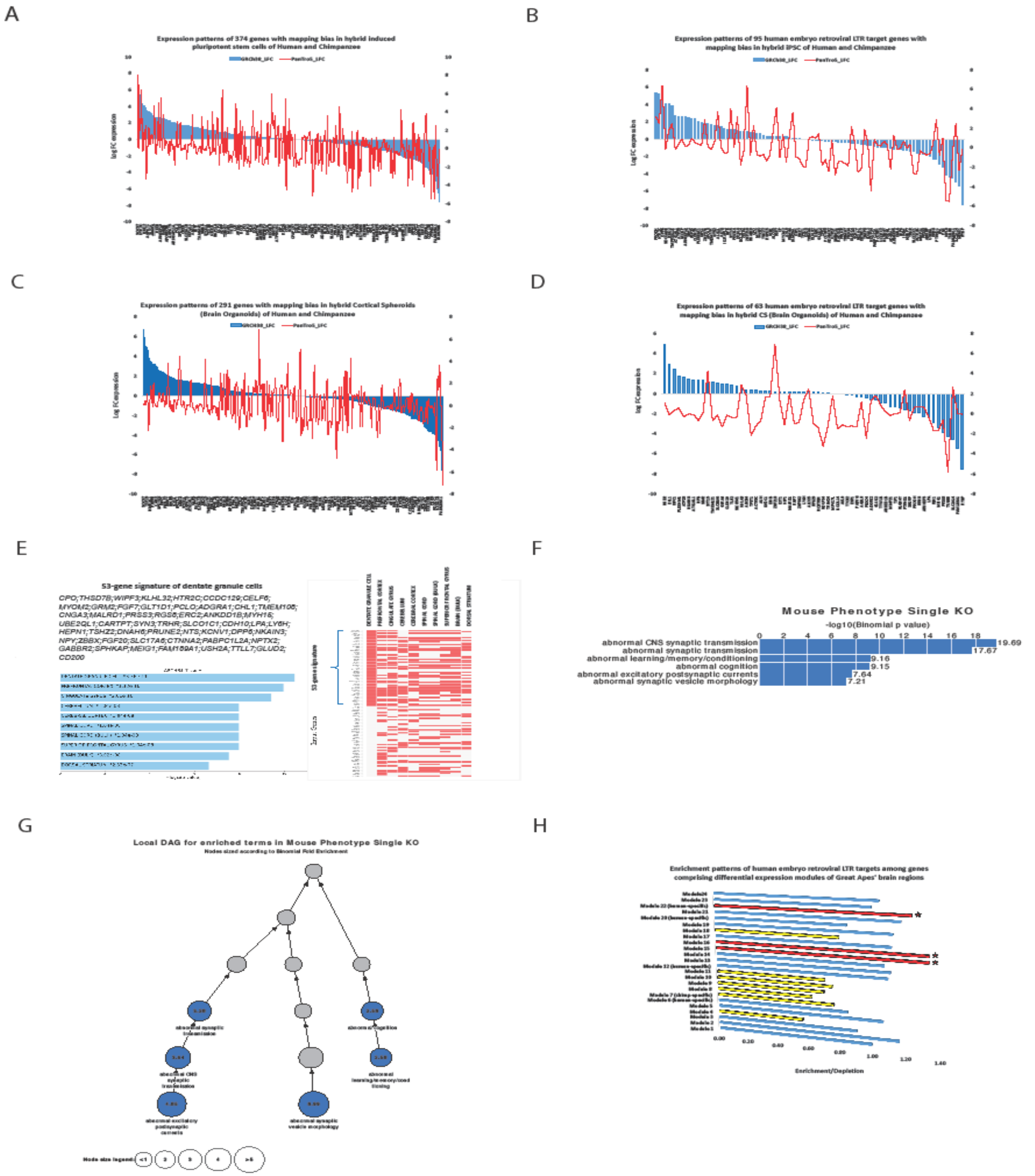
A
Expression patterns of 374 genes with mapping bias in hybrid induced pluripotent stem cells of Human and Chimpanzee
B
Expression patterns of 95 human embryo retroviral LTR target genes with mapping bias in hybrid iPSC of Human and Chimpanzee
C
Expression patterns of 291 genes with mapping bias in hybrid Cortical Spheroids (Brain Organoids) of Human and Chimpanzee
D
Expression patterns of 63 human embryo retroviral LTR target genes with mapping bias in hybrid CS (Brain Organoids) of Human and Chimpanzee
E
53-gene signature of dentate granule cells
CPO;THSD7B;WIPF3;KLHL32;HTR2C;CCDC129;CELF6;MYOM2;GRM2;FGF7;GLT1D1;PCLO;ADGRA1;CHL1;TMEM108;CNGA3;MALRD1;PRSS3;RGS8;ERC2;ANKDD1B;MYH15;UBE2QL1;CARTPT;SYN3;TRHR;SLCO1C1;CDH10;LPA;LY6H;HEPN1;TSHZ2;DNAH6;PRUNE2;NTS;KCNV1;DPP6;NKAIN3;NPY;ZBBX;FGF20;SLC17A6;CTNNA2;PABPC1L2A;NPTX2;GABBR2;SPHKAP;MEIG1;FAM189A1;USH2A;TTLL7;GLUD2;CD200
F
Mouse Phenotype Single KO
-log10(Binomial p value)
abnormal CNS synaptic transmission 19.69
abnormal synaptic transmission 17.67
abnormal learning/memory/conditioning 9.16
abnormal cognition 9.15
abnormal excitatory postsynaptic currents 7.64
abnormal synaptic vesicle morphology 7.21
G
Local DAG for enriched terms in Mouse Phenotype Single KO
Nodes sized according to Binomial Fold Enrichment
H
Enrichment patterns of human embryo retroviral LTR targets among genes comprising differential expression modules of Great Apes' brain regions
Enrichment/Depletion

Figure 16.